\documentclass[apjl,iop]{emulateapj}
\usepackage{comment}
\usepackage{ifthen}
\usepackage[switch]{lineno}

\newcommand{\forloop}[5][1]%
{%
\setcounter{#2}{#3}%
\ifthenelse{#4}%
	{%
	#5%
	\addtocounter{#2}{#1}%
	\forloop[#1]{#2}{\value{#2}}{#4}{#5}%
	}%
	{%
	}%
}%

\newcommand{\ctbd}[1]{}

\newcommand{\lc}{light curve}
\newcommand{\lcs}{light curves}
\newcommand{\Lc}{Light curve}

\newcommand{\band}[1]{\ensuremath{#1}~band}

\newcommand{\masy}{\ensuremath{\rm mas\,yr^{-1}}}
\newcommand{\kms}{\ensuremath{\rm km\,s^{-1}}}
\newcommand{\ms}{\ensuremath{\rm m\,s^{-1}}}

\newcommand{\gcmc}{\ensuremath{\rm g\,cm^{-3}}}

\newcommand{\vsini}{\ensuremath{v \sin{i}}}
\newcommand{\feh}{\ensuremath{\rm [Fe/H]}}

\newcommand{\vmac}{\ensuremath{v_{\rm mac}}}
\newcommand{\vmic}{\ensuremath{v_{\rm mic}}}

\newcommand{\rsun}{\ensuremath{R_\sun}}
\newcommand{\msun}{\ensuremath{M_\sun}}
\newcommand{\lsun}{\ensuremath{L_\sun}}

\newcommand{\rstar}{\ensuremath{R_\star}}
\newcommand{\mstar}{\ensuremath{M_\star}}
\newcommand{\lstar}{\ensuremath{L_\star}}

\newcommand{\teffstar}{\ensuremath{T_{\rm eff\star}}}
\newcommand{\rhostar}{\ensuremath{\rho_\star}}
\newcommand{\loggstar}{\ensuremath{\log{g_{\star}}}}

\newcommand{\rpl}{\ensuremath{R_{p}}}
\newcommand{\mpl}{\ensuremath{M_{p}}}

\newcommand{\rhopl}{\ensuremath{\rho_{p}}}

\newcommand{\arstar}{\ensuremath{a/\rstar}}
\newcommand{\zrstar}{\ensuremath{\zeta/\rstar}}

\newcommand{\rjup}{\ensuremath{R_{\rm J}}}
\newcommand{\mjup}{\ensuremath{M_{\rm J}}}

\newcommand{\reffigl}[1]{Figure~\ref{fig:#1}}
\newcommand{\refsecl}[1]{\mbox{Section \ref{sec:#1}}}

\newcommand{\reftabl}[1]{Table~\ref{tab:#1}}

\newcommand{\loopand}{\ifnum\value{planetcounter}=2 and \else\fi}
\newcommand{\loopcomma}{\ifnum\value{planetcounter}<2 ,\else. \fi}
\newcommand{\loopcommanoperiod}{\ifnum\value{planetcounter}<2 ,\else \space\fi}
\newcommand{\loopcommanospace}{\ifnum\value{planetcounter}<2 ,\else \fi}

\newcommand{\hatcurhtrxxxxxA}{HATS579-017}                             
\newcommand{\hatcurfieldxxxxxA}{579}                                   
\newcommand{\hatcurCCraxxxxxA}{\ensuremath{19^{\mathrm h}23^{\mathrm m}14.28{\mathrm s}}}                            
\newcommand{\hatcurCCdecxxxxxA}{\ensuremath{-20{\arcdeg}09{\arcmin}58.7{\arcsec}}}                           
\newcommand{\hatcurCCmagxxxxxA}{13.276}                                
\newcommand{\hatcurCCtwomassxxxxxA}{2MASS~19231442-2009587}            
\newcommand{\hatcurCCgscxxxxxA}{GSC~6305-02502}                        
\newcommand{\hatcurCCtassmvxxxxxA}{\ensuremath{13.276\pm0.010}}        
\newcommand{\hatcurCCtassmvshortxxxxxA}{\ensuremath{13.3}}             
\newcommand{\hatcurCCtassmBxxxxxA}{\ensuremath{14.080\pm0.010}}        
\newcommand{\hatcurCCtassmBshortxxxxxA}{\ensuremath{14.1}}             
\newcommand{\hatcurCCtassmIxxxxxA}{\ensuremath{100\pm1000}}            
\newcommand{\hatcurCCtassmIshortxxxxxA}{\ensuremath{100.0}}            
\newcommand{\hatcurCCtassmgxxxxxA}{\ensuremath{13.629\pm0.010}}        
\newcommand{\hatcurCCtassmgshortxxxxxA}{\ensuremath{13.6}}             
\newcommand{\hatcurCCtassmrxxxxxA}{\ensuremath{13.072\pm0.030}}        
\newcommand{\hatcurCCtassmrshortxxxxxA}{\ensuremath{13.1}}             
\newcommand{\hatcurCCtassmixxxxxA}{\ensuremath{12.865\pm0.010}}        
\newcommand{\hatcurCCtassmishortxxxxxA}{\ensuremath{12.9}}             
\newcommand{\hatcurCCtwomassJmagxxxxxA}{\ensuremath{11.885\pm0.022}}   
\newcommand{\hatcurCCtwomassHmagxxxxxA}{\ensuremath{11.558\pm0.027}}   
\newcommand{\hatcurCCtwomassKmagxxxxxA}{\ensuremath{11.479\pm0.022}}   
\newcommand{\hatcurCCcitJmagxxxxxA}{\ensuremath{11.900\pm0.023}}       
\newcommand{\hatcurCCcitHmagxxxxxA}{\ensuremath{11.553\pm0.027}}       
\newcommand{\hatcurCCcitKmagxxxxxA}{\ensuremath{11.503\pm0.022}}       
\newcommand{\hatcurCCbbJmagxxxxxA}{\ensuremath{11.952\pm0.024}}        
\newcommand{\hatcurCCbbHmagxxxxxA}{\ensuremath{11.574\pm0.028}}        
\newcommand{\hatcurCCbbKmagxxxxxA}{\ensuremath{11.523\pm0.022}}        
\newcommand{\hatcurCCesoJmagxxxxxA}{\ensuremath{11.955\pm0.025}}       
\newcommand{\hatcurCCesoHmagxxxxxA}{\ensuremath{11.569\pm0.031}}       
\newcommand{\hatcurCCesoKmagxxxxxA}{\ensuremath{11.522\pm0.023}}       
\newcommand{\hatcurCCesoJHmagxxxxxA}{\ensuremath{0.386\pm0.039}}       
\newcommand{\hatcurCCesoJKmagxxxxxA}{\ensuremath{0.433\pm0.010}}       
\newcommand{\hatcurCCesoHKmagxxxxxA}{\ensuremath{0.047\pm0.038}}       
\newcommand{\hatcurLCdipxxxxxA}{\ensuremath{8.2}}                      
\newcommand{\hatcurLCrprstarxxxxxA}{\ensuremath{0.0725\pm0.0041}}      
\newcommand{\hatcurLCbsqxxxxxA}{\ensuremath{0.071_{-0.050}^{+0.112}}}  
\newcommand{\hatcurLCimpxxxxxA}{\ensuremath{0.27_{-0.12}^{+0.16}}}     
\newcommand{\hatcurLCzetaxxxxxA}{\ensuremath{14.84\pm0.26}}            
\newcommand{\hatcurLCdurxxxxxA}{\ensuremath{0.1457\pm0.0024}}          
\newcommand{\hatcurLCdurshortxxxxxA}{\ensuremath{0.1457}}              
\newcommand{\hatcurLCdurhrxxxxxA}{\ensuremath{3.496\pm0.057}}          
\newcommand{\hatcurLCdurhrshortxxxxxA}{\ensuremath{3.496}}             
\newcommand{\hatcurLCqxxxxxA}{\ensuremath{0.0761\pm0.0012}}            
\newcommand{\hatcurLCqshortxxxxxA}{\ensuremath{0.076}}                 
\newcommand{\hatcurLCingdurxxxxxA}{\ensuremath{0.0106\pm0.0015}}       
\newcommand{\hatcurLCPxxxxxA}{\ensuremath{1.9153073\pm0.0000052}}      
\newcommand{\hatcurLCPprecxxxxxA}{\ensuremath{1.9153073}}              
\newcommand{\hatcurLCPshortxxxxxA}{\ensuremath{1.9153}}                
\newcommand{\hatcurLCTxxxxxA}{\ensuremath{2456124.25896\pm0.00086}}    
\newcommand{\hatcurLCTAxxxxxA}{\ensuremath{2455270.0321\pm0.0022}}     
\newcommand{\hatcurLCTBxxxxxA}{\ensuremath{2456444.1153\pm0.0014}}     
\newcommand{\hatcurLChatnetmxxxxxA}{\ensuremath{13.020710\pm0.000069}} 
\newcommand{\hatcurLCiblendxxxxxA}{\ensuremath{1.01\pm0.12}}           
\newcommand{\hatcurSMEiteffxxxxxA}{\ensuremath{5363\pm90}}             
\newcommand{\hatcurSMEizfehxxxxxA}{\ensuremath{0.330\pm0.090}}         
\newcommand{\hatcurSMEizfehshortxxxxxA}{\ensuremath{0.33}}             
\newcommand{\hatcurSMEiloggxxxxxA}{\ensuremath{3.97\pm0.20}}           
\newcommand{\hatcurSMEivsinxxxxxA}{\ensuremath{4.67\pm0.50}}           
\newcommand{\hatcurSMEivmacxxxxxA}{\ensuremath{0.0}}                   
\newcommand{\hatcurSMEivmicxxxxxA}{\ensuremath{0.0}}                   
\newcommand{\hatcurSMEiiteffxxxxxA}{\ensuremath{5366\pm70}}            
\newcommand{\hatcurSMEiizfehxxxxxA}{\ensuremath{0.340\pm0.050}}        
\newcommand{\hatcurSMEiizfehshortxxxxxA}{\ensuremath{0.34}}            
\newcommand{\hatcurSMEiiloggxxxxxA}{\ensuremath{4.120\pm0.040}}        
\newcommand{\hatcurSMEiivsinxxxxxA}{\ensuremath{4.58\pm0.90}}          
\newcommand{\hatcurLBizxxxxxA}{\ensuremath{0.2713}}                    
\newcommand{\hatcurLBiizxxxxxA}{\ensuremath{0.3052}}                   
\newcommand{\hatcurLBiixxxxxA}{\ensuremath{0.3533}}                    
\newcommand{\hatcurLBiiixxxxxA}{\ensuremath{0.2892}}                   
\newcommand{\hatcurLBiIxxxxxA}{\ensuremath{0.3265}}                    
\newcommand{\hatcurLBiiIxxxxxA}{\ensuremath{0.2945}}                   
\newcommand{\hatcurLBigxxxxxA}{\ensuremath{0.7006}}                    
\newcommand{\hatcurLBiigxxxxxA}{\ensuremath{0.1231}}                   
\newcommand{\hatcurLBirxxxxxA}{\ensuremath{0.4688}}                    
\newcommand{\hatcurLBiirxxxxxA}{\ensuremath{0.2596}}                   
\newcommand{\hatcurLBiRxxxxxA}{\ensuremath{0.4369}}                    
\newcommand{\hatcurLBiiRxxxxxA}{\ensuremath{0.2687}}                   
\newcommand{\hatcurISOmxxxxxA}{\ensuremath{1.030\pm0.039}}             
\newcommand{\hatcurISOmshortxxxxxA}{\ensuremath{1.03}}                 
\newcommand{\hatcurISOmlongxxxxxA}{\ensuremath{1.030\pm0.039}}         
\newcommand{\hatcurISOrxxxxxA}{\ensuremath{1.503_{-0.043}^{+0.101}}}   
\newcommand{\hatcurISOrshortxxxxxA}{\ensuremath{1.50}}                 
\newcommand{\hatcurISOrlongxxxxxA}{\ensuremath{1.503_{-0.043}^{+0.101}}} 
\newcommand{\hatcurISOrhoxxxxxA}{\ensuremath{0.427_{-0.070}^{+0.030}}} 
\newcommand{\hatcurISOrholongxxxxxA}{\ensuremath{0.427_{-0.070}^{+0.030}}} 
\newcommand{\hatcurISOloggxxxxxA}{\ensuremath{4.095\pm0.038}}          
\newcommand{\hatcurISOlumxxxxxA}{\ensuremath{1.70_{-0.16}^{+0.24}}}    
\newcommand{\hatcurISOlumshortxxxxxA}{\ensuremath{1.70}}               
\newcommand{\hatcurISOmvxxxxxA}{\ensuremath{4.33\pm0.15}}              
\newcommand{\hatcurISOvixxxxxA}{\ensuremath{0.826\pm0.018}}            
\newcommand{\hatcurISOagexxxxxA}{\ensuremath{10.8\pm1.5}}              
\newcommand{\hatcurISOsigmaxxxxxA}{\ensuremath{0.000500\pm0.000046}}   
\newcommand{\hatcurISOMJxxxxxA}{\ensuremath{2.96\pm0.13}}              
\newcommand{\hatcurISOMHxxxxxA}{\ensuremath{2.56\pm0.13}}              
\newcommand{\hatcurISOMKxxxxxA}{\ensuremath{2.49\pm0.13}}              
\newcommand{\hatcurISOJKxxxxxA}{\ensuremath{0.480\pm0.020}}            
\newcommand{\hatcurISOspecxxxxxA}{G}                                   
\newcommand{\hatcurRVKxxxxxA}{\ensuremath{133.5\pm3.4}}                
\newcommand{\hatcurRVrkxxxxxA}{\ensuremath{0\pm0}}                     
\newcommand{\hatcurRVrhxxxxxA}{\ensuremath{0\pm0}}                     
\newcommand{\hatcurRVkxxxxxA}{\ensuremath{0\pm0}}                      
\newcommand{\hatcurRVhxxxxxA}{\ensuremath{0\pm0}}                      
\newcommand{\hatcurRVtronexxxxxA}{\ensuremath{0\pm0}}                  
\newcommand{\hatcurRVtrtwoxxxxxA}{\ensuremath{0\pm0}}                  
\newcommand{\hatcurRVgammaAxxxxxA}{\ensuremath{-42.1\pm4.3}}           
\newcommand{\hatcurRVjitterAxxxxxA}{\ensuremath{0.1\pm5.2}}            
\newcommand{\hatcurRVfitrmsAxxxxxA}{\ensuremath{0.0}}                  
\newcommand{\hatcurRVgammaBxxxxxA}{\ensuremath{-10653.0\pm8.6}}        
\newcommand{\hatcurRVjitterBxxxxxA}{\ensuremath{0.0\pm1.7}}            
\newcommand{\hatcurRVfitrmsBxxxxxA}{\ensuremath{0.0}}                  
\newcommand{\hatcurRVgammaCxxxxxA}{\ensuremath{-10634\pm16}}           
\newcommand{\hatcurRVjitterCxxxxxA}{\ensuremath{0.0\pm1.1}}            
\newcommand{\hatcurRVfitrmsCxxxxxA}{\ensuremath{0.0}}                  
\newcommand{\hatcurRVeccenxxxxxA}{\ensuremath{0\pm0}}                  
\newcommand{\hatcurRVeccentwosiglimxxxxxA}{\ensuremath{<0.000}}        
\newcommand{\hatcurRVomegaxxxxxA}{\ensuremath{0\pm0}}                  
\newcommand{\hatcurPPixxxxxA}{\ensuremath{86.5_{-2.5}^{+1.6}}}         
\newcommand{\hatcurPPgxxxxxA}{\ensuremath{17.9_{-2.0}^{+2.7}}}         
\newcommand{\hatcurPPloggxxxxxA}{\ensuremath{3.253\pm0.068}}           
\newcommand{\hatcurPParxxxxxA}{\ensuremath{4.36_{-0.25}^{+0.10}}}      
\newcommand{\hatcurPParelxxxxxA}{\ensuremath{0.03048\pm0.00038}}       
\newcommand{\hatcurPPrhoxxxxxA}{\ensuremath{0.85\pm0.19}}              
\newcommand{\hatcurPPmxxxxxA}{\ensuremath{0.837\pm0.029}}              
\newcommand{\hatcurPPmshortxxxxxA}{\ensuremath{0.84}}                  
\newcommand{\hatcurPPmlongxxxxxA}{\ensuremath{0.837\pm0.029}}          
\newcommand{\hatcurPPmexxxxxA}{\ensuremath{266.1\pm9.3}}               
\newcommand{\hatcurPPmeshortxxxxxA}{\ensuremath{266.1}}                
\newcommand{\hatcurPPmelongxxxxxA}{\ensuremath{266.1\pm9.3}}           
\newcommand{\hatcurPPrxxxxxA}{\ensuremath{1.065\pm0.098}}              
\newcommand{\hatcurPPrshortxxxxxA}{\ensuremath{1.07}}                  
\newcommand{\hatcurPPrlongxxxxxA}{\ensuremath{1.065\pm0.098}}          
\newcommand{\hatcurPPrexxxxxA}{\ensuremath{11.9\pm1.1}}                
\newcommand{\hatcurPPreshortxxxxxA}{\ensuremath{11.9}}                 
\newcommand{\hatcurPPrelongxxxxxA}{\ensuremath{11.9\pm1.1}}            
\newcommand{\hatcurPPmrcorrxxxxxA}{\ensuremath{0.48}}                  
\newcommand{\hatcurPPteffxxxxxA}{\ensuremath{1823_{-35}^{+52}}}        
\newcommand{\hatcurPPthetaxxxxxA}{\ensuremath{0.0460\pm0.0039}}        
\newcommand{\hatcurPPfluxperixxxxxA}{\ensuremath{2.49_{-0.19}^{+0.30}}} 
\newcommand{\hatcurPPfluxperidimxxxxxA}{\ensuremath{9}}                
\newcommand{\hatcurPPfluxapxxxxxA}{\ensuremath{2.49_{-0.19}^{+0.30}}}  
\newcommand{\hatcurPPfluxapdimxxxxxA}{\ensuremath{9}}                  
\newcommand{\hatcurPPfluxavgxxxxxA}{\ensuremath{2.49_{-0.19}^{+0.30}}} 
\newcommand{\hatcurPPfluxavgdimxxxxxA}{\ensuremath{9}}                 
\newcommand{\hatcurPPfluxavglogxxxxxA}{\ensuremath{9.397_{-0.033}^{+0.049}}} 
\newcommand{\hatcurXsecphasexxxxxA}{\ensuremath{0\pm0}}                
\newcommand{\hatcurXsecondaryxxxxxA}{\ensuremath{2456125.21661\pm0.00086}} 
\newcommand{\hatcurXsecdurxxxxxA}{\ensuremath{0.1457\pm0.0024}}        
\newcommand{\hatcurXsecingdurxxxxxA}{\ensuremath{0.0106\pm0.0015}}     
\newcommand{\hatcurPPphiconjxxxxxA}{\ensuremath{0\pm0}}                
\newcommand{\hatcurPPperixxxxxA}{\ensuremath{2456123.78013\pm0.00086}} 
\newcommand{\hatcurPPaequivxxxxxA}{\ensuremath{0.02340_{-0.00130}^{+0.00090}}} 
\newcommand{\hatcurPPtcircxxxxxA}{\ensuremath{31.0_{-8.7}^{+11.5}}}    
\newcommand{\hatcurPPtinfallxxxxxA}{\ensuremath{34.7_{-8.2}^{+4.4}}}   
\newcommand{\hatcurXdistxxxxxA}{\ensuremath{642_{-22}^{+42}}}          
\newcommand{\hatcurXAvxxxxxA}{\ensuremath{0.000\pm0.011}}              
\newcommand{\hatcurXdistredxxxxxA}{\ensuremath{622_{-30}^{+42}}}       
\newcommand{\hatcurXEBVxxxxxA}{\ensuremath{0.0000\pm0.0037}}           
\newcommand{\hatcurXmvisoredxxxxxA}{\ensuremath{13.301\pm0.017}}       
\newcommand{\hatcurXmiisoredxxxxxA}{\ensuremath{12.4740\pm0.0089}}     
\newcommand{\hatcurXmjisoredxxxxxA}{\ensuremath{11.927\pm0.025}}       
\newcommand{\hatcurXmhisoredxxxxxA}{\ensuremath{11.516\pm0.038}}       
\newcommand{\hatcurXmkisoredxxxxxA}{\ensuremath{11.445\pm0.040}}       
\newcommand{\hatcurXviisoredxxxxxA}{\ensuremath{0.826\pm0.016}}        
\newcommand{\hatcurXvkisoredxxxxxA}{\ensuremath{1.855\pm0.054}}        
\newcommand{\hatcurXjhisoredxxxxxA}{\ensuremath{0.411\pm0.014}}        
\newcommand{\hatcurXjkisoredxxxxxA}{\ensuremath{0.481\pm0.016}}        
\newcommand{\hatcurCCpmraxxxxxA}{\ensuremath{0.3\pm4.3}}               
\newcommand{\hatcurCCpmdecxxxxxA}{\ensuremath{-1.9\pm2.8}}             
\newcommand{\hatcurCCpmxxxxxA}{\ensuremath{1.9\pm5.1}}                 

\newcommand{\hatcurhtrxxxxxB}{HATS579-026}                             
\newcommand{\hatcurfieldxxxxxB}{\ensuremath{string}}                   
\newcommand{\hatcurCCraxxxxxB}{\ensuremath{19^{\mathrm h}37^{\mathrm m}13.80{\mathrm s}}}                            
\newcommand{\hatcurCCdecxxxxxB}{\ensuremath{-22{\arcdeg}12{\arcmin}16.1{\arcsec}}}                           
\newcommand{\hatcurCCmagxxxxxB}{13.113}                                
\newcommand{\hatcurCCtwomassxxxxxB}{2MASS~19371363-2212161}            
\newcommand{\hatcurCCgscxxxxxB}{GSC~6311-00085}                        
\newcommand{\hatcurCCtassmvxxxxxB}{\ensuremath{13.113\pm0.010}}        
\newcommand{\hatcurCCtassmvshortxxxxxB}{\ensuremath{13.1}}             
\newcommand{\hatcurCCtassmBxxxxxB}{\ensuremath{13.820\pm0.010}}        
\newcommand{\hatcurCCtassmBshortxxxxxB}{\ensuremath{13.8}}             
\newcommand{\hatcurCCtassmIxxxxxB}{\ensuremath{nff\pmnff}}             
\newcommand{\hatcurCCtassmIshortxxxxxB}{\ensuremath{0.0}}              
\newcommand{\hatcurCCtassmgxxxxxB}{\ensuremath{13.448\pm0.010}}        
\newcommand{\hatcurCCtassmgshortxxxxxB}{\ensuremath{13.4}}             
\newcommand{\hatcurCCtassmrxxxxxB}{\ensuremath{12.967\pm0.010}}        
\newcommand{\hatcurCCtassmrshortxxxxxB}{\ensuremath{13.0}}             
\newcommand{\hatcurCCtassmixxxxxB}{\ensuremath{12.781\pm0.010}}        
\newcommand{\hatcurCCtassmishortxxxxxB}{\ensuremath{12.8}}             
\newcommand{\hatcurCCtwomassJmagxxxxxB}{\ensuremath{11.866\pm0.024}}   
\newcommand{\hatcurCCtwomassHmagxxxxxB}{\ensuremath{11.568\pm0.024}}   
\newcommand{\hatcurCCtwomassKmagxxxxxB}{\ensuremath{11.511\pm0.025}}   
\newcommand{\hatcurCCcitJmagxxxxxB}{\ensuremath{11.883\pm0.024}}       
\newcommand{\hatcurCCcitHmagxxxxxB}{\ensuremath{11.563\pm0.025}}       
\newcommand{\hatcurCCcitKmagxxxxxB}{\ensuremath{11.535\pm0.025}}       
\newcommand{\hatcurCCbbJmagxxxxxB}{\ensuremath{11.931\pm0.026}}        
\newcommand{\hatcurCCbbHmagxxxxxB}{\ensuremath{11.584\pm0.025}}        
\newcommand{\hatcurCCbbKmagxxxxxB}{\ensuremath{11.555\pm0.025}}        
\newcommand{\hatcurCCesoJmagxxxxxB}{\ensuremath{11.934\pm0.027}}       
\newcommand{\hatcurCCesoHmagxxxxxB}{\ensuremath{11.578\pm0.028}}       
\newcommand{\hatcurCCesoKmagxxxxxB}{\ensuremath{11.554\pm0.025}}       
\newcommand{\hatcurCCesoJHmagxxxxxB}{\ensuremath{0.3550\pm0.0090}}     
\newcommand{\hatcurCCesoJKmagxxxxxB}{\ensuremath{0.379\pm0.037}}       
\newcommand{\hatcurCCesoHKmagxxxxxB}{\ensuremath{0.024\pm0.038}}       
\newcommand{\hatcurLCdipxxxxxB}{\ensuremath{10.7}}                     
\newcommand{\hatcurLCrprstarxxxxxB}{\ensuremath{0.0903\pm0.0013}}      
\newcommand{\hatcurLCbsqxxxxxB}{\ensuremath{0.113_{-0.059}^{+0.087}}}  
\newcommand{\hatcurLCimpxxxxxB}{\ensuremath{0.34_{-0.10}^{+0.11}}}     
\newcommand{\hatcurLCzetaxxxxxB}{\ensuremath{17.588\pm0.067}}          
\newcommand{\hatcurLCdurxxxxxB}{\ensuremath{0.1253\pm0.0011}}          
\newcommand{\hatcurLCdurshortxxxxxB}{\ensuremath{0.1253}}              
\newcommand{\hatcurLCdurhrxxxxxB}{\ensuremath{3.007\pm0.026}}          
\newcommand{\hatcurLCdurhrshortxxxxxB}{\ensuremath{3.007}}             
\newcommand{\hatcurLCqxxxxxB}{\ensuremath{0.03780\pm0.00032}}          
\newcommand{\hatcurLCqshortxxxxxB}{\ensuremath{0.038}}                 
\newcommand{\hatcurLCingdurxxxxxB}{\ensuremath{0.01157\pm0.00100}}     
\newcommand{\hatcurLCPxxxxxB}{\ensuremath{3.3128460\pm0.0000058}}      
\newcommand{\hatcurLCPprecxxxxxB}{\ensuremath{3.3128460}}              
\newcommand{\hatcurLCPshortxxxxxB}{\ensuremath{3.3128}}                
\newcommand{\hatcurLCTxxxxxB}{\ensuremath{2456457.88193\pm0.00022}}    
\newcommand{\hatcurLCTAxxxxxB}{\ensuremath{2455106.2407\pm0.0023}}     
\newcommand{\hatcurLCTBxxxxxB}{\ensuremath{2456471.13332\pm0.00023}}   
\newcommand{\hatcurLChatnetmxxxxxB}{\ensuremath{12.919640\pm0.000071}} 
\newcommand{\hatcurLCiblendxxxxxB}{\ensuremath{0.934\pm0.045}}         
\newcommand{\hatcurSMEiteffxxxxxB}{\ensuremath{5970\pm110}}            
\newcommand{\hatcurSMEizfehxxxxxB}{\ensuremath{0.190\pm0.070}}         
\newcommand{\hatcurSMEizfehshortxxxxxB}{\ensuremath{0.19}}             
\newcommand{\hatcurSMEiloggxxxxxB}{\ensuremath{4.44\pm0.13}}           
\newcommand{\hatcurSMEivsinxxxxxB}{\ensuremath{5.66\pm0.50}}           
\newcommand{\hatcurSMEivmacxxxxxB}{\ensuremath{0.0}}                   
\newcommand{\hatcurSMEivmicxxxxxB}{\ensuremath{0.0}}                   
\newcommand{\hatcurSMEiiteffxxxxxB}{\ensuremath{5880\pm120}}           
\newcommand{\hatcurSMEiizfehxxxxxB}{\ensuremath{0.15\pm0.10}}          
\newcommand{\hatcurSMEiizfehshortxxxxxB}{\ensuremath{0.15}}            
\newcommand{\hatcurSMEiiloggxxxxxB}{\ensuremath{4.380\pm0.030}}        
\newcommand{\hatcurSMEiivsinxxxxxB}{\ensuremath{5.68\pm0.70}}          
\newcommand{\hatcurLBizxxxxxB}{\ensuremath{0.1978}}                    
\newcommand{\hatcurLBiizxxxxxB}{\ensuremath{0.3360}}                   
\newcommand{\hatcurLBiixxxxxB}{\ensuremath{0.2587}}                    
\newcommand{\hatcurLBiiixxxxxB}{\ensuremath{0.3388}}                   
\newcommand{\hatcurLBiIxxxxxB}{\ensuremath{0.2380}}                    
\newcommand{\hatcurLBiiIxxxxxB}{\ensuremath{0.3387}}                   
\newcommand{\hatcurLBigxxxxxB}{\ensuremath{0.5380}}                    
\newcommand{\hatcurLBiigxxxxxB}{\ensuremath{0.2487}}                   
\newcommand{\hatcurLBirxxxxxB}{\ensuremath{0.3459}}                    
\newcommand{\hatcurLBiirxxxxxB}{\ensuremath{0.3349}}                   
\newcommand{\hatcurLBiRxxxxxB}{\ensuremath{0.3216}}                    
\newcommand{\hatcurLBiiRxxxxxB}{\ensuremath{0.3371}}                   
\newcommand{\hatcurLBikepxxxxxB}{\ensuremath{0.1000}}                  
\newcommand{\hatcurLBiikepxxxxxB}{\ensuremath{0.1000}}                 
\newcommand{\hatcurISOmxxxxxB}{\ensuremath{1.101\pm0.054}}             
\newcommand{\hatcurISOmshortxxxxxB}{\ensuremath{1.10}}                 
\newcommand{\hatcurISOmlongxxxxxB}{\ensuremath{1.101\pm0.054}}         
\newcommand{\hatcurISOrxxxxxB}{\ensuremath{1.105_{-0.040}^{+0.055}}}   
\newcommand{\hatcurISOrshortxxxxxB}{\ensuremath{1.10}}                 
\newcommand{\hatcurISOrlongxxxxxB}{\ensuremath{1.105_{-0.040}^{+0.055}}} 
\newcommand{\hatcurISOrhoxxxxxB}{\ensuremath{1.15_{-0.16}^{+0.12}}}    
\newcommand{\hatcurISOrholongxxxxxB}{\ensuremath{1.15_{-0.16}^{+0.12}}} 
\newcommand{\hatcurISOloggxxxxxB}{\ensuremath{4.392\pm0.032}}          
\newcommand{\hatcurISOlumxxxxxB}{\ensuremath{1.31\pm0.18}}             
\newcommand{\hatcurISOlumshortxxxxxB}{\ensuremath{1.31}}               
\newcommand{\hatcurISOmvxxxxxB}{\ensuremath{4.52\pm0.16}}              
\newcommand{\hatcurISOvixxxxxB}{\ensuremath{0.669\pm0.035}}            
\newcommand{\hatcurISOagexxxxxB}{\ensuremath{3.3\pm1.7}}               
\newcommand{\hatcurISOsigmaxxxxxB}{\ensuremath{0.00060\pm0.00010}}     
\newcommand{\hatcurISOMJxxxxxB}{\ensuremath{3.42\pm0.12}}              
\newcommand{\hatcurISOMHxxxxxB}{\ensuremath{3.10\pm0.11}}              
\newcommand{\hatcurISOMKxxxxxB}{\ensuremath{3.05\pm0.10}}              
\newcommand{\hatcurISOJKxxxxxB}{\ensuremath{0.380\pm0.020}}            
\newcommand{\hatcurISOspecxxxxxB}{G}                                   
\newcommand{\hatcurRVKxxxxxB}{\ensuremath{67\pm10}}                    
\newcommand{\hatcurRVrkxxxxxB}{\ensuremath{0\pm0}}                     
\newcommand{\hatcurRVrhxxxxxB}{\ensuremath{0\pm0}}                     
\newcommand{\hatcurRVkxxxxxB}{\ensuremath{0\pm0}}                      
\newcommand{\hatcurRVhxxxxxB}{\ensuremath{0\pm0}}                      
\newcommand{\hatcurRVtronexxxxxB}{\ensuremath{0\pm0}}                  
\newcommand{\hatcurRVtrtwoxxxxxB}{\ensuremath{0\pm0}}                  
\newcommand{\hatcurRVgammaAxxxxxB}{\ensuremath{-28131\pm18}}           
\newcommand{\hatcurRVjitterAxxxxxB}{\ensuremath{45\pm23}}              
\newcommand{\hatcurRVfitrmsAxxxxxB}{\ensuremath{0.0}}                  
\newcommand{\hatcurRVgammaBxxxxxB}{\ensuremath{-28044\pm28}}           
\newcommand{\hatcurRVjitterBxxxxxB}{\ensuremath{38\pm28}}              
\newcommand{\hatcurRVfitrmsBxxxxxB}{\ensuremath{0.0}}                  
\newcommand{\hatcurRVgammaCxxxxxB}{\ensuremath{9.2\pm6.8}}             
\newcommand{\hatcurRVjitterCxxxxxB}{\ensuremath{0.00\pm0.53}}          
\newcommand{\hatcurRVfitrmsCxxxxxB}{\ensuremath{0.0}}                  
\newcommand{\hatcurRVeccenxxxxxB}{\ensuremath{0\pm0}}                  
\newcommand{\hatcurRVeccentwosiglimxxxxxB}{\ensuremath{<0.000}}        
\newcommand{\hatcurRVomegaxxxxxB}{\ensuremath{0\pm0}}                  
\newcommand{\hatcurPPixxxxxB}{\ensuremath{87.79\pm0.72}}               
\newcommand{\hatcurPPgxxxxxB}{\ensuremath{13.8\pm2.5}}                 
\newcommand{\hatcurPPloggxxxxxB}{\ensuremath{3.140\pm0.082}}           
\newcommand{\hatcurPParxxxxxB}{\ensuremath{8.73_{-0.44}^{+0.29}}}      
\newcommand{\hatcurPParelxxxxxB}{\ensuremath{0.04491\pm0.00074}}       
\newcommand{\hatcurPPrhoxxxxxB}{\ensuremath{0.70\pm0.15}}              
\newcommand{\hatcurPPmxxxxxB}{\ensuremath{0.526\pm0.081}}              
\newcommand{\hatcurPPmshortxxxxxB}{\ensuremath{0.53}}                  
\newcommand{\hatcurPPmlongxxxxxB}{\ensuremath{0.526\pm0.081}}          
\newcommand{\hatcurPPmexxxxxB}{\ensuremath{167\pm26}}                  
\newcommand{\hatcurPPmeshortxxxxxB}{\ensuremath{167.1}}                
\newcommand{\hatcurPPmelongxxxxxB}{\ensuremath{167\pm26}}              
\newcommand{\hatcurPPrxxxxxB}{\ensuremath{0.969_{-0.045}^{+0.061}}}    
\newcommand{\hatcurPPrshortxxxxxB}{\ensuremath{0.97}}                  
\newcommand{\hatcurPPrlongxxxxxB}{\ensuremath{0.969_{-0.045}^{+0.061}}} 
\newcommand{\hatcurPPrexxxxxB}{\ensuremath{10.86_{-0.50}^{+0.68}}}     
\newcommand{\hatcurPPreshortxxxxxB}{\ensuremath{10.9}}                 
\newcommand{\hatcurPPrelongxxxxxB}{\ensuremath{10.86_{-0.50}^{+0.68}}} 
\newcommand{\hatcurPPmrcorrxxxxxB}{\ensuremath{0.02}}                  
\newcommand{\hatcurPPteffxxxxxB}{\ensuremath{1407\pm39}}               
\newcommand{\hatcurPPthetaxxxxxB}{\ensuremath{0.0440\pm0.0071}}        
\newcommand{\hatcurPPfluxperixxxxxB}{\ensuremath{8.86\pm0.99}}         
\newcommand{\hatcurPPfluxperidimxxxxxB}{\ensuremath{8}}                
\newcommand{\hatcurPPfluxapxxxxxB}{\ensuremath{8.86\pm0.99}}           
\newcommand{\hatcurPPfluxapdimxxxxxB}{\ensuremath{8}}                  
\newcommand{\hatcurPPfluxavgxxxxxB}{\ensuremath{8.86\pm0.99}}          
\newcommand{\hatcurPPfluxavgdimxxxxxB}{\ensuremath{8}}                 
\newcommand{\hatcurPPfluxavglogxxxxxB}{\ensuremath{8.947\pm0.047}}     
\newcommand{\hatcurXsecphasexxxxxB}{\ensuremath{0\pm0}}                
\newcommand{\hatcurXsecondaryxxxxxB}{\ensuremath{2456459.53836\pm0.00022}} 
\newcommand{\hatcurXsecdurxxxxxB}{\ensuremath{0.1253\pm0.0011}}        
\newcommand{\hatcurXsecingdurxxxxxB}{\ensuremath{0.01157\pm0.00100}}   
\newcommand{\hatcurPPphiconjxxxxxB}{\ensuremath{0\pm0}}                
\newcommand{\hatcurPPperixxxxxB}{\ensuremath{2456457.05372\pm0.00022}} 
\newcommand{\hatcurPPaequivxxxxxB}{\ensuremath{0.0393\pm0.0021}}       
\newcommand{\hatcurPPtcircxxxxxB}{\ensuremath{345_{-91}^{+122}}}       
\newcommand{\hatcurPPtinfallxxxxxB}{\ensuremath{3320_{-690}^{+930}}}   
\newcommand{\hatcurXdistxxxxxB}{\ensuremath{503\pm25}}                 
\newcommand{\hatcurXAvxxxxxB}{\ensuremath{0.112\pm0.075}}              
\newcommand{\hatcurXdistredxxxxxB}{\ensuremath{496\pm24}}              
\newcommand{\hatcurXEBVxxxxxB}{\ensuremath{0.036\pm0.024}}             
\newcommand{\hatcurXmvisoredxxxxxB}{\ensuremath{13.114\pm0.011}}       
\newcommand{\hatcurXmiisoredxxxxxB}{\ensuremath{12.383\pm0.012}}       
\newcommand{\hatcurXmjisoredxxxxxB}{\ensuremath{11.931\pm0.015}}       
\newcommand{\hatcurXmhisoredxxxxxB}{\ensuremath{11.598\pm0.020}}       
\newcommand{\hatcurXmkisoredxxxxxB}{\ensuremath{11.537\pm0.020}}       
\newcommand{\hatcurXviisoredxxxxxB}{\ensuremath{0.730\pm0.014}}        
\newcommand{\hatcurXvkisoredxxxxxB}{\ensuremath{1.576\pm0.025}}        
\newcommand{\hatcurXjhisoredxxxxxB}{\ensuremath{0.333\pm0.012}}        
\newcommand{\hatcurXjkisoredxxxxxB}{\ensuremath{0.394\pm0.010}}        
\newcommand{\hatcurCCpmraxxxxxB}{\ensuremath{3.1\pm1.3}}               
\newcommand{\hatcurCCpmdecxxxxxB}{\ensuremath{-3.2\pm1.6}}             
\newcommand{\hatcurCCpmxxxxxB}{\ensuremath{4.5\pm2.1}}                 

\newcommand{\hatcurCCbbHmag}[1]{\ifnum#1=9 %
\hatcurCCbbHmagxxxxxA
\else
\ifnum#1=10 %
\hatcurCCbbHmagxxxxxB
\else
??????\fi
\fi
}
\newcommand{\hatcurCCbbJmag}[1]{\ifnum#1=9 %
\hatcurCCbbJmagxxxxxA
\else
\ifnum#1=10 %
\hatcurCCbbJmagxxxxxB
\else
??????\fi
\fi
}
\newcommand{\hatcurCCbbKmag}[1]{\ifnum#1=9 %
\hatcurCCbbKmagxxxxxA
\else
\ifnum#1=10 %
\hatcurCCbbKmagxxxxxB
\else
??????\fi
\fi
}
\newcommand{\hatcurCCcitHmag}[1]{\ifnum#1=9 %
\hatcurCCcitHmagxxxxxA
\else
\ifnum#1=10 %
\hatcurCCcitHmagxxxxxB
\else
??????\fi
\fi
}
\newcommand{\hatcurCCcitJmag}[1]{\ifnum#1=9 %
\hatcurCCcitJmagxxxxxA
\else
\ifnum#1=10 %
\hatcurCCcitJmagxxxxxB
\else
??????\fi
\fi
}
\newcommand{\hatcurCCcitKmag}[1]{\ifnum#1=9 %
\hatcurCCcitKmagxxxxxA
\else
\ifnum#1=10 %
\hatcurCCcitKmagxxxxxB
\else
??????\fi
\fi
}
\newcommand{\hatcurCCdec}[1]{\ifnum#1=9 %
\hatcurCCdecxxxxxA
\else
\ifnum#1=10 %
\hatcurCCdecxxxxxB
\else
??????\fi
\fi
}
\newcommand{\hatcurCCesoHKmag}[1]{\ifnum#1=9 %
\hatcurCCesoHKmagxxxxxA
\else
\ifnum#1=10 %
\hatcurCCesoHKmagxxxxxB
\else
??????\fi
\fi
}
\newcommand{\hatcurCCesoHmag}[1]{\ifnum#1=9 %
\hatcurCCesoHmagxxxxxA
\else
\ifnum#1=10 %
\hatcurCCesoHmagxxxxxB
\else
??????\fi
\fi
}
\newcommand{\hatcurCCesoJHmag}[1]{\ifnum#1=9 %
\hatcurCCesoJHmagxxxxxA
\else
\ifnum#1=10 %
\hatcurCCesoJHmagxxxxxB
\else
??????\fi
\fi
}
\newcommand{\hatcurCCesoJKmag}[1]{\ifnum#1=9 %
\hatcurCCesoJKmagxxxxxA
\else
\ifnum#1=10 %
\hatcurCCesoJKmagxxxxxB
\else
??????\fi
\fi
}
\newcommand{\hatcurCCesoJmag}[1]{\ifnum#1=9 %
\hatcurCCesoJmagxxxxxA
\else
\ifnum#1=10 %
\hatcurCCesoJmagxxxxxB
\else
??????\fi
\fi
}
\newcommand{\hatcurCCesoKmag}[1]{\ifnum#1=9 %
\hatcurCCesoKmagxxxxxA
\else
\ifnum#1=10 %
\hatcurCCesoKmagxxxxxB
\else
??????\fi
\fi
}
\newcommand{\hatcurCCgsc}[1]{\ifnum#1=9 %
\hatcurCCgscxxxxxA
\else
\ifnum#1=10 %
\hatcurCCgscxxxxxB
\else
??????\fi
\fi
}
\newcommand{\hatcurCCmag}[1]{\ifnum#1=9 %
\hatcurCCmagxxxxxA
\else
\ifnum#1=10 %
\hatcurCCmagxxxxxB
\else
??????\fi
\fi
}
\newcommand{\hatcurCCpm}[1]{\ifnum#1=9 %
\hatcurCCpmxxxxxA
\else
\ifnum#1=10 %
\hatcurCCpmxxxxxB
\else
??????\fi
\fi
}
\newcommand{\hatcurCCpmdec}[1]{\ifnum#1=9 %
\hatcurCCpmdecxxxxxA
\else
\ifnum#1=10 %
\hatcurCCpmdecxxxxxB
\else
??????\fi
\fi
}
\newcommand{\hatcurCCpmra}[1]{\ifnum#1=9 %
\hatcurCCpmraxxxxxA
\else
\ifnum#1=10 %
\hatcurCCpmraxxxxxB
\else
??????\fi
\fi
}
\newcommand{\hatcurCCra}[1]{\ifnum#1=9 %
\hatcurCCraxxxxxA
\else
\ifnum#1=10 %
\hatcurCCraxxxxxB
\else
??????\fi
\fi
}
\newcommand{\hatcurCCtassmB}[1]{\ifnum#1=9 %
\hatcurCCtassmBxxxxxA
\else
\ifnum#1=10 %
\hatcurCCtassmBxxxxxB
\else
??????\fi
\fi
}
\newcommand{\hatcurCCtassmBshort}[1]{\ifnum#1=9 %
\hatcurCCtassmBshortxxxxxA
\else
\ifnum#1=10 %
\hatcurCCtassmBshortxxxxxB
\else
??????\fi
\fi
}
\newcommand{\hatcurCCtassmg}[1]{\ifnum#1=9 %
\hatcurCCtassmgxxxxxA
\else
\ifnum#1=10 %
\hatcurCCtassmgxxxxxB
\else
??????\fi
\fi
}
\newcommand{\hatcurCCtassmgshort}[1]{\ifnum#1=9 %
\hatcurCCtassmgshortxxxxxA
\else
\ifnum#1=10 %
\hatcurCCtassmgshortxxxxxB
\else
??????\fi
\fi
}
\newcommand{\hatcurCCtassmi}[1]{\ifnum#1=9 %
\hatcurCCtassmixxxxxA
\else
\ifnum#1=10 %
\hatcurCCtassmixxxxxB
\else
??????\fi
\fi
}
\newcommand{\hatcurCCtassmI}[1]{\ifnum#1=9 %
\hatcurCCtassmIxxxxxA
\else
\ifnum#1=10 %
\hatcurCCtassmIxxxxxB
\else
??????\fi
\fi
}
\newcommand{\hatcurCCtassmishort}[1]{\ifnum#1=9 %
\hatcurCCtassmishortxxxxxA
\else
\ifnum#1=10 %
\hatcurCCtassmishortxxxxxB
\else
??????\fi
\fi
}
\newcommand{\hatcurCCtassmIshort}[1]{\ifnum#1=9 %
\hatcurCCtassmIshortxxxxxA
\else
\ifnum#1=10 %
\hatcurCCtassmIshortxxxxxB
\else
??????\fi
\fi
}
\newcommand{\hatcurCCtassmr}[1]{\ifnum#1=9 %
\hatcurCCtassmrxxxxxA
\else
\ifnum#1=10 %
\hatcurCCtassmrxxxxxB
\else
??????\fi
\fi
}
\newcommand{\hatcurCCtassmrshort}[1]{\ifnum#1=9 %
\hatcurCCtassmrshortxxxxxA
\else
\ifnum#1=10 %
\hatcurCCtassmrshortxxxxxB
\else
??????\fi
\fi
}
\newcommand{\hatcurCCtassmv}[1]{\ifnum#1=9 %
\hatcurCCtassmvxxxxxA
\else
\ifnum#1=10 %
\hatcurCCtassmvxxxxxB
\else
??????\fi
\fi
}
\newcommand{\hatcurCCtassmvshort}[1]{\ifnum#1=9 %
\hatcurCCtassmvshortxxxxxA
\else
\ifnum#1=10 %
\hatcurCCtassmvshortxxxxxB
\else
??????\fi
\fi
}
\newcommand{\hatcurCCtwomass}[1]{\ifnum#1=9 %
\hatcurCCtwomassxxxxxA
\else
\ifnum#1=10 %
\hatcurCCtwomassxxxxxB
\else
??????\fi
\fi
}
\newcommand{\hatcurCCtwomassHmag}[1]{\ifnum#1=9 %
\hatcurCCtwomassHmagxxxxxA
\else
\ifnum#1=10 %
\hatcurCCtwomassHmagxxxxxB
\else
??????\fi
\fi
}
\newcommand{\hatcurCCtwomassJmag}[1]{\ifnum#1=9 %
\hatcurCCtwomassJmagxxxxxA
\else
\ifnum#1=10 %
\hatcurCCtwomassJmagxxxxxB
\else
??????\fi
\fi
}
\newcommand{\hatcurCCtwomassKmag}[1]{\ifnum#1=9 %
\hatcurCCtwomassKmagxxxxxA
\else
\ifnum#1=10 %
\hatcurCCtwomassKmagxxxxxB
\else
??????\fi
\fi
}
\newcommand{\hatcurfield}[1]{\ifnum#1=9 %
\hatcurfieldxxxxxA
\else
\ifnum#1=10 %
\hatcurfieldxxxxxB
\else
??????\fi
\fi
}
\newcommand{\hatcurhtr}[1]{\ifnum#1=9 %
\hatcurhtrxxxxxA
\else
\ifnum#1=10 %
\hatcurhtrxxxxxB
\else
??????\fi
\fi
}
\newcommand{\hatcurISOage}[1]{\ifnum#1=9 %
\hatcurISOagexxxxxA
\else
\ifnum#1=10 %
\hatcurISOagexxxxxB
\else
??????\fi
\fi
}
\newcommand{\hatcurISOJK}[1]{\ifnum#1=9 %
\hatcurISOJKxxxxxA
\else
\ifnum#1=10 %
\hatcurISOJKxxxxxB
\else
??????\fi
\fi
}
\newcommand{\hatcurISOlogg}[1]{\ifnum#1=9 %
\hatcurISOloggxxxxxA
\else
\ifnum#1=10 %
\hatcurISOloggxxxxxB
\else
??????\fi
\fi
}
\newcommand{\hatcurISOlum}[1]{\ifnum#1=9 %
\hatcurISOlumxxxxxA
\else
\ifnum#1=10 %
\hatcurISOlumxxxxxB
\else
??????\fi
\fi
}
\newcommand{\hatcurISOlumshort}[1]{\ifnum#1=9 %
\hatcurISOlumshortxxxxxA
\else
\ifnum#1=10 %
\hatcurISOlumshortxxxxxB
\else
??????\fi
\fi
}
\newcommand{\hatcurISOm}[1]{\ifnum#1=9 %
\hatcurISOmxxxxxA
\else
\ifnum#1=10 %
\hatcurISOmxxxxxB
\else
??????\fi
\fi
}
\newcommand{\hatcurISOMH}[1]{\ifnum#1=9 %
\hatcurISOMHxxxxxA
\else
\ifnum#1=10 %
\hatcurISOMHxxxxxB
\else
??????\fi
\fi
}
\newcommand{\hatcurISOMJ}[1]{\ifnum#1=9 %
\hatcurISOMJxxxxxA
\else
\ifnum#1=10 %
\hatcurISOMJxxxxxB
\else
??????\fi
\fi
}
\newcommand{\hatcurISOMK}[1]{\ifnum#1=9 %
\hatcurISOMKxxxxxA
\else
\ifnum#1=10 %
\hatcurISOMKxxxxxB
\else
??????\fi
\fi
}
\newcommand{\hatcurISOmlong}[1]{\ifnum#1=9 %
\hatcurISOmlongxxxxxA
\else
\ifnum#1=10 %
\hatcurISOmlongxxxxxB
\else
??????\fi
\fi
}
\newcommand{\hatcurISOmshort}[1]{\ifnum#1=9 %
\hatcurISOmshortxxxxxA
\else
\ifnum#1=10 %
\hatcurISOmshortxxxxxB
\else
??????\fi
\fi
}
\newcommand{\hatcurISOmv}[1]{\ifnum#1=9 %
\hatcurISOmvxxxxxA
\else
\ifnum#1=10 %
\hatcurISOmvxxxxxB
\else
??????\fi
\fi
}
\newcommand{\hatcurISOr}[1]{\ifnum#1=9 %
\hatcurISOrxxxxxA
\else
\ifnum#1=10 %
\hatcurISOrxxxxxB
\else
??????\fi
\fi
}
\newcommand{\hatcurISOrho}[1]{\ifnum#1=9 %
\hatcurISOrhoxxxxxA
\else
\ifnum#1=10 %
\hatcurISOrhoxxxxxB
\else
??????\fi
\fi
}
\newcommand{\hatcurISOrholong}[1]{\ifnum#1=9 %
\hatcurISOrholongxxxxxA
\else
\ifnum#1=10 %
\hatcurISOrholongxxxxxB
\else
??????\fi
\fi
}
\newcommand{\hatcurISOrlong}[1]{\ifnum#1=9 %
\hatcurISOrlongxxxxxA
\else
\ifnum#1=10 %
\hatcurISOrlongxxxxxB
\else
??????\fi
\fi
}
\newcommand{\hatcurISOrshort}[1]{\ifnum#1=9 %
\hatcurISOrshortxxxxxA
\else
\ifnum#1=10 %
\hatcurISOrshortxxxxxB
\else
??????\fi
\fi
}
\newcommand{\hatcurISOsigma}[1]{\ifnum#1=9 %
\hatcurISOsigmaxxxxxA
\else
\ifnum#1=10 %
\hatcurISOsigmaxxxxxB
\else
??????\fi
\fi
}
\newcommand{\hatcurISOspec}[1]{\ifnum#1=9 %
\hatcurISOspecxxxxxA
\else
\ifnum#1=10 %
\hatcurISOspecxxxxxB
\else
??????\fi
\fi
}
\newcommand{\hatcurISOvi}[1]{\ifnum#1=9 %
\hatcurISOvixxxxxA
\else
\ifnum#1=10 %
\hatcurISOvixxxxxB
\else
??????\fi
\fi
}
\newcommand{\hatcurLBig}[1]{\ifnum#1=9 %
\hatcurLBigxxxxxA
\else
\ifnum#1=10 %
\hatcurLBigxxxxxB
\else
??????\fi
\fi
}
\newcommand{\hatcurLBii}[1]{\ifnum#1=9 %
\hatcurLBiixxxxxA
\else
\ifnum#1=10 %
\hatcurLBiixxxxxB
\else
??????\fi
\fi
}
\newcommand{\hatcurLBiI}[1]{\ifnum#1=9 %
\hatcurLBiIxxxxxA
\else
\ifnum#1=10 %
\hatcurLBiIxxxxxB
\else
??????\fi
\fi
}
\newcommand{\hatcurLBiig}[1]{\ifnum#1=9 %
\hatcurLBiigxxxxxA
\else
\ifnum#1=10 %
\hatcurLBiigxxxxxB
\else
??????\fi
\fi
}
\newcommand{\hatcurLBiii}[1]{\ifnum#1=9 %
\hatcurLBiiixxxxxA
\else
\ifnum#1=10 %
\hatcurLBiiixxxxxB
\else
??????\fi
\fi
}
\newcommand{\hatcurLBiiI}[1]{\ifnum#1=9 %
\hatcurLBiiIxxxxxA
\else
\ifnum#1=10 %
\hatcurLBiiIxxxxxB
\else
??????\fi
\fi
}
\newcommand{\hatcurLBiikep}[1]{\ifnum#1=10 %
\hatcurLBiikepxxxxxB
\else
??????\fi
}
\newcommand{\hatcurLBiir}[1]{\ifnum#1=9 %
\hatcurLBiirxxxxxA
\else
\ifnum#1=10 %
\hatcurLBiirxxxxxB
\else
??????\fi
\fi
}
\newcommand{\hatcurLBiiR}[1]{\ifnum#1=9 %
\hatcurLBiiRxxxxxA
\else
\ifnum#1=10 %
\hatcurLBiiRxxxxxB
\else
??????\fi
\fi
}
\newcommand{\hatcurLBiiz}[1]{\ifnum#1=9 %
\hatcurLBiizxxxxxA
\else
\ifnum#1=10 %
\hatcurLBiizxxxxxB
\else
??????\fi
\fi
}
\newcommand{\hatcurLBikep}[1]{\ifnum#1=10 %
\hatcurLBikepxxxxxB
\else
??????\fi
}
\newcommand{\hatcurLBir}[1]{\ifnum#1=9 %
\hatcurLBirxxxxxA
\else
\ifnum#1=10 %
\hatcurLBirxxxxxB
\else
??????\fi
\fi
}
\newcommand{\hatcurLBiR}[1]{\ifnum#1=9 %
\hatcurLBiRxxxxxA
\else
\ifnum#1=10 %
\hatcurLBiRxxxxxB
\else
??????\fi
\fi
}
\newcommand{\hatcurLBiz}[1]{\ifnum#1=9 %
\hatcurLBizxxxxxA
\else
\ifnum#1=10 %
\hatcurLBizxxxxxB
\else
??????\fi
\fi
}
\newcommand{\hatcurLCbsq}[1]{\ifnum#1=9 %
\hatcurLCbsqxxxxxA
\else
\ifnum#1=10 %
\hatcurLCbsqxxxxxB
\else
??????\fi
\fi
}
\newcommand{\hatcurLCdip}[1]{\ifnum#1=9 %
\hatcurLCdipxxxxxA
\else
\ifnum#1=10 %
\hatcurLCdipxxxxxB
\else
??????\fi
\fi
}
\newcommand{\hatcurLCdur}[1]{\ifnum#1=9 %
\hatcurLCdurxxxxxA
\else
\ifnum#1=10 %
\hatcurLCdurxxxxxB
\else
??????\fi
\fi
}
\newcommand{\hatcurLCdurhr}[1]{\ifnum#1=9 %
\hatcurLCdurhrxxxxxA
\else
\ifnum#1=10 %
\hatcurLCdurhrxxxxxB
\else
??????\fi
\fi
}
\newcommand{\hatcurLCdurhrshort}[1]{\ifnum#1=9 %
\hatcurLCdurhrshortxxxxxA
\else
\ifnum#1=10 %
\hatcurLCdurhrshortxxxxxB
\else
??????\fi
\fi
}
\newcommand{\hatcurLCdurshort}[1]{\ifnum#1=9 %
\hatcurLCdurshortxxxxxA
\else
\ifnum#1=10 %
\hatcurLCdurshortxxxxxB
\else
??????\fi
\fi
}
\newcommand{\hatcurLChatnetm}[1]{\ifnum#1=9 %
\hatcurLChatnetmxxxxxA
\else
\ifnum#1=10 %
\hatcurLChatnetmxxxxxB
\else
??????\fi
\fi
}
\newcommand{\hatcurLCiblend}[1]{\ifnum#1=9 %
\hatcurLCiblendxxxxxA
\else
\ifnum#1=10 %
\hatcurLCiblendxxxxxB
\else
??????\fi
\fi
}
\newcommand{\hatcurLCimp}[1]{\ifnum#1=9 %
\hatcurLCimpxxxxxA
\else
\ifnum#1=10 %
\hatcurLCimpxxxxxB
\else
??????\fi
\fi
}
\newcommand{\hatcurLCingdur}[1]{\ifnum#1=9 %
\hatcurLCingdurxxxxxA
\else
\ifnum#1=10 %
\hatcurLCingdurxxxxxB
\else
??????\fi
\fi
}
\newcommand{\hatcurLCP}[1]{\ifnum#1=9 %
\hatcurLCPxxxxxA
\else
\ifnum#1=10 %
\hatcurLCPxxxxxB
\else
??????\fi
\fi
}
\newcommand{\hatcurLCPprec}[1]{\ifnum#1=9 %
\hatcurLCPprecxxxxxA
\else
\ifnum#1=10 %
\hatcurLCPprecxxxxxB
\else
??????\fi
\fi
}
\newcommand{\hatcurLCPshort}[1]{\ifnum#1=9 %
\hatcurLCPshortxxxxxA
\else
\ifnum#1=10 %
\hatcurLCPshortxxxxxB
\else
??????\fi
\fi
}
\newcommand{\hatcurLCq}[1]{\ifnum#1=9 %
\hatcurLCqxxxxxA
\else
\ifnum#1=10 %
\hatcurLCqxxxxxB
\else
??????\fi
\fi
}
\newcommand{\hatcurLCqshort}[1]{\ifnum#1=9 %
\hatcurLCqshortxxxxxA
\else
\ifnum#1=10 %
\hatcurLCqshortxxxxxB
\else
??????\fi
\fi
}
\newcommand{\hatcurLCrprstar}[1]{\ifnum#1=9 %
\hatcurLCrprstarxxxxxA
\else
\ifnum#1=10 %
\hatcurLCrprstarxxxxxB
\else
??????\fi
\fi
}
\newcommand{\hatcurLCT}[1]{\ifnum#1=9 %
\hatcurLCTxxxxxA
\else
\ifnum#1=10 %
\hatcurLCTxxxxxB
\else
??????\fi
\fi
}
\newcommand{\hatcurLCTA}[1]{\ifnum#1=9 %
\hatcurLCTAxxxxxA
\else
\ifnum#1=10 %
\hatcurLCTAxxxxxB
\else
??????\fi
\fi
}
\newcommand{\hatcurLCTB}[1]{\ifnum#1=9 %
\hatcurLCTBxxxxxA
\else
\ifnum#1=10 %
\hatcurLCTBxxxxxB
\else
??????\fi
\fi
}
\newcommand{\hatcurLCzeta}[1]{\ifnum#1=9 %
\hatcurLCzetaxxxxxA
\else
\ifnum#1=10 %
\hatcurLCzetaxxxxxB
\else
??????\fi
\fi
}
\newcommand{\hatcurPPaequiv}[1]{\ifnum#1=9 %
\hatcurPPaequivxxxxxA
\else
\ifnum#1=10 %
\hatcurPPaequivxxxxxB
\else
??????\fi
\fi
}
\newcommand{\hatcurPPar}[1]{\ifnum#1=9 %
\hatcurPParxxxxxA
\else
\ifnum#1=10 %
\hatcurPParxxxxxB
\else
??????\fi
\fi
}
\newcommand{\hatcurPParel}[1]{\ifnum#1=9 %
\hatcurPParelxxxxxA
\else
\ifnum#1=10 %
\hatcurPParelxxxxxB
\else
??????\fi
\fi
}
\newcommand{\hatcurPPfluxap}[1]{\ifnum#1=9 %
\hatcurPPfluxapxxxxxA
\else
\ifnum#1=10 %
\hatcurPPfluxapxxxxxB
\else
??????\fi
\fi
}
\newcommand{\hatcurPPfluxapdim}[1]{\ifnum#1=9 %
\hatcurPPfluxapdimxxxxxA
\else
\ifnum#1=10 %
\hatcurPPfluxapdimxxxxxB
\else
??????\fi
\fi
}
\newcommand{\hatcurPPfluxavg}[1]{\ifnum#1=9 %
\hatcurPPfluxavgxxxxxA
\else
\ifnum#1=10 %
\hatcurPPfluxavgxxxxxB
\else
??????\fi
\fi
}
\newcommand{\hatcurPPfluxavgdim}[1]{\ifnum#1=9 %
\hatcurPPfluxavgdimxxxxxA
\else
\ifnum#1=10 %
\hatcurPPfluxavgdimxxxxxB
\else
??????\fi
\fi
}
\newcommand{\hatcurPPfluxavglog}[1]{\ifnum#1=9 %
\hatcurPPfluxavglogxxxxxA
\else
\ifnum#1=10 %
\hatcurPPfluxavglogxxxxxB
\else
??????\fi
\fi
}
\newcommand{\hatcurPPfluxperi}[1]{\ifnum#1=9 %
\hatcurPPfluxperixxxxxA
\else
\ifnum#1=10 %
\hatcurPPfluxperixxxxxB
\else
??????\fi
\fi
}
\newcommand{\hatcurPPfluxperidim}[1]{\ifnum#1=9 %
\hatcurPPfluxperidimxxxxxA
\else
\ifnum#1=10 %
\hatcurPPfluxperidimxxxxxB
\else
??????\fi
\fi
}
\newcommand{\hatcurPPg}[1]{\ifnum#1=9 %
\hatcurPPgxxxxxA
\else
\ifnum#1=10 %
\hatcurPPgxxxxxB
\else
??????\fi
\fi
}
\newcommand{\hatcurPPi}[1]{\ifnum#1=9 %
\hatcurPPixxxxxA
\else
\ifnum#1=10 %
\hatcurPPixxxxxB
\else
??????\fi
\fi
}
\newcommand{\hatcurPPlogg}[1]{\ifnum#1=9 %
\hatcurPPloggxxxxxA
\else
\ifnum#1=10 %
\hatcurPPloggxxxxxB
\else
??????\fi
\fi
}
\newcommand{\hatcurPPm}[1]{\ifnum#1=9 %
\hatcurPPmxxxxxA
\else
\ifnum#1=10 %
\hatcurPPmxxxxxB
\else
??????\fi
\fi
}
\newcommand{\hatcurPPme}[1]{\ifnum#1=9 %
\hatcurPPmexxxxxA
\else
\ifnum#1=10 %
\hatcurPPmexxxxxB
\else
??????\fi
\fi
}
\newcommand{\hatcurPPmelong}[1]{\ifnum#1=9 %
\hatcurPPmelongxxxxxA
\else
\ifnum#1=10 %
\hatcurPPmelongxxxxxB
\else
??????\fi
\fi
}
\newcommand{\hatcurPPmeshort}[1]{\ifnum#1=9 %
\hatcurPPmeshortxxxxxA
\else
\ifnum#1=10 %
\hatcurPPmeshortxxxxxB
\else
??????\fi
\fi
}
\newcommand{\hatcurPPmlong}[1]{\ifnum#1=9 %
\hatcurPPmlongxxxxxA
\else
\ifnum#1=10 %
\hatcurPPmlongxxxxxB
\else
??????\fi
\fi
}
\newcommand{\hatcurPPmrcorr}[1]{\ifnum#1=9 %
\hatcurPPmrcorrxxxxxA
\else
\ifnum#1=10 %
\hatcurPPmrcorrxxxxxB
\else
??????\fi
\fi
}
\newcommand{\hatcurPPmshort}[1]{\ifnum#1=9 %
\hatcurPPmshortxxxxxA
\else
\ifnum#1=10 %
\hatcurPPmshortxxxxxB
\else
??????\fi
\fi
}
\newcommand{\hatcurPPperi}[1]{\ifnum#1=9 %
\hatcurPPperixxxxxA
\else
\ifnum#1=10 %
\hatcurPPperixxxxxB
\else
??????\fi
\fi
}
\newcommand{\hatcurPPphiconj}[1]{\ifnum#1=9 %
\hatcurPPphiconjxxxxxA
\else
\ifnum#1=10 %
\hatcurPPphiconjxxxxxB
\else
??????\fi
\fi
}
\newcommand{\hatcurPPr}[1]{\ifnum#1=9 %
\hatcurPPrxxxxxA
\else
\ifnum#1=10 %
\hatcurPPrxxxxxB
\else
??????\fi
\fi
}
\newcommand{\hatcurPPre}[1]{\ifnum#1=9 %
\hatcurPPrexxxxxA
\else
\ifnum#1=10 %
\hatcurPPrexxxxxB
\else
??????\fi
\fi
}
\newcommand{\hatcurPPrelong}[1]{\ifnum#1=9 %
\hatcurPPrelongxxxxxA
\else
\ifnum#1=10 %
\hatcurPPrelongxxxxxB
\else
??????\fi
\fi
}
\newcommand{\hatcurPPreshort}[1]{\ifnum#1=9 %
\hatcurPPreshortxxxxxA
\else
\ifnum#1=10 %
\hatcurPPreshortxxxxxB
\else
??????\fi
\fi
}
\newcommand{\hatcurPPrho}[1]{\ifnum#1=9 %
\hatcurPPrhoxxxxxA
\else
\ifnum#1=10 %
\hatcurPPrhoxxxxxB
\else
??????\fi
\fi
}
\newcommand{\hatcurPPrlong}[1]{\ifnum#1=9 %
\hatcurPPrlongxxxxxA
\else
\ifnum#1=10 %
\hatcurPPrlongxxxxxB
\else
??????\fi
\fi
}
\newcommand{\hatcurPPrshort}[1]{\ifnum#1=9 %
\hatcurPPrshortxxxxxA
\else
\ifnum#1=10 %
\hatcurPPrshortxxxxxB
\else
??????\fi
\fi
}
\newcommand{\hatcurPPtcirc}[1]{\ifnum#1=9 %
\hatcurPPtcircxxxxxA
\else
\ifnum#1=10 %
\hatcurPPtcircxxxxxB
\else
??????\fi
\fi
}
\newcommand{\hatcurPPteff}[1]{\ifnum#1=9 %
\hatcurPPteffxxxxxA
\else
\ifnum#1=10 %
\hatcurPPteffxxxxxB
\else
??????\fi
\fi
}
\newcommand{\hatcurPPtheta}[1]{\ifnum#1=9 %
\hatcurPPthetaxxxxxA
\else
\ifnum#1=10 %
\hatcurPPthetaxxxxxB
\else
??????\fi
\fi
}
\newcommand{\hatcurPPtinfall}[1]{\ifnum#1=9 %
\hatcurPPtinfallxxxxxA
\else
\ifnum#1=10 %
\hatcurPPtinfallxxxxxB
\else
??????\fi
\fi
}
\newcommand{\hatcurRVeccen}[1]{\ifnum#1=9 %
\hatcurRVeccenxxxxxA
\else
\ifnum#1=10 %
\hatcurRVeccenxxxxxB
\else
??????\fi
\fi
}
\newcommand{\hatcurRVeccentwosiglim}[1]{\ifnum#1=9 %
\hatcurRVeccentwosiglimxxxxxA
\else
\ifnum#1=10 %
\hatcurRVeccentwosiglimxxxxxB
\else
??????\fi
\fi
}
\newcommand{\hatcurRVfitrmsA}[1]{\ifnum#1=9 %
\hatcurRVfitrmsAxxxxxA
\else
\ifnum#1=10 %
\hatcurRVfitrmsAxxxxxB
\else
??????\fi
\fi
}
\newcommand{\hatcurRVfitrmsB}[1]{\ifnum#1=9 %
\hatcurRVfitrmsBxxxxxA
\else
\ifnum#1=10 %
\hatcurRVfitrmsBxxxxxB
\else
??????\fi
\fi
}
\newcommand{\hatcurRVfitrmsC}[1]{\ifnum#1=9 %
\hatcurRVfitrmsCxxxxxA
\else
\ifnum#1=10 %
\hatcurRVfitrmsCxxxxxB
\else
??????\fi
\fi
}
\newcommand{\hatcurRVgammaA}[1]{\ifnum#1=9 %
\hatcurRVgammaAxxxxxA
\else
\ifnum#1=10 %
\hatcurRVgammaAxxxxxB
\else
??????\fi
\fi
}
\newcommand{\hatcurRVgammaB}[1]{\ifnum#1=9 %
\hatcurRVgammaBxxxxxA
\else
\ifnum#1=10 %
\hatcurRVgammaBxxxxxB
\else
??????\fi
\fi
}
\newcommand{\hatcurRVgammaC}[1]{\ifnum#1=9 %
\hatcurRVgammaCxxxxxA
\else
\ifnum#1=10 %
\hatcurRVgammaCxxxxxB
\else
??????\fi
\fi
}
\newcommand{\hatcurRVh}[1]{\ifnum#1=9 %
\hatcurRVhxxxxxA
\else
\ifnum#1=10 %
\hatcurRVhxxxxxB
\else
??????\fi
\fi
}
\newcommand{\hatcurRVjitterA}[1]{\ifnum#1=9 %
\hatcurRVjitterAxxxxxA
\else
\ifnum#1=10 %
\hatcurRVjitterAxxxxxB
\else
??????\fi
\fi
}
\newcommand{\hatcurRVjitterB}[1]{\ifnum#1=9 %
\hatcurRVjitterBxxxxxA
\else
\ifnum#1=10 %
\hatcurRVjitterBxxxxxB
\else
??????\fi
\fi
}
\newcommand{\hatcurRVjitterC}[1]{\ifnum#1=9 %
\hatcurRVjitterCxxxxxA
\else
\ifnum#1=10 %
\hatcurRVjitterCxxxxxB
\else
??????\fi
\fi
}
\newcommand{\hatcurRVk}[1]{\ifnum#1=9 %
\hatcurRVkxxxxxA
\else
\ifnum#1=10 %
\hatcurRVkxxxxxB
\else
??????\fi
\fi
}
\newcommand{\hatcurRVK}[1]{\ifnum#1=9 %
\hatcurRVKxxxxxA
\else
\ifnum#1=10 %
\hatcurRVKxxxxxB
\else
??????\fi
\fi
}
\newcommand{\hatcurRVomega}[1]{\ifnum#1=9 %
\hatcurRVomegaxxxxxA
\else
\ifnum#1=10 %
\hatcurRVomegaxxxxxB
\else
??????\fi
\fi
}
\newcommand{\hatcurRVrh}[1]{\ifnum#1=9 %
\hatcurRVrhxxxxxA
\else
\ifnum#1=10 %
\hatcurRVrhxxxxxB
\else
??????\fi
\fi
}
\newcommand{\hatcurRVrk}[1]{\ifnum#1=9 %
\hatcurRVrkxxxxxA
\else
\ifnum#1=10 %
\hatcurRVrkxxxxxB
\else
??????\fi
\fi
}
\newcommand{\hatcurRVtrone}[1]{\ifnum#1=9 %
\hatcurRVtronexxxxxA
\else
\ifnum#1=10 %
\hatcurRVtronexxxxxB
\else
??????\fi
\fi
}
\newcommand{\hatcurRVtrtwo}[1]{\ifnum#1=9 %
\hatcurRVtrtwoxxxxxA
\else
\ifnum#1=10 %
\hatcurRVtrtwoxxxxxB
\else
??????\fi
\fi
}
\newcommand{\hatcurSMEiilogg}[1]{\ifnum#1=9 %
\hatcurSMEiiloggxxxxxA
\else
\ifnum#1=10 %
\hatcurSMEiiloggxxxxxB
\else
??????\fi
\fi
}
\newcommand{\hatcurSMEiiteff}[1]{\ifnum#1=9 %
\hatcurSMEiiteffxxxxxA
\else
\ifnum#1=10 %
\hatcurSMEiiteffxxxxxB
\else
??????\fi
\fi
}
\newcommand{\hatcurSMEiivsin}[1]{\ifnum#1=9 %
\hatcurSMEiivsinxxxxxA
\else
\ifnum#1=10 %
\hatcurSMEiivsinxxxxxB
\else
??????\fi
\fi
}
\newcommand{\hatcurSMEiizfeh}[1]{\ifnum#1=9 %
\hatcurSMEiizfehxxxxxA
\else
\ifnum#1=10 %
\hatcurSMEiizfehxxxxxB
\else
??????\fi
\fi
}
\newcommand{\hatcurSMEiizfehshort}[1]{\ifnum#1=9 %
\hatcurSMEiizfehshortxxxxxA
\else
\ifnum#1=10 %
\hatcurSMEiizfehshortxxxxxB
\else
??????\fi
\fi
}
\newcommand{\hatcurSMEilogg}[1]{\ifnum#1=9 %
\hatcurSMEiloggxxxxxA
\else
\ifnum#1=10 %
\hatcurSMEiloggxxxxxB
\else
??????\fi
\fi
}
\newcommand{\hatcurSMEiteff}[1]{\ifnum#1=9 %
\hatcurSMEiteffxxxxxA
\else
\ifnum#1=10 %
\hatcurSMEiteffxxxxxB
\else
??????\fi
\fi
}
\newcommand{\hatcurSMEivmac}[1]{\ifnum#1=9 %
\hatcurSMEivmacxxxxxA
\else
\ifnum#1=10 %
\hatcurSMEivmacxxxxxB
\else
??????\fi
\fi
}
\newcommand{\hatcurSMEivmic}[1]{\ifnum#1=9 %
\hatcurSMEivmicxxxxxA
\else
\ifnum#1=10 %
\hatcurSMEivmicxxxxxB
\else
??????\fi
\fi
}
\newcommand{\hatcurSMEivsin}[1]{\ifnum#1=9 %
\hatcurSMEivsinxxxxxA
\else
\ifnum#1=10 %
\hatcurSMEivsinxxxxxB
\else
??????\fi
\fi
}
\newcommand{\hatcurSMEizfeh}[1]{\ifnum#1=9 %
\hatcurSMEizfehxxxxxA
\else
\ifnum#1=10 %
\hatcurSMEizfehxxxxxB
\else
??????\fi
\fi
}
\newcommand{\hatcurSMEizfehshort}[1]{\ifnum#1=9 %
\hatcurSMEizfehshortxxxxxA
\else
\ifnum#1=10 %
\hatcurSMEizfehshortxxxxxB
\else
??????\fi
\fi
}
\newcommand{\hatcurXAv}[1]{\ifnum#1=9 %
\hatcurXAvxxxxxA
\else
\ifnum#1=10 %
\hatcurXAvxxxxxB
\else
??????\fi
\fi
}
\newcommand{\hatcurXdist}[1]{\ifnum#1=9 %
\hatcurXdistxxxxxA
\else
\ifnum#1=10 %
\hatcurXdistxxxxxB
\else
??????\fi
\fi
}
\newcommand{\hatcurXdistred}[1]{\ifnum#1=9 %
\hatcurXdistredxxxxxA
\else
\ifnum#1=10 %
\hatcurXdistredxxxxxB
\else
??????\fi
\fi
}
\newcommand{\hatcurXEBV}[1]{\ifnum#1=9 %
\hatcurXEBVxxxxxA
\else
\ifnum#1=10 %
\hatcurXEBVxxxxxB
\else
??????\fi
\fi
}
\newcommand{\hatcurXjhisored}[1]{\ifnum#1=9 %
\hatcurXjhisoredxxxxxA
\else
\ifnum#1=10 %
\hatcurXjhisoredxxxxxB
\else
??????\fi
\fi
}
\newcommand{\hatcurXjkisored}[1]{\ifnum#1=9 %
\hatcurXjkisoredxxxxxA
\else
\ifnum#1=10 %
\hatcurXjkisoredxxxxxB
\else
??????\fi
\fi
}
\newcommand{\hatcurXmhisored}[1]{\ifnum#1=9 %
\hatcurXmhisoredxxxxxA
\else
\ifnum#1=10 %
\hatcurXmhisoredxxxxxB
\else
??????\fi
\fi
}
\newcommand{\hatcurXmiisored}[1]{\ifnum#1=9 %
\hatcurXmiisoredxxxxxA
\else
\ifnum#1=10 %
\hatcurXmiisoredxxxxxB
\else
??????\fi
\fi
}
\newcommand{\hatcurXmjisored}[1]{\ifnum#1=9 %
\hatcurXmjisoredxxxxxA
\else
\ifnum#1=10 %
\hatcurXmjisoredxxxxxB
\else
??????\fi
\fi
}
\newcommand{\hatcurXmkisored}[1]{\ifnum#1=9 %
\hatcurXmkisoredxxxxxA
\else
\ifnum#1=10 %
\hatcurXmkisoredxxxxxB
\else
??????\fi
\fi
}
\newcommand{\hatcurXmvisored}[1]{\ifnum#1=9 %
\hatcurXmvisoredxxxxxA
\else
\ifnum#1=10 %
\hatcurXmvisoredxxxxxB
\else
??????\fi
\fi
}
\newcommand{\hatcurXsecdur}[1]{\ifnum#1=9 %
\hatcurXsecdurxxxxxA
\else
\ifnum#1=10 %
\hatcurXsecdurxxxxxB
\else
??????\fi
\fi
}
\newcommand{\hatcurXsecingdur}[1]{\ifnum#1=9 %
\hatcurXsecingdurxxxxxA
\else
\ifnum#1=10 %
\hatcurXsecingdurxxxxxB
\else
??????\fi
\fi
}
\newcommand{\hatcurXsecondary}[1]{\ifnum#1=9 %
\hatcurXsecondaryxxxxxA
\else
\ifnum#1=10 %
\hatcurXsecondaryxxxxxB
\else
??????\fi
\fi
}
\newcommand{\hatcurXsecphase}[1]{\ifnum#1=9 %
\hatcurXsecphasexxxxxA
\else
\ifnum#1=10 %
\hatcurXsecphasexxxxxB
\else
??????\fi
\fi
}
\newcommand{\hatcurXviisored}[1]{\ifnum#1=9 %
\hatcurXviisoredxxxxxA
\else
\ifnum#1=10 %
\hatcurXviisoredxxxxxB
\else
??????\fi
\fi
}
\newcommand{\hatcurXvkisored}[1]{\ifnum#1=9 %
\hatcurXvkisoredxxxxxA
\else
\ifnum#1=10 %
\hatcurXvkisoredxxxxxB
\else
??????\fi
\fi
}

\newcommand{\hatcurhtreccenxxxxxA}{HATS579-017}                        
\newcommand{\hatcurfieldeccenxxxxxA}{579}                              
\newcommand{\hatcurCCraeccenxxxxxA}{\ensuremath{19^{\mathrm h}23^{\mathrm m}14.28{\mathrm s}}}                       
\newcommand{\hatcurCCdececcenxxxxxA}{\ensuremath{-20{\arcdeg}09{\arcmin}58.7{\arcsec}}}                      
\newcommand{\hatcurCCmageccenxxxxxA}{13.276}                           
\newcommand{\hatcurCCtwomasseccenxxxxxA}{2MASS~19231442-2009587}       
\newcommand{\hatcurCCgsceccenxxxxxA}{GSC~6305-02502}                   
\newcommand{\hatcurCCtassmveccenxxxxxA}{\ensuremath{13.276\pm0.010}}   
\newcommand{\hatcurCCtassmvshorteccenxxxxxA}{\ensuremath{13.3}}        
\newcommand{\hatcurCCtassmBeccenxxxxxA}{\ensuremath{14.080\pm0.010}}   
\newcommand{\hatcurCCtassmBshorteccenxxxxxA}{\ensuremath{14.1}}        
\newcommand{\hatcurCCtassmIeccenxxxxxA}{\ensuremath{100\pm1000}}       
\newcommand{\hatcurCCtassmIshorteccenxxxxxA}{\ensuremath{100.0}}       
\newcommand{\hatcurCCtassmgeccenxxxxxA}{\ensuremath{13.629\pm0.010}}   
\newcommand{\hatcurCCtassmgshorteccenxxxxxA}{\ensuremath{13.6}}        
\newcommand{\hatcurCCtassmreccenxxxxxA}{\ensuremath{13.072\pm0.030}}   
\newcommand{\hatcurCCtassmrshorteccenxxxxxA}{\ensuremath{13.1}}        
\newcommand{\hatcurCCtassmieccenxxxxxA}{\ensuremath{12.865\pm0.010}}   
\newcommand{\hatcurCCtassmishorteccenxxxxxA}{\ensuremath{12.9}}        
\newcommand{\hatcurCCtwomassJmageccenxxxxxA}{\ensuremath{11.885\pm0.022}} 
\newcommand{\hatcurCCtwomassHmageccenxxxxxA}{\ensuremath{11.558\pm0.027}} 
\newcommand{\hatcurCCtwomassKmageccenxxxxxA}{\ensuremath{11.479\pm0.022}} 
\newcommand{\hatcurCCcitJmageccenxxxxxA}{\ensuremath{11.900\pm0.023}}  
\newcommand{\hatcurCCcitHmageccenxxxxxA}{\ensuremath{11.553\pm0.027}}  
\newcommand{\hatcurCCcitKmageccenxxxxxA}{\ensuremath{11.503\pm0.022}}  
\newcommand{\hatcurCCbbJmageccenxxxxxA}{\ensuremath{11.952\pm0.024}}   
\newcommand{\hatcurCCbbHmageccenxxxxxA}{\ensuremath{11.574\pm0.028}}   
\newcommand{\hatcurCCbbKmageccenxxxxxA}{\ensuremath{11.523\pm0.022}}   
\newcommand{\hatcurCCesoJmageccenxxxxxA}{\ensuremath{11.955\pm0.025}}  
\newcommand{\hatcurCCesoHmageccenxxxxxA}{\ensuremath{11.569\pm0.031}}  
\newcommand{\hatcurCCesoKmageccenxxxxxA}{\ensuremath{11.522\pm0.023}}  
\newcommand{\hatcurCCesoJHmageccenxxxxxA}{\ensuremath{0.386\pm0.039}}  
\newcommand{\hatcurCCesoJKmageccenxxxxxA}{\ensuremath{0.433\pm0.010}}  
\newcommand{\hatcurCCesoHKmageccenxxxxxA}{\ensuremath{0.047\pm0.038}}  
\newcommand{\hatcurLCdipeccenxxxxxA}{\ensuremath{8.2}}                 
\newcommand{\hatcurLCrprstareccenxxxxxA}{\ensuremath{0.0715\pm0.0044}} 
\newcommand{\hatcurLCbsqeccenxxxxxA}{\ensuremath{0.064_{-0.052}^{+0.120}}} 
\newcommand{\hatcurLCimpeccenxxxxxA}{\ensuremath{0.25_{-0.14}^{+0.18}}} 
\newcommand{\hatcurLCzetaeccenxxxxxA}{\ensuremath{14.75_{-0.14}^{+0.23}}} 
\newcommand{\hatcurLCdureccenxxxxxA}{\ensuremath{0.1460\pm0.0026}}     
\newcommand{\hatcurLCdurshorteccenxxxxxA}{\ensuremath{0.1460}}         
\newcommand{\hatcurLCdurhreccenxxxxxA}{\ensuremath{3.505\pm0.062}}     
\newcommand{\hatcurLCdurhrshorteccenxxxxxA}{\ensuremath{3.505}}        
\newcommand{\hatcurLCqeccenxxxxxA}{\ensuremath{0.0762\pm0.0014}}       
\newcommand{\hatcurLCqshorteccenxxxxxA}{\ensuremath{0.076}}            
\newcommand{\hatcurLCingdureccenxxxxxA}{\ensuremath{0.0105\pm0.0016}}  
\newcommand{\hatcurLCPeccenxxxxxA}{\ensuremath{1.9153076\pm0.0000053}} 
\newcommand{\hatcurLCPprececcenxxxxxA}{\ensuremath{1.9153076}}         
\newcommand{\hatcurLCPshorteccenxxxxxA}{\ensuremath{1.9153}}           
\newcommand{\hatcurLCTeccenxxxxxA}{\ensuremath{2456156.81944\pm0.00093}} 
\newcommand{\hatcurLCTAeccenxxxxxA}{\ensuremath{2455270.0320\pm0.0024}} 
\newcommand{\hatcurLCTBeccenxxxxxA}{\ensuremath{2456444.1156\pm0.0014}} 
\newcommand{\hatcurLChatnetmeccenxxxxxA}{\ensuremath{13.020710\pm0.000062}} 
\newcommand{\hatcurLCiblendeccenxxxxxA}{\ensuremath{1.03\pm0.16}}      
\newcommand{\hatcurSMEiteffeccenxxxxxA}{\ensuremath{5363\pm90}}        
\newcommand{\hatcurSMEizfeheccenxxxxxA}{\ensuremath{0.330\pm0.090}}    
\newcommand{\hatcurSMEizfehshorteccenxxxxxA}{\ensuremath{0.33}}        
\newcommand{\hatcurSMEiloggeccenxxxxxA}{\ensuremath{3.97\pm0.20}}      
\newcommand{\hatcurSMEivsineccenxxxxxA}{\ensuremath{4.67\pm0.50}}      
\newcommand{\hatcurSMEivmaceccenxxxxxA}{\ensuremath{0.0}}              
\newcommand{\hatcurSMEivmiceccenxxxxxA}{\ensuremath{0.0}}              
\newcommand{\hatcurSMEiiteffeccenxxxxxA}{\ensuremath{5366\pm70}}       
\newcommand{\hatcurSMEiizfeheccenxxxxxA}{\ensuremath{0.340\pm0.050}}   
\newcommand{\hatcurSMEiizfehshorteccenxxxxxA}{\ensuremath{0.34}}       
\newcommand{\hatcurSMEiiloggeccenxxxxxA}{\ensuremath{4.120\pm0.040}}   
\newcommand{\hatcurSMEiivsineccenxxxxxA}{\ensuremath{4.58\pm0.90}}     
\newcommand{\hatcurLBizeccenxxxxxA}{\ensuremath{0.2713}}               
\newcommand{\hatcurLBiizeccenxxxxxA}{\ensuremath{0.3052}}              
\newcommand{\hatcurLBiieccenxxxxxA}{\ensuremath{0.3533}}               
\newcommand{\hatcurLBiiieccenxxxxxA}{\ensuremath{0.2892}}              
\newcommand{\hatcurLBiIeccenxxxxxA}{\ensuremath{0.3265}}               
\newcommand{\hatcurLBiiIeccenxxxxxA}{\ensuremath{0.2945}}              
\newcommand{\hatcurLBigeccenxxxxxA}{\ensuremath{0.7006}}               
\newcommand{\hatcurLBiigeccenxxxxxA}{\ensuremath{0.1231}}              
\newcommand{\hatcurLBireccenxxxxxA}{\ensuremath{0.4688}}               
\newcommand{\hatcurLBiireccenxxxxxA}{\ensuremath{0.2596}}              
\newcommand{\hatcurLBiReccenxxxxxA}{\ensuremath{0.4369}}               
\newcommand{\hatcurLBiiReccenxxxxxA}{\ensuremath{0.2687}}              
\newcommand{\hatcurISOmeccenxxxxxA}{\ensuremath{1.014\pm0.045}}        
\newcommand{\hatcurISOmshorteccenxxxxxA}{\ensuremath{1.01}}            
\newcommand{\hatcurISOmlongeccenxxxxxA}{\ensuremath{1.014\pm0.045}}    
\newcommand{\hatcurISOreccenxxxxxA}{\ensuremath{1.444_{-0.099}^{+0.149}}} 
\newcommand{\hatcurISOrshorteccenxxxxxA}{\ensuremath{1.44}}            
\newcommand{\hatcurISOrlongeccenxxxxxA}{\ensuremath{1.444_{-0.099}^{+0.149}}} 
\newcommand{\hatcurISOrhoeccenxxxxxA}{\ensuremath{0.472\pm0.098}}      
\newcommand{\hatcurISOrholongeccenxxxxxA}{\ensuremath{0.472\pm0.098}}  
\newcommand{\hatcurISOloggeccenxxxxxA}{\ensuremath{4.122\pm0.060}}     
\newcommand{\hatcurISOlumeccenxxxxxA}{\ensuremath{1.56_{-0.23}^{+0.36}}} 
\newcommand{\hatcurISOlumshorteccenxxxxxA}{\ensuremath{1.56}}          
\newcommand{\hatcurISOmveccenxxxxxA}{\ensuremath{4.42\pm0.21}}         
\newcommand{\hatcurISOvieccenxxxxxA}{\ensuremath{0.826\pm0.018}}       
\newcommand{\hatcurISOageeccenxxxxxA}{\ensuremath{11.2\pm1.7}}         
\newcommand{\hatcurISOsigmaeccenxxxxxA}{\ensuremath{0.000500\pm0.000080}} 
\newcommand{\hatcurISOMJeccenxxxxxA}{\ensuremath{3.05\pm0.19}}         
\newcommand{\hatcurISOMHeccenxxxxxA}{\ensuremath{2.65\pm0.19}}         
\newcommand{\hatcurISOMKeccenxxxxxA}{\ensuremath{2.58\pm0.19}}         
\newcommand{\hatcurISOJKeccenxxxxxA}{\ensuremath{0.480\pm0.020}}       
\newcommand{\hatcurISOspececcenxxxxxA}{G}                              
\newcommand{\hatcurRVKeccenxxxxxA}{\ensuremath{131.5\pm3.4}}           
\newcommand{\hatcurRVrkeccenxxxxxA}{\ensuremath{0.117_{-0.166}^{+0.085}}} 
\newcommand{\hatcurRVrheccenxxxxxA}{\ensuremath{-0.17_{-0.11}^{+0.19}}} 
\newcommand{\hatcurRVkeccenxxxxxA}{\ensuremath{0.026\pm0.034}}         
\newcommand{\hatcurRVheccenxxxxxA}{\ensuremath{-0.038_{-0.052}^{+0.039}}} 
\newcommand{\hatcurRVtroneeccenxxxxxA}{\ensuremath{0\pm0}}             
\newcommand{\hatcurRVtrtwoeccenxxxxxA}{\ensuremath{0\pm0}}             
\newcommand{\hatcurRVgammaAeccenxxxxxA}{\ensuremath{-42.2\pm3.2}}      
\newcommand{\hatcurRVjitterAeccenxxxxxA}{\ensuremath{0.2\pm3.6}}       
\newcommand{\hatcurRVfitrmsAeccenxxxxxA}{\ensuremath{0.0}}             
\newcommand{\hatcurRVgammaBeccenxxxxxA}{\ensuremath{-10649.7\pm9.5}}   
\newcommand{\hatcurRVjitterBeccenxxxxxA}{\ensuremath{0.0\pm2.0}}       
\newcommand{\hatcurRVfitrmsBeccenxxxxxA}{\ensuremath{0.0}}             
\newcommand{\hatcurRVgammaCeccenxxxxxA}{\ensuremath{-10639\pm16}}      
\newcommand{\hatcurRVjitterCeccenxxxxxA}{\ensuremath{0.0\pm1.5}}       
\newcommand{\hatcurRVfitrmsCeccenxxxxxA}{\ensuremath{0.0}}             
\newcommand{\hatcurRVecceneccenxxxxxA}{\ensuremath{0.060\pm0.040}}     
\newcommand{\hatcurRVeccentwosiglimeccenxxxxxA}{\ensuremath{<0.129}}   
\newcommand{\hatcurRVomegaeccenxxxxxA}{\ensuremath{290\pm100}}         
\newcommand{\hatcurPPieccenxxxxxA}{\ensuremath{86.9_{-2.8}^{+1.9}}}    
\newcommand{\hatcurPPgeccenxxxxxA}{\ensuremath{20.3_{-4.7}^{+2.9}}}    
\newcommand{\hatcurPPloggeccenxxxxxA}{\ensuremath{3.307_{-0.113}^{+0.058}}} 
\newcommand{\hatcurPPareccenxxxxxA}{\ensuremath{4.51\pm0.32}}          
\newcommand{\hatcurPPareleccenxxxxxA}{\ensuremath{0.03033\pm0.00044}}  
\newcommand{\hatcurPPrhoeccenxxxxxA}{\ensuremath{1.02_{-0.34}^{+0.24}}} 
\newcommand{\hatcurPPmeccenxxxxxA}{\ensuremath{0.812\pm0.035}}         
\newcommand{\hatcurPPmshorteccenxxxxxA}{\ensuremath{0.81}}             
\newcommand{\hatcurPPmlongeccenxxxxxA}{\ensuremath{0.812\pm0.035}}     
\newcommand{\hatcurPPmeeccenxxxxxA}{\ensuremath{258\pm11}}             
\newcommand{\hatcurPPmeshorteccenxxxxxA}{\ensuremath{258.0}}           
\newcommand{\hatcurPPmelongeccenxxxxxA}{\ensuremath{258\pm11}}         
\newcommand{\hatcurPPreccenxxxxxA}{\ensuremath{0.997_{-0.078}^{+0.152}}} 
\newcommand{\hatcurPPrshorteccenxxxxxA}{\ensuremath{1.00}}             
\newcommand{\hatcurPPrlongeccenxxxxxA}{\ensuremath{0.997_{-0.078}^{+0.152}}} 
\newcommand{\hatcurPPreeccenxxxxxA}{\ensuremath{11.18_{-0.87}^{+1.70}}} 
\newcommand{\hatcurPPreshorteccenxxxxxA}{\ensuremath{11.2}}            
\newcommand{\hatcurPPrelongeccenxxxxxA}{\ensuremath{11.18_{-0.87}^{+1.70}}} 
\newcommand{\hatcurPPmrcorreccenxxxxxA}{\ensuremath{0.67}}             
\newcommand{\hatcurPPteffeccenxxxxxA}{\ensuremath{1790_{-60}^{+80}}}   
\newcommand{\hatcurPPthetaeccenxxxxxA}{\ensuremath{0.0486_{-0.0061}^{+0.0038}}} 
\newcommand{\hatcurPPfluxperieccenxxxxxA}{\ensuremath{2.61_{-0.22}^{+0.40}}} 
\newcommand{\hatcurPPfluxperidimeccenxxxxxA}{\ensuremath{9}}           
\newcommand{\hatcurPPfluxapeccenxxxxxA}{\ensuremath{2.10\pm0.46}}      
\newcommand{\hatcurPPfluxapdimeccenxxxxxA}{\ensuremath{9}}             
\newcommand{\hatcurPPfluxavgeccenxxxxxA}{\ensuremath{2.32_{-0.30}^{+0.44}}} 
\newcommand{\hatcurPPfluxavgdimeccenxxxxxA}{\ensuremath{9}}            
\newcommand{\hatcurPPfluxavglogeccenxxxxxA}{\ensuremath{9.365\pm0.068}} 
\newcommand{\hatcurXsecphaseeccenxxxxxA}{\ensuremath{0.517\pm0.022}}   
\newcommand{\hatcurXsecondaryeccenxxxxxA}{\ensuremath{2456157.810\pm0.042}} 
\newcommand{\hatcurXsecdureccenxxxxxA}{\ensuremath{0.136\pm0.011}}     
\newcommand{\hatcurXsecingdureccenxxxxxA}{\ensuremath{0.0095\pm0.0019}} 
\newcommand{\hatcurPPphiconjeccenxxxxxA}{\ensuremath{0.346_{-0.585}^{+0.077}}} 
\newcommand{\hatcurPPperieccenxxxxxA}{\ensuremath{2456156.16\pm0.63}}  
\newcommand{\hatcurPPaequiveccenxxxxxA}{\ensuremath{0.0243\pm0.0019}}  
\newcommand{\hatcurPPtcirceccenxxxxxA}{\ensuremath{41\pm22}}           
\newcommand{\hatcurPPtinfalleccenxxxxxA}{\ensuremath{42\pm14}}         
\newcommand{\hatcurXdisteccenxxxxxA}{\ensuremath{616_{-43}^{+64}}}     
\newcommand{\hatcurXAveccenxxxxxA}{\ensuremath{0.000\pm0.011}}         
\newcommand{\hatcurXdistredeccenxxxxxA}{\ensuremath{595_{-46}^{+64}}}  
\newcommand{\hatcurXEBVeccenxxxxxA}{\ensuremath{0.0000\pm0.0036}}      
\newcommand{\hatcurXmvisoredeccenxxxxxA}{\ensuremath{13.301\pm0.017}}  
\newcommand{\hatcurXmiisoredeccenxxxxxA}{\ensuremath{12.4740\pm0.0089}} 
\newcommand{\hatcurXmjisoredeccenxxxxxA}{\ensuremath{11.926\pm0.025}}  
\newcommand{\hatcurXmhisoredeccenxxxxxA}{\ensuremath{11.515\pm0.038}}  
\newcommand{\hatcurXmkisoredeccenxxxxxA}{\ensuremath{11.445\pm0.040}}  
\newcommand{\hatcurXviisoredeccenxxxxxA}{\ensuremath{0.826\pm0.016}}   
\newcommand{\hatcurXvkisoredeccenxxxxxA}{\ensuremath{1.856\pm0.055}}   
\newcommand{\hatcurXjhisoredeccenxxxxxA}{\ensuremath{0.411\pm0.015}}   
\newcommand{\hatcurXjkisoredeccenxxxxxA}{\ensuremath{0.482\pm0.016}}   
\newcommand{\hatcurCCpmraeccenxxxxxA}{\ensuremath{0.3\pm4.3}}          
\newcommand{\hatcurCCpmdececcenxxxxxA}{\ensuremath{-1.9\pm2.8}}        
\newcommand{\hatcurCCpmeccenxxxxxA}{\ensuremath{1.9\pm5.1}}            

\newcommand{\hatcurhtreccenxxxxxB}{HATS579-026}                        
\newcommand{\hatcurfieldeccenxxxxxB}{\ensuremath{string}}              
\newcommand{\hatcurCCraeccenxxxxxB}{\ensuremath{19^{\mathrm h}37^{\mathrm m}13.80{\mathrm s}}}                       
\newcommand{\hatcurCCdececcenxxxxxB}{\ensuremath{-22{\arcdeg}12{\arcmin}16.1{\arcsec}}}                      
\newcommand{\hatcurCCmageccenxxxxxB}{13.113}                           
\newcommand{\hatcurCCtwomasseccenxxxxxB}{2MASS~19371363-2212161}       
\newcommand{\hatcurCCgsceccenxxxxxB}{GSC~6311-00085}                   
\newcommand{\hatcurCCtassmveccenxxxxxB}{\ensuremath{13.113\pm0.010}}   
\newcommand{\hatcurCCtassmvshorteccenxxxxxB}{\ensuremath{13.1}}        
\newcommand{\hatcurCCtassmBeccenxxxxxB}{\ensuremath{13.820\pm0.010}}   
\newcommand{\hatcurCCtassmBshorteccenxxxxxB}{\ensuremath{13.8}}        
\newcommand{\hatcurCCtassmIeccenxxxxxB}{\ensuremath{nff\pmnff}}        
\newcommand{\hatcurCCtassmIshorteccenxxxxxB}{\ensuremath{0.0}}         
\newcommand{\hatcurCCtassmgeccenxxxxxB}{\ensuremath{13.448\pm0.010}}   
\newcommand{\hatcurCCtassmgshorteccenxxxxxB}{\ensuremath{13.4}}        
\newcommand{\hatcurCCtassmreccenxxxxxB}{\ensuremath{12.967\pm0.010}}   
\newcommand{\hatcurCCtassmrshorteccenxxxxxB}{\ensuremath{13.0}}        
\newcommand{\hatcurCCtassmieccenxxxxxB}{\ensuremath{12.781\pm0.010}}   
\newcommand{\hatcurCCtassmishorteccenxxxxxB}{\ensuremath{12.8}}        
\newcommand{\hatcurCCtwomassJmageccenxxxxxB}{\ensuremath{11.866\pm0.024}} 
\newcommand{\hatcurCCtwomassHmageccenxxxxxB}{\ensuremath{11.568\pm0.024}} 
\newcommand{\hatcurCCtwomassKmageccenxxxxxB}{\ensuremath{11.511\pm0.025}} 
\newcommand{\hatcurCCcitJmageccenxxxxxB}{\ensuremath{11.883\pm0.024}}  
\newcommand{\hatcurCCcitHmageccenxxxxxB}{\ensuremath{11.563\pm0.025}}  
\newcommand{\hatcurCCcitKmageccenxxxxxB}{\ensuremath{11.535\pm0.025}}  
\newcommand{\hatcurCCbbJmageccenxxxxxB}{\ensuremath{11.931\pm0.026}}   
\newcommand{\hatcurCCbbHmageccenxxxxxB}{\ensuremath{11.584\pm0.025}}   
\newcommand{\hatcurCCbbKmageccenxxxxxB}{\ensuremath{11.555\pm0.025}}   
\newcommand{\hatcurCCesoJmageccenxxxxxB}{\ensuremath{11.934\pm0.027}}  
\newcommand{\hatcurCCesoHmageccenxxxxxB}{\ensuremath{11.578\pm0.028}}  
\newcommand{\hatcurCCesoKmageccenxxxxxB}{\ensuremath{11.554\pm0.025}}  
\newcommand{\hatcurCCesoJHmageccenxxxxxB}{\ensuremath{0.3550\pm0.0090}} 
\newcommand{\hatcurCCesoJKmageccenxxxxxB}{\ensuremath{0.379\pm0.037}}  
\newcommand{\hatcurCCesoHKmageccenxxxxxB}{\ensuremath{0.024\pm0.038}}  
\newcommand{\hatcurLCdipeccenxxxxxB}{\ensuremath{10.7}}                
\newcommand{\hatcurLCrprstareccenxxxxxB}{\ensuremath{0.0898\pm0.0014}} 
\newcommand{\hatcurLCbsqeccenxxxxxB}{\ensuremath{0.093_{-0.056}^{+0.079}}} 
\newcommand{\hatcurLCimpeccenxxxxxB}{\ensuremath{0.30_{-0.11}^{+0.11}}} 
\newcommand{\hatcurLCzetaeccenxxxxxB}{\ensuremath{17.640_{-0.068}^{+0.092}}} 
\newcommand{\hatcurLCdureccenxxxxxB}{\ensuremath{0.1245\pm0.0011}}     
\newcommand{\hatcurLCdurshorteccenxxxxxB}{\ensuremath{0.1245}}         
\newcommand{\hatcurLCdurhreccenxxxxxB}{\ensuremath{2.987\pm0.027}}     
\newcommand{\hatcurLCdurhrshorteccenxxxxxB}{\ensuremath{2.987}}        
\newcommand{\hatcurLCqeccenxxxxxB}{\ensuremath{0.03760\pm0.00035}}     
\newcommand{\hatcurLCqshorteccenxxxxxB}{\ensuremath{0.038}}            
\newcommand{\hatcurLCingdureccenxxxxxB}{\ensuremath{0.0112\pm0.0010}}  
\newcommand{\hatcurLCPeccenxxxxxB}{\ensuremath{3.3128475\pm0.0000060}} 
\newcommand{\hatcurLCPprececcenxxxxxB}{\ensuremath{3.3128475}}         
\newcommand{\hatcurLCPshorteccenxxxxxB}{\ensuremath{3.3128}}           
\newcommand{\hatcurLCTeccenxxxxxB}{\ensuremath{2456431.37929\pm0.00040}} 
\newcommand{\hatcurLCTAeccenxxxxxB}{\ensuremath{2455106.2404\pm0.0023}} 
\newcommand{\hatcurLCTBeccenxxxxxB}{\ensuremath{2456471.13347\pm0.00043}} 
\newcommand{\hatcurLChatnetmeccenxxxxxB}{\ensuremath{12.919640\pm0.000070}} 
\newcommand{\hatcurLCiblendeccenxxxxxB}{\ensuremath{0.938\pm0.043}}    
\newcommand{\hatcurSMEiteffeccenxxxxxB}{\ensuremath{5970\pm110}}       
\newcommand{\hatcurSMEizfeheccenxxxxxB}{\ensuremath{0.190\pm0.070}}    
\newcommand{\hatcurSMEizfehshorteccenxxxxxB}{\ensuremath{0.19}}        
\newcommand{\hatcurSMEiloggeccenxxxxxB}{\ensuremath{4.44\pm0.13}}      
\newcommand{\hatcurSMEivsineccenxxxxxB}{\ensuremath{5.66\pm0.50}}      
\newcommand{\hatcurSMEivmaceccenxxxxxB}{\ensuremath{0.0}}              
\newcommand{\hatcurSMEivmiceccenxxxxxB}{\ensuremath{0.0}}              
\newcommand{\hatcurSMEiiteffeccenxxxxxB}{\ensuremath{5880\pm120}}      
\newcommand{\hatcurSMEiizfeheccenxxxxxB}{\ensuremath{0.15\pm0.10}}     
\newcommand{\hatcurSMEiizfehshorteccenxxxxxB}{\ensuremath{0.15}}       
\newcommand{\hatcurSMEiiloggeccenxxxxxB}{\ensuremath{4.380\pm0.030}}   
\newcommand{\hatcurSMEiivsineccenxxxxxB}{\ensuremath{5.68\pm0.70}}     
\newcommand{\hatcurLBizeccenxxxxxB}{\ensuremath{0.1978}}               
\newcommand{\hatcurLBiizeccenxxxxxB}{\ensuremath{0.3360}}              
\newcommand{\hatcurLBiieccenxxxxxB}{\ensuremath{0.2587}}               
\newcommand{\hatcurLBiiieccenxxxxxB}{\ensuremath{0.3388}}              
\newcommand{\hatcurLBiIeccenxxxxxB}{\ensuremath{0.2380}}               
\newcommand{\hatcurLBiiIeccenxxxxxB}{\ensuremath{0.3387}}              
\newcommand{\hatcurLBigeccenxxxxxB}{\ensuremath{0.5380}}               
\newcommand{\hatcurLBiigeccenxxxxxB}{\ensuremath{0.2487}}              
\newcommand{\hatcurLBireccenxxxxxB}{\ensuremath{0.3459}}               
\newcommand{\hatcurLBiireccenxxxxxB}{\ensuremath{0.3349}}              
\newcommand{\hatcurLBiReccenxxxxxB}{\ensuremath{0.3216}}               
\newcommand{\hatcurLBiiReccenxxxxxB}{\ensuremath{0.3371}}              
\newcommand{\hatcurLBikepeccenxxxxxB}{\ensuremath{0.1000}}             
\newcommand{\hatcurLBiikepeccenxxxxxB}{\ensuremath{0.1000}}            
\newcommand{\hatcurISOmeccenxxxxxB}{\ensuremath{1.143\pm0.086}}        
\newcommand{\hatcurISOmshorteccenxxxxxB}{\ensuremath{1.14}}            
\newcommand{\hatcurISOmlongeccenxxxxxB}{\ensuremath{1.143\pm0.086}}    
\newcommand{\hatcurISOreccenxxxxxB}{\ensuremath{1.34\pm0.22}}          
\newcommand{\hatcurISOrshorteccenxxxxxB}{\ensuremath{1.34}}            
\newcommand{\hatcurISOrlongeccenxxxxxB}{\ensuremath{1.34\pm0.22}}      
\newcommand{\hatcurISOrhoeccenxxxxxB}{\ensuremath{0.67_{-0.23}^{+0.39}}} 
\newcommand{\hatcurISOrholongeccenxxxxxB}{\ensuremath{0.67_{-0.23}^{+0.39}}} 
\newcommand{\hatcurISOloggeccenxxxxxB}{\ensuremath{4.24\pm0.11}}       
\newcommand{\hatcurISOlumeccenxxxxxB}{\ensuremath{1.90_{-0.53}^{+0.78}}} 
\newcommand{\hatcurISOlumshorteccenxxxxxB}{\ensuremath{1.90}}          
\newcommand{\hatcurISOmveccenxxxxxB}{\ensuremath{4.11\pm0.36}}         
\newcommand{\hatcurISOvieccenxxxxxB}{\ensuremath{0.668\pm0.036}}       
\newcommand{\hatcurISOageeccenxxxxxB}{\ensuremath{4.4_{-1.2}^{+2.4}}}  
\newcommand{\hatcurISOsigmaeccenxxxxxB}{\ensuremath{0.00050\pm0.00011}} 
\newcommand{\hatcurISOMJeccenxxxxxB}{\ensuremath{3.01\pm0.34}}         
\newcommand{\hatcurISOMHeccenxxxxxB}{\ensuremath{2.69\pm0.34}}         
\newcommand{\hatcurISOMKeccenxxxxxB}{\ensuremath{2.64\pm0.34}}         
\newcommand{\hatcurISOJKeccenxxxxxB}{\ensuremath{0.360\pm0.080}}       
\newcommand{\hatcurISOspececcenxxxxxB}{G}                              
\newcommand{\hatcurRVKeccenxxxxxB}{\ensuremath{82\pm21}}               
\newcommand{\hatcurRVrkeccenxxxxxB}{\ensuremath{0.33_{-0.19}^{+0.14}}} 
\newcommand{\hatcurRVrheccenxxxxxB}{\ensuremath{0.34_{-0.22}^{+0.12}}} 
\newcommand{\hatcurRVkeccenxxxxxB}{\ensuremath{0.16\pm0.12}}           
\newcommand{\hatcurRVheccenxxxxxB}{\ensuremath{0.17\pm0.12}}           
\newcommand{\hatcurRVtroneeccenxxxxxB}{\ensuremath{0\pm0}}             
\newcommand{\hatcurRVtrtwoeccenxxxxxB}{\ensuremath{0\pm0}}             
\newcommand{\hatcurRVgammaAeccenxxxxxB}{\ensuremath{-28122\pm20}}      
\newcommand{\hatcurRVjitterAeccenxxxxxB}{\ensuremath{40\pm25}}         
\newcommand{\hatcurRVfitrmsAeccenxxxxxB}{\ensuremath{0.0}}             
\newcommand{\hatcurRVgammaBeccenxxxxxB}{\ensuremath{-28050\pm26}}      
\newcommand{\hatcurRVjitterBeccenxxxxxB}{\ensuremath{41\pm34}}         
\newcommand{\hatcurRVfitrmsBeccenxxxxxB}{\ensuremath{0.0}}             
\newcommand{\hatcurRVgammaCeccenxxxxxB}{\ensuremath{18.3\pm9.4}}       
\newcommand{\hatcurRVjitterCeccenxxxxxB}{\ensuremath{0.00\pm0.65}}     
\newcommand{\hatcurRVfitrmsCeccenxxxxxB}{\ensuremath{0.0}}             
\newcommand{\hatcurRVecceneccenxxxxxB}{\ensuremath{0.24\pm0.15}}       
\newcommand{\hatcurRVeccentwosiglimeccenxxxxxB}{\ensuremath{<0.501}}   
\newcommand{\hatcurRVomegaeccenxxxxxB}{\ensuremath{47\pm80}}           
\newcommand{\hatcurPPieccenxxxxxB}{\ensuremath{87.28_{-1.78}^{+0.84}}} 
\newcommand{\hatcurPPgeccenxxxxxB}{\ensuremath{11.4\pm2.6}}            
\newcommand{\hatcurPPloggeccenxxxxxB}{\ensuremath{3.056\pm0.100}}      
\newcommand{\hatcurPPareccenxxxxxB}{\ensuremath{7.31\pm1.00}}          
\newcommand{\hatcurPPareleccenxxxxxB}{\ensuremath{0.0455\pm0.0011}}    
\newcommand{\hatcurPPrhoeccenxxxxxB}{\ensuremath{0.48_{-0.15}^{+0.21}}} 
\newcommand{\hatcurPPmeccenxxxxxB}{\ensuremath{0.63\pm0.15}}           
\newcommand{\hatcurPPmshorteccenxxxxxB}{\ensuremath{0.63}}             
\newcommand{\hatcurPPmlongeccenxxxxxB}{\ensuremath{0.63\pm0.15}}       
\newcommand{\hatcurPPmeeccenxxxxxB}{\ensuremath{201\pm48}}             
\newcommand{\hatcurPPmeshorteccenxxxxxB}{\ensuremath{201.2}}           
\newcommand{\hatcurPPmelongeccenxxxxxB}{\ensuremath{201\pm48}}         
\newcommand{\hatcurPPreccenxxxxxB}{\ensuremath{1.17\pm0.20}}           
\newcommand{\hatcurPPrshorteccenxxxxxB}{\ensuremath{1.17}}             
\newcommand{\hatcurPPrlongeccenxxxxxB}{\ensuremath{1.17\pm0.20}}       
\newcommand{\hatcurPPreeccenxxxxxB}{\ensuremath{13.1\pm2.2}}           
\newcommand{\hatcurPPreshorteccenxxxxxB}{\ensuremath{13.1}}            
\newcommand{\hatcurPPrelongeccenxxxxxB}{\ensuremath{13.1\pm2.2}}       
\newcommand{\hatcurPPmrcorreccenxxxxxB}{\ensuremath{0.69}}             
\newcommand{\hatcurPPteffeccenxxxxxB}{\ensuremath{1550\pm130}}         
\newcommand{\hatcurPPthetaeccenxxxxxB}{\ensuremath{0.0424\pm0.0071}}   
\newcommand{\hatcurPPfluxperieccenxxxxxB}{\ensuremath{2.3_{-1.2}^{+2.5}}} 
\newcommand{\hatcurPPfluxperidimeccenxxxxxB}{\ensuremath{9}}           
\newcommand{\hatcurPPfluxapeccenxxxxxB}{\ensuremath{8.3\pm1.2}}        
\newcommand{\hatcurPPfluxapdimeccenxxxxxB}{\ensuremath{8}}             
\newcommand{\hatcurPPfluxavgeccenxxxxxB}{\ensuremath{1.30_{-0.37}^{+0.55}}} 
\newcommand{\hatcurPPfluxavgdimeccenxxxxxB}{\ensuremath{9}}            
\newcommand{\hatcurPPfluxavglogeccenxxxxxB}{\ensuremath{9.12\pm0.15}}  
\newcommand{\hatcurXsecphaseeccenxxxxxB}{\ensuremath{0.605\pm0.080}}   
\newcommand{\hatcurXsecondaryeccenxxxxxB}{\ensuremath{2456433.38\pm0.26}} 
\newcommand{\hatcurXsecdureccenxxxxxB}{\ensuremath{0.164\pm0.033}}     
\newcommand{\hatcurXsecingdureccenxxxxxB}{\ensuremath{0.017\pm0.010}}  
\newcommand{\hatcurPPphiconjeccenxxxxxB}{\ensuremath{0.065_{-0.027}^{+0.055}}} 
\newcommand{\hatcurPPperieccenxxxxxB}{\ensuremath{2456431.16\pm0.37}}  
\newcommand{\hatcurPPaequiveccenxxxxxB}{\ensuremath{0.0329\pm0.0047}}  
\newcommand{\hatcurPPtcirceccenxxxxxB}{\ensuremath{112_{-84}^{+179}}}  
\newcommand{\hatcurPPtinfalleccenxxxxxB}{\ensuremath{1190_{-660}^{+1780}}} 
\newcommand{\hatcurXdisteccenxxxxxB}{\ensuremath{610\pm100}}           
\newcommand{\hatcurXAveccenxxxxxB}{\ensuremath{0.114\pm0.078}}         
\newcommand{\hatcurXdistredeccenxxxxxB}{\ensuremath{599\pm99}}         
\newcommand{\hatcurXEBVeccenxxxxxB}{\ensuremath{0.037\pm0.025}}        
\newcommand{\hatcurXmvisoredeccenxxxxxB}{\ensuremath{13.114\pm0.011}}  
\newcommand{\hatcurXmiisoredeccenxxxxxB}{\ensuremath{12.383\pm0.012}}  
\newcommand{\hatcurXmjisoredeccenxxxxxB}{\ensuremath{11.931\pm0.015}}  
\newcommand{\hatcurXmhisoredeccenxxxxxB}{\ensuremath{11.599\pm0.020}}  
\newcommand{\hatcurXmkisoredeccenxxxxxB}{\ensuremath{11.537\pm0.020}}  
\newcommand{\hatcurXviisoredeccenxxxxxB}{\ensuremath{0.730\pm0.014}}   
\newcommand{\hatcurXvkisoredeccenxxxxxB}{\ensuremath{1.576\pm0.025}}   
\newcommand{\hatcurXjhisoredeccenxxxxxB}{\ensuremath{0.331\pm0.013}}   
\newcommand{\hatcurXjkisoredeccenxxxxxB}{\ensuremath{0.393\pm0.011}}   
\newcommand{\hatcurCCpmraeccenxxxxxB}{\ensuremath{3.1\pm1.3}}          
\newcommand{\hatcurCCpmdececcenxxxxxB}{\ensuremath{-3.2\pm1.6}}        
\newcommand{\hatcurCCpmeccenxxxxxB}{\ensuremath{4.5\pm2.1}}            

\newcommand{\hatcurCCbbHmageccen}[1]{\ifnum#1=9 %
\hatcurCCbbHmageccenxxxxxA
\else
\ifnum#1=10 %
\hatcurCCbbHmageccenxxxxxB
\else
??????\fi
\fi
}
\newcommand{\hatcurCCbbJmageccen}[1]{\ifnum#1=9 %
\hatcurCCbbJmageccenxxxxxA
\else
\ifnum#1=10 %
\hatcurCCbbJmageccenxxxxxB
\else
??????\fi
\fi
}
\newcommand{\hatcurCCbbKmageccen}[1]{\ifnum#1=9 %
\hatcurCCbbKmageccenxxxxxA
\else
\ifnum#1=10 %
\hatcurCCbbKmageccenxxxxxB
\else
??????\fi
\fi
}
\newcommand{\hatcurCCcitHmageccen}[1]{\ifnum#1=9 %
\hatcurCCcitHmageccenxxxxxA
\else
\ifnum#1=10 %
\hatcurCCcitHmageccenxxxxxB
\else
??????\fi
\fi
}
\newcommand{\hatcurCCcitJmageccen}[1]{\ifnum#1=9 %
\hatcurCCcitJmageccenxxxxxA
\else
\ifnum#1=10 %
\hatcurCCcitJmageccenxxxxxB
\else
??????\fi
\fi
}
\newcommand{\hatcurCCcitKmageccen}[1]{\ifnum#1=9 %
\hatcurCCcitKmageccenxxxxxA
\else
\ifnum#1=10 %
\hatcurCCcitKmageccenxxxxxB
\else
??????\fi
\fi
}
\newcommand{\hatcurCCdececcen}[1]{\ifnum#1=9 %
\hatcurCCdececcenxxxxxA
\else
\ifnum#1=10 %
\hatcurCCdececcenxxxxxB
\else
??????\fi
\fi
}
\newcommand{\hatcurCCesoHKmageccen}[1]{\ifnum#1=9 %
\hatcurCCesoHKmageccenxxxxxA
\else
\ifnum#1=10 %
\hatcurCCesoHKmageccenxxxxxB
\else
??????\fi
\fi
}
\newcommand{\hatcurCCesoHmageccen}[1]{\ifnum#1=9 %
\hatcurCCesoHmageccenxxxxxA
\else
\ifnum#1=10 %
\hatcurCCesoHmageccenxxxxxB
\else
??????\fi
\fi
}
\newcommand{\hatcurCCesoJHmageccen}[1]{\ifnum#1=9 %
\hatcurCCesoJHmageccenxxxxxA
\else
\ifnum#1=10 %
\hatcurCCesoJHmageccenxxxxxB
\else
??????\fi
\fi
}
\newcommand{\hatcurCCesoJKmageccen}[1]{\ifnum#1=9 %
\hatcurCCesoJKmageccenxxxxxA
\else
\ifnum#1=10 %
\hatcurCCesoJKmageccenxxxxxB
\else
??????\fi
\fi
}
\newcommand{\hatcurCCesoJmageccen}[1]{\ifnum#1=9 %
\hatcurCCesoJmageccenxxxxxA
\else
\ifnum#1=10 %
\hatcurCCesoJmageccenxxxxxB
\else
??????\fi
\fi
}
\newcommand{\hatcurCCesoKmageccen}[1]{\ifnum#1=9 %
\hatcurCCesoKmageccenxxxxxA
\else
\ifnum#1=10 %
\hatcurCCesoKmageccenxxxxxB
\else
??????\fi
\fi
}
\newcommand{\hatcurCCgsceccen}[1]{\ifnum#1=9 %
\hatcurCCgsceccenxxxxxA
\else
\ifnum#1=10 %
\hatcurCCgsceccenxxxxxB
\else
??????\fi
\fi
}
\newcommand{\hatcurCCmageccen}[1]{\ifnum#1=9 %
\hatcurCCmageccenxxxxxA
\else
\ifnum#1=10 %
\hatcurCCmageccenxxxxxB
\else
??????\fi
\fi
}
\newcommand{\hatcurCCpmdececcen}[1]{\ifnum#1=9 %
\hatcurCCpmdececcenxxxxxA
\else
\ifnum#1=10 %
\hatcurCCpmdececcenxxxxxB
\else
??????\fi
\fi
}
\newcommand{\hatcurCCpmeccen}[1]{\ifnum#1=9 %
\hatcurCCpmeccenxxxxxA
\else
\ifnum#1=10 %
\hatcurCCpmeccenxxxxxB
\else
??????\fi
\fi
}
\newcommand{\hatcurCCpmraeccen}[1]{\ifnum#1=9 %
\hatcurCCpmraeccenxxxxxA
\else
\ifnum#1=10 %
\hatcurCCpmraeccenxxxxxB
\else
??????\fi
\fi
}
\newcommand{\hatcurCCraeccen}[1]{\ifnum#1=9 %
\hatcurCCraeccenxxxxxA
\else
\ifnum#1=10 %
\hatcurCCraeccenxxxxxB
\else
??????\fi
\fi
}
\newcommand{\hatcurCCtassmBeccen}[1]{\ifnum#1=9 %
\hatcurCCtassmBeccenxxxxxA
\else
\ifnum#1=10 %
\hatcurCCtassmBeccenxxxxxB
\else
??????\fi
\fi
}
\newcommand{\hatcurCCtassmBshorteccen}[1]{\ifnum#1=9 %
\hatcurCCtassmBshorteccenxxxxxA
\else
\ifnum#1=10 %
\hatcurCCtassmBshorteccenxxxxxB
\else
??????\fi
\fi
}
\newcommand{\hatcurCCtassmgeccen}[1]{\ifnum#1=9 %
\hatcurCCtassmgeccenxxxxxA
\else
\ifnum#1=10 %
\hatcurCCtassmgeccenxxxxxB
\else
??????\fi
\fi
}
\newcommand{\hatcurCCtassmgshorteccen}[1]{\ifnum#1=9 %
\hatcurCCtassmgshorteccenxxxxxA
\else
\ifnum#1=10 %
\hatcurCCtassmgshorteccenxxxxxB
\else
??????\fi
\fi
}
\newcommand{\hatcurCCtassmieccen}[1]{\ifnum#1=9 %
\hatcurCCtassmieccenxxxxxA
\else
\ifnum#1=10 %
\hatcurCCtassmieccenxxxxxB
\else
??????\fi
\fi
}
\newcommand{\hatcurCCtassmIeccen}[1]{\ifnum#1=9 %
\hatcurCCtassmIeccenxxxxxA
\else
\ifnum#1=10 %
\hatcurCCtassmIeccenxxxxxB
\else
??????\fi
\fi
}
\newcommand{\hatcurCCtassmishorteccen}[1]{\ifnum#1=9 %
\hatcurCCtassmishorteccenxxxxxA
\else
\ifnum#1=10 %
\hatcurCCtassmishorteccenxxxxxB
\else
??????\fi
\fi
}
\newcommand{\hatcurCCtassmIshorteccen}[1]{\ifnum#1=9 %
\hatcurCCtassmIshorteccenxxxxxA
\else
\ifnum#1=10 %
\hatcurCCtassmIshorteccenxxxxxB
\else
??????\fi
\fi
}
\newcommand{\hatcurCCtassmreccen}[1]{\ifnum#1=9 %
\hatcurCCtassmreccenxxxxxA
\else
\ifnum#1=10 %
\hatcurCCtassmreccenxxxxxB
\else
??????\fi
\fi
}
\newcommand{\hatcurCCtassmrshorteccen}[1]{\ifnum#1=9 %
\hatcurCCtassmrshorteccenxxxxxA
\else
\ifnum#1=10 %
\hatcurCCtassmrshorteccenxxxxxB
\else
??????\fi
\fi
}
\newcommand{\hatcurCCtassmveccen}[1]{\ifnum#1=9 %
\hatcurCCtassmveccenxxxxxA
\else
\ifnum#1=10 %
\hatcurCCtassmveccenxxxxxB
\else
??????\fi
\fi
}
\newcommand{\hatcurCCtassmvshorteccen}[1]{\ifnum#1=9 %
\hatcurCCtassmvshorteccenxxxxxA
\else
\ifnum#1=10 %
\hatcurCCtassmvshorteccenxxxxxB
\else
??????\fi
\fi
}
\newcommand{\hatcurCCtwomasseccen}[1]{\ifnum#1=9 %
\hatcurCCtwomasseccenxxxxxA
\else
\ifnum#1=10 %
\hatcurCCtwomasseccenxxxxxB
\else
??????\fi
\fi
}
\newcommand{\hatcurCCtwomassHmageccen}[1]{\ifnum#1=9 %
\hatcurCCtwomassHmageccenxxxxxA
\else
\ifnum#1=10 %
\hatcurCCtwomassHmageccenxxxxxB
\else
??????\fi
\fi
}
\newcommand{\hatcurCCtwomassJmageccen}[1]{\ifnum#1=9 %
\hatcurCCtwomassJmageccenxxxxxA
\else
\ifnum#1=10 %
\hatcurCCtwomassJmageccenxxxxxB
\else
??????\fi
\fi
}
\newcommand{\hatcurCCtwomassKmageccen}[1]{\ifnum#1=9 %
\hatcurCCtwomassKmageccenxxxxxA
\else
\ifnum#1=10 %
\hatcurCCtwomassKmageccenxxxxxB
\else
??????\fi
\fi
}
\newcommand{\hatcurfieldeccen}[1]{\ifnum#1=9 %
\hatcurfieldeccenxxxxxA
\else
\ifnum#1=10 %
\hatcurfieldeccenxxxxxB
\else
??????\fi
\fi
}
\newcommand{\hatcurhtreccen}[1]{\ifnum#1=9 %
\hatcurhtreccenxxxxxA
\else
\ifnum#1=10 %
\hatcurhtreccenxxxxxB
\else
??????\fi
\fi
}
\newcommand{\hatcurISOageeccen}[1]{\ifnum#1=9 %
\hatcurISOageeccenxxxxxA
\else
\ifnum#1=10 %
\hatcurISOageeccenxxxxxB
\else
??????\fi
\fi
}
\newcommand{\hatcurISOJKeccen}[1]{\ifnum#1=9 %
\hatcurISOJKeccenxxxxxA
\else
\ifnum#1=10 %
\hatcurISOJKeccenxxxxxB
\else
??????\fi
\fi
}
\newcommand{\hatcurISOloggeccen}[1]{\ifnum#1=9 %
\hatcurISOloggeccenxxxxxA
\else
\ifnum#1=10 %
\hatcurISOloggeccenxxxxxB
\else
??????\fi
\fi
}
\newcommand{\hatcurISOlumeccen}[1]{\ifnum#1=9 %
\hatcurISOlumeccenxxxxxA
\else
\ifnum#1=10 %
\hatcurISOlumeccenxxxxxB
\else
??????\fi
\fi
}
\newcommand{\hatcurISOlumshorteccen}[1]{\ifnum#1=9 %
\hatcurISOlumshorteccenxxxxxA
\else
\ifnum#1=10 %
\hatcurISOlumshorteccenxxxxxB
\else
??????\fi
\fi
}
\newcommand{\hatcurISOmeccen}[1]{\ifnum#1=9 %
\hatcurISOmeccenxxxxxA
\else
\ifnum#1=10 %
\hatcurISOmeccenxxxxxB
\else
??????\fi
\fi
}
\newcommand{\hatcurISOMHeccen}[1]{\ifnum#1=9 %
\hatcurISOMHeccenxxxxxA
\else
\ifnum#1=10 %
\hatcurISOMHeccenxxxxxB
\else
??????\fi
\fi
}
\newcommand{\hatcurISOMJeccen}[1]{\ifnum#1=9 %
\hatcurISOMJeccenxxxxxA
\else
\ifnum#1=10 %
\hatcurISOMJeccenxxxxxB
\else
??????\fi
\fi
}
\newcommand{\hatcurISOMKeccen}[1]{\ifnum#1=9 %
\hatcurISOMKeccenxxxxxA
\else
\ifnum#1=10 %
\hatcurISOMKeccenxxxxxB
\else
??????\fi
\fi
}
\newcommand{\hatcurISOmlongeccen}[1]{\ifnum#1=9 %
\hatcurISOmlongeccenxxxxxA
\else
\ifnum#1=10 %
\hatcurISOmlongeccenxxxxxB
\else
??????\fi
\fi
}
\newcommand{\hatcurISOmshorteccen}[1]{\ifnum#1=9 %
\hatcurISOmshorteccenxxxxxA
\else
\ifnum#1=10 %
\hatcurISOmshorteccenxxxxxB
\else
??????\fi
\fi
}
\newcommand{\hatcurISOmveccen}[1]{\ifnum#1=9 %
\hatcurISOmveccenxxxxxA
\else
\ifnum#1=10 %
\hatcurISOmveccenxxxxxB
\else
??????\fi
\fi
}
\newcommand{\hatcurISOreccen}[1]{\ifnum#1=9 %
\hatcurISOreccenxxxxxA
\else
\ifnum#1=10 %
\hatcurISOreccenxxxxxB
\else
??????\fi
\fi
}
\newcommand{\hatcurISOrhoeccen}[1]{\ifnum#1=9 %
\hatcurISOrhoeccenxxxxxA
\else
\ifnum#1=10 %
\hatcurISOrhoeccenxxxxxB
\else
??????\fi
\fi
}
\newcommand{\hatcurISOrholongeccen}[1]{\ifnum#1=9 %
\hatcurISOrholongeccenxxxxxA
\else
\ifnum#1=10 %
\hatcurISOrholongeccenxxxxxB
\else
??????\fi
\fi
}
\newcommand{\hatcurISOrlongeccen}[1]{\ifnum#1=9 %
\hatcurISOrlongeccenxxxxxA
\else
\ifnum#1=10 %
\hatcurISOrlongeccenxxxxxB
\else
??????\fi
\fi
}
\newcommand{\hatcurISOrshorteccen}[1]{\ifnum#1=9 %
\hatcurISOrshorteccenxxxxxA
\else
\ifnum#1=10 %
\hatcurISOrshorteccenxxxxxB
\else
??????\fi
\fi
}
\newcommand{\hatcurISOsigmaeccen}[1]{\ifnum#1=9 %
\hatcurISOsigmaeccenxxxxxA
\else
\ifnum#1=10 %
\hatcurISOsigmaeccenxxxxxB
\else
??????\fi
\fi
}
\newcommand{\hatcurISOspececcen}[1]{\ifnum#1=9 %
\hatcurISOspececcenxxxxxA
\else
\ifnum#1=10 %
\hatcurISOspececcenxxxxxB
\else
??????\fi
\fi
}
\newcommand{\hatcurISOvieccen}[1]{\ifnum#1=9 %
\hatcurISOvieccenxxxxxA
\else
\ifnum#1=10 %
\hatcurISOvieccenxxxxxB
\else
??????\fi
\fi
}
\newcommand{\hatcurLBigeccen}[1]{\ifnum#1=9 %
\hatcurLBigeccenxxxxxA
\else
\ifnum#1=10 %
\hatcurLBigeccenxxxxxB
\else
??????\fi
\fi
}
\newcommand{\hatcurLBiieccen}[1]{\ifnum#1=9 %
\hatcurLBiieccenxxxxxA
\else
\ifnum#1=10 %
\hatcurLBiieccenxxxxxB
\else
??????\fi
\fi
}
\newcommand{\hatcurLBiIeccen}[1]{\ifnum#1=9 %
\hatcurLBiIeccenxxxxxA
\else
\ifnum#1=10 %
\hatcurLBiIeccenxxxxxB
\else
??????\fi
\fi
}
\newcommand{\hatcurLBiigeccen}[1]{\ifnum#1=9 %
\hatcurLBiigeccenxxxxxA
\else
\ifnum#1=10 %
\hatcurLBiigeccenxxxxxB
\else
??????\fi
\fi
}
\newcommand{\hatcurLBiiieccen}[1]{\ifnum#1=9 %
\hatcurLBiiieccenxxxxxA
\else
\ifnum#1=10 %
\hatcurLBiiieccenxxxxxB
\else
??????\fi
\fi
}
\newcommand{\hatcurLBiiIeccen}[1]{\ifnum#1=9 %
\hatcurLBiiIeccenxxxxxA
\else
\ifnum#1=10 %
\hatcurLBiiIeccenxxxxxB
\else
??????\fi
\fi
}
\newcommand{\hatcurLBiikepeccen}[1]{\ifnum#1=10 %
\hatcurLBiikepeccenxxxxxB
\else
??????\fi
}
\newcommand{\hatcurLBiireccen}[1]{\ifnum#1=9 %
\hatcurLBiireccenxxxxxA
\else
\ifnum#1=10 %
\hatcurLBiireccenxxxxxB
\else
??????\fi
\fi
}
\newcommand{\hatcurLBiiReccen}[1]{\ifnum#1=9 %
\hatcurLBiiReccenxxxxxA
\else
\ifnum#1=10 %
\hatcurLBiiReccenxxxxxB
\else
??????\fi
\fi
}
\newcommand{\hatcurLBiizeccen}[1]{\ifnum#1=9 %
\hatcurLBiizeccenxxxxxA
\else
\ifnum#1=10 %
\hatcurLBiizeccenxxxxxB
\else
??????\fi
\fi
}
\newcommand{\hatcurLBikepeccen}[1]{\ifnum#1=10 %
\hatcurLBikepeccenxxxxxB
\else
??????\fi
}
\newcommand{\hatcurLBireccen}[1]{\ifnum#1=9 %
\hatcurLBireccenxxxxxA
\else
\ifnum#1=10 %
\hatcurLBireccenxxxxxB
\else
??????\fi
\fi
}
\newcommand{\hatcurLBiReccen}[1]{\ifnum#1=9 %
\hatcurLBiReccenxxxxxA
\else
\ifnum#1=10 %
\hatcurLBiReccenxxxxxB
\else
??????\fi
\fi
}
\newcommand{\hatcurLBizeccen}[1]{\ifnum#1=9 %
\hatcurLBizeccenxxxxxA
\else
\ifnum#1=10 %
\hatcurLBizeccenxxxxxB
\else
??????\fi
\fi
}
\newcommand{\hatcurLCbsqeccen}[1]{\ifnum#1=9 %
\hatcurLCbsqeccenxxxxxA
\else
\ifnum#1=10 %
\hatcurLCbsqeccenxxxxxB
\else
??????\fi
\fi
}
\newcommand{\hatcurLCdipeccen}[1]{\ifnum#1=9 %
\hatcurLCdipeccenxxxxxA
\else
\ifnum#1=10 %
\hatcurLCdipeccenxxxxxB
\else
??????\fi
\fi
}
\newcommand{\hatcurLCdureccen}[1]{\ifnum#1=9 %
\hatcurLCdureccenxxxxxA
\else
\ifnum#1=10 %
\hatcurLCdureccenxxxxxB
\else
??????\fi
\fi
}
\newcommand{\hatcurLCdurhreccen}[1]{\ifnum#1=9 %
\hatcurLCdurhreccenxxxxxA
\else
\ifnum#1=10 %
\hatcurLCdurhreccenxxxxxB
\else
??????\fi
\fi
}
\newcommand{\hatcurLCdurhrshorteccen}[1]{\ifnum#1=9 %
\hatcurLCdurhrshorteccenxxxxxA
\else
\ifnum#1=10 %
\hatcurLCdurhrshorteccenxxxxxB
\else
??????\fi
\fi
}
\newcommand{\hatcurLCdurshorteccen}[1]{\ifnum#1=9 %
\hatcurLCdurshorteccenxxxxxA
\else
\ifnum#1=10 %
\hatcurLCdurshorteccenxxxxxB
\else
??????\fi
\fi
}
\newcommand{\hatcurLChatnetmeccen}[1]{\ifnum#1=9 %
\hatcurLChatnetmeccenxxxxxA
\else
\ifnum#1=10 %
\hatcurLChatnetmeccenxxxxxB
\else
??????\fi
\fi
}
\newcommand{\hatcurLCiblendeccen}[1]{\ifnum#1=9 %
\hatcurLCiblendeccenxxxxxA
\else
\ifnum#1=10 %
\hatcurLCiblendeccenxxxxxB
\else
??????\fi
\fi
}
\newcommand{\hatcurLCimpeccen}[1]{\ifnum#1=9 %
\hatcurLCimpeccenxxxxxA
\else
\ifnum#1=10 %
\hatcurLCimpeccenxxxxxB
\else
??????\fi
\fi
}
\newcommand{\hatcurLCingdureccen}[1]{\ifnum#1=9 %
\hatcurLCingdureccenxxxxxA
\else
\ifnum#1=10 %
\hatcurLCingdureccenxxxxxB
\else
??????\fi
\fi
}
\newcommand{\hatcurLCPeccen}[1]{\ifnum#1=9 %
\hatcurLCPeccenxxxxxA
\else
\ifnum#1=10 %
\hatcurLCPeccenxxxxxB
\else
??????\fi
\fi
}
\newcommand{\hatcurLCPprececcen}[1]{\ifnum#1=9 %
\hatcurLCPprececcenxxxxxA
\else
\ifnum#1=10 %
\hatcurLCPprececcenxxxxxB
\else
??????\fi
\fi
}
\newcommand{\hatcurLCPshorteccen}[1]{\ifnum#1=9 %
\hatcurLCPshorteccenxxxxxA
\else
\ifnum#1=10 %
\hatcurLCPshorteccenxxxxxB
\else
??????\fi
\fi
}
\newcommand{\hatcurLCqeccen}[1]{\ifnum#1=9 %
\hatcurLCqeccenxxxxxA
\else
\ifnum#1=10 %
\hatcurLCqeccenxxxxxB
\else
??????\fi
\fi
}
\newcommand{\hatcurLCqshorteccen}[1]{\ifnum#1=9 %
\hatcurLCqshorteccenxxxxxA
\else
\ifnum#1=10 %
\hatcurLCqshorteccenxxxxxB
\else
??????\fi
\fi
}
\newcommand{\hatcurLCrprstareccen}[1]{\ifnum#1=9 %
\hatcurLCrprstareccenxxxxxA
\else
\ifnum#1=10 %
\hatcurLCrprstareccenxxxxxB
\else
??????\fi
\fi
}
\newcommand{\hatcurLCTAeccen}[1]{\ifnum#1=9 %
\hatcurLCTAeccenxxxxxA
\else
\ifnum#1=10 %
\hatcurLCTAeccenxxxxxB
\else
??????\fi
\fi
}
\newcommand{\hatcurLCTBeccen}[1]{\ifnum#1=9 %
\hatcurLCTBeccenxxxxxA
\else
\ifnum#1=10 %
\hatcurLCTBeccenxxxxxB
\else
??????\fi
\fi
}
\newcommand{\hatcurLCTeccen}[1]{\ifnum#1=9 %
\hatcurLCTeccenxxxxxA
\else
\ifnum#1=10 %
\hatcurLCTeccenxxxxxB
\else
??????\fi
\fi
}
\newcommand{\hatcurLCzetaeccen}[1]{\ifnum#1=9 %
\hatcurLCzetaeccenxxxxxA
\else
\ifnum#1=10 %
\hatcurLCzetaeccenxxxxxB
\else
??????\fi
\fi
}
\newcommand{\hatcurPPaequiveccen}[1]{\ifnum#1=9 %
\hatcurPPaequiveccenxxxxxA
\else
\ifnum#1=10 %
\hatcurPPaequiveccenxxxxxB
\else
??????\fi
\fi
}
\newcommand{\hatcurPPareccen}[1]{\ifnum#1=9 %
\hatcurPPareccenxxxxxA
\else
\ifnum#1=10 %
\hatcurPPareccenxxxxxB
\else
??????\fi
\fi
}
\newcommand{\hatcurPPareleccen}[1]{\ifnum#1=9 %
\hatcurPPareleccenxxxxxA
\else
\ifnum#1=10 %
\hatcurPPareleccenxxxxxB
\else
??????\fi
\fi
}
\newcommand{\hatcurPPfluxapdimeccen}[1]{\ifnum#1=9 %
\hatcurPPfluxapdimeccenxxxxxA
\else
\ifnum#1=10 %
\hatcurPPfluxapdimeccenxxxxxB
\else
??????\fi
\fi
}
\newcommand{\hatcurPPfluxapeccen}[1]{\ifnum#1=9 %
\hatcurPPfluxapeccenxxxxxA
\else
\ifnum#1=10 %
\hatcurPPfluxapeccenxxxxxB
\else
??????\fi
\fi
}
\newcommand{\hatcurPPfluxavgdimeccen}[1]{\ifnum#1=9 %
\hatcurPPfluxavgdimeccenxxxxxA
\else
\ifnum#1=10 %
\hatcurPPfluxavgdimeccenxxxxxB
\else
??????\fi
\fi
}
\newcommand{\hatcurPPfluxavgeccen}[1]{\ifnum#1=9 %
\hatcurPPfluxavgeccenxxxxxA
\else
\ifnum#1=10 %
\hatcurPPfluxavgeccenxxxxxB
\else
??????\fi
\fi
}
\newcommand{\hatcurPPfluxavglogeccen}[1]{\ifnum#1=9 %
\hatcurPPfluxavglogeccenxxxxxA
\else
\ifnum#1=10 %
\hatcurPPfluxavglogeccenxxxxxB
\else
??????\fi
\fi
}
\newcommand{\hatcurPPfluxperidimeccen}[1]{\ifnum#1=9 %
\hatcurPPfluxperidimeccenxxxxxA
\else
\ifnum#1=10 %
\hatcurPPfluxperidimeccenxxxxxB
\else
??????\fi
\fi
}
\newcommand{\hatcurPPfluxperieccen}[1]{\ifnum#1=9 %
\hatcurPPfluxperieccenxxxxxA
\else
\ifnum#1=10 %
\hatcurPPfluxperieccenxxxxxB
\else
??????\fi
\fi
}
\newcommand{\hatcurPPgeccen}[1]{\ifnum#1=9 %
\hatcurPPgeccenxxxxxA
\else
\ifnum#1=10 %
\hatcurPPgeccenxxxxxB
\else
??????\fi
\fi
}
\newcommand{\hatcurPPieccen}[1]{\ifnum#1=9 %
\hatcurPPieccenxxxxxA
\else
\ifnum#1=10 %
\hatcurPPieccenxxxxxB
\else
??????\fi
\fi
}
\newcommand{\hatcurPPloggeccen}[1]{\ifnum#1=9 %
\hatcurPPloggeccenxxxxxA
\else
\ifnum#1=10 %
\hatcurPPloggeccenxxxxxB
\else
??????\fi
\fi
}
\newcommand{\hatcurPPmeccen}[1]{\ifnum#1=9 %
\hatcurPPmeccenxxxxxA
\else
\ifnum#1=10 %
\hatcurPPmeccenxxxxxB
\else
??????\fi
\fi
}
\newcommand{\hatcurPPmeeccen}[1]{\ifnum#1=9 %
\hatcurPPmeeccenxxxxxA
\else
\ifnum#1=10 %
\hatcurPPmeeccenxxxxxB
\else
??????\fi
\fi
}
\newcommand{\hatcurPPmelongeccen}[1]{\ifnum#1=9 %
\hatcurPPmelongeccenxxxxxA
\else
\ifnum#1=10 %
\hatcurPPmelongeccenxxxxxB
\else
??????\fi
\fi
}
\newcommand{\hatcurPPmeshorteccen}[1]{\ifnum#1=9 %
\hatcurPPmeshorteccenxxxxxA
\else
\ifnum#1=10 %
\hatcurPPmeshorteccenxxxxxB
\else
??????\fi
\fi
}
\newcommand{\hatcurPPmlongeccen}[1]{\ifnum#1=9 %
\hatcurPPmlongeccenxxxxxA
\else
\ifnum#1=10 %
\hatcurPPmlongeccenxxxxxB
\else
??????\fi
\fi
}
\newcommand{\hatcurPPmrcorreccen}[1]{\ifnum#1=9 %
\hatcurPPmrcorreccenxxxxxA
\else
\ifnum#1=10 %
\hatcurPPmrcorreccenxxxxxB
\else
??????\fi
\fi
}
\newcommand{\hatcurPPmshorteccen}[1]{\ifnum#1=9 %
\hatcurPPmshorteccenxxxxxA
\else
\ifnum#1=10 %
\hatcurPPmshorteccenxxxxxB
\else
??????\fi
\fi
}
\newcommand{\hatcurPPperieccen}[1]{\ifnum#1=9 %
\hatcurPPperieccenxxxxxA
\else
\ifnum#1=10 %
\hatcurPPperieccenxxxxxB
\else
??????\fi
\fi
}
\newcommand{\hatcurPPphiconjeccen}[1]{\ifnum#1=9 %
\hatcurPPphiconjeccenxxxxxA
\else
\ifnum#1=10 %
\hatcurPPphiconjeccenxxxxxB
\else
??????\fi
\fi
}
\newcommand{\hatcurPPreccen}[1]{\ifnum#1=9 %
\hatcurPPreccenxxxxxA
\else
\ifnum#1=10 %
\hatcurPPreccenxxxxxB
\else
??????\fi
\fi
}
\newcommand{\hatcurPPreeccen}[1]{\ifnum#1=9 %
\hatcurPPreeccenxxxxxA
\else
\ifnum#1=10 %
\hatcurPPreeccenxxxxxB
\else
??????\fi
\fi
}
\newcommand{\hatcurPPrelongeccen}[1]{\ifnum#1=9 %
\hatcurPPrelongeccenxxxxxA
\else
\ifnum#1=10 %
\hatcurPPrelongeccenxxxxxB
\else
??????\fi
\fi
}
\newcommand{\hatcurPPreshorteccen}[1]{\ifnum#1=9 %
\hatcurPPreshorteccenxxxxxA
\else
\ifnum#1=10 %
\hatcurPPreshorteccenxxxxxB
\else
??????\fi
\fi
}
\newcommand{\hatcurPPrhoeccen}[1]{\ifnum#1=9 %
\hatcurPPrhoeccenxxxxxA
\else
\ifnum#1=10 %
\hatcurPPrhoeccenxxxxxB
\else
??????\fi
\fi
}
\newcommand{\hatcurPPrlongeccen}[1]{\ifnum#1=9 %
\hatcurPPrlongeccenxxxxxA
\else
\ifnum#1=10 %
\hatcurPPrlongeccenxxxxxB
\else
??????\fi
\fi
}
\newcommand{\hatcurPPrshorteccen}[1]{\ifnum#1=9 %
\hatcurPPrshorteccenxxxxxA
\else
\ifnum#1=10 %
\hatcurPPrshorteccenxxxxxB
\else
??????\fi
\fi
}
\newcommand{\hatcurPPtcirceccen}[1]{\ifnum#1=9 %
\hatcurPPtcirceccenxxxxxA
\else
\ifnum#1=10 %
\hatcurPPtcirceccenxxxxxB
\else
??????\fi
\fi
}
\newcommand{\hatcurPPteffeccen}[1]{\ifnum#1=9 %
\hatcurPPteffeccenxxxxxA
\else
\ifnum#1=10 %
\hatcurPPteffeccenxxxxxB
\else
??????\fi
\fi
}
\newcommand{\hatcurPPthetaeccen}[1]{\ifnum#1=9 %
\hatcurPPthetaeccenxxxxxA
\else
\ifnum#1=10 %
\hatcurPPthetaeccenxxxxxB
\else
??????\fi
\fi
}
\newcommand{\hatcurPPtinfalleccen}[1]{\ifnum#1=9 %
\hatcurPPtinfalleccenxxxxxA
\else
\ifnum#1=10 %
\hatcurPPtinfalleccenxxxxxB
\else
??????\fi
\fi
}
\newcommand{\hatcurRVecceneccen}[1]{\ifnum#1=9 %
\hatcurRVecceneccenxxxxxA
\else
\ifnum#1=10 %
\hatcurRVecceneccenxxxxxB
\else
??????\fi
\fi
}
\newcommand{\hatcurRVeccentwosiglimeccen}[1]{\ifnum#1=9 %
\hatcurRVeccentwosiglimeccenxxxxxA
\else
\ifnum#1=10 %
\hatcurRVeccentwosiglimeccenxxxxxB
\else
??????\fi
\fi
}
\newcommand{\hatcurRVfitrmsAeccen}[1]{\ifnum#1=9 %
\hatcurRVfitrmsAeccenxxxxxA
\else
\ifnum#1=10 %
\hatcurRVfitrmsAeccenxxxxxB
\else
??????\fi
\fi
}
\newcommand{\hatcurRVfitrmsBeccen}[1]{\ifnum#1=9 %
\hatcurRVfitrmsBeccenxxxxxA
\else
\ifnum#1=10 %
\hatcurRVfitrmsBeccenxxxxxB
\else
??????\fi
\fi
}
\newcommand{\hatcurRVfitrmsCeccen}[1]{\ifnum#1=9 %
\hatcurRVfitrmsCeccenxxxxxA
\else
\ifnum#1=10 %
\hatcurRVfitrmsCeccenxxxxxB
\else
??????\fi
\fi
}
\newcommand{\hatcurRVgammaAeccen}[1]{\ifnum#1=9 %
\hatcurRVgammaAeccenxxxxxA
\else
\ifnum#1=10 %
\hatcurRVgammaAeccenxxxxxB
\else
??????\fi
\fi
}
\newcommand{\hatcurRVgammaBeccen}[1]{\ifnum#1=9 %
\hatcurRVgammaBeccenxxxxxA
\else
\ifnum#1=10 %
\hatcurRVgammaBeccenxxxxxB
\else
??????\fi
\fi
}
\newcommand{\hatcurRVgammaCeccen}[1]{\ifnum#1=9 %
\hatcurRVgammaCeccenxxxxxA
\else
\ifnum#1=10 %
\hatcurRVgammaCeccenxxxxxB
\else
??????\fi
\fi
}
\newcommand{\hatcurRVheccen}[1]{\ifnum#1=9 %
\hatcurRVheccenxxxxxA
\else
\ifnum#1=10 %
\hatcurRVheccenxxxxxB
\else
??????\fi
\fi
}
\newcommand{\hatcurRVjitterAeccen}[1]{\ifnum#1=9 %
\hatcurRVjitterAeccenxxxxxA
\else
\ifnum#1=10 %
\hatcurRVjitterAeccenxxxxxB
\else
??????\fi
\fi
}
\newcommand{\hatcurRVjitterBeccen}[1]{\ifnum#1=9 %
\hatcurRVjitterBeccenxxxxxA
\else
\ifnum#1=10 %
\hatcurRVjitterBeccenxxxxxB
\else
??????\fi
\fi
}
\newcommand{\hatcurRVjitterCeccen}[1]{\ifnum#1=9 %
\hatcurRVjitterCeccenxxxxxA
\else
\ifnum#1=10 %
\hatcurRVjitterCeccenxxxxxB
\else
??????\fi
\fi
}
\newcommand{\hatcurRVkeccen}[1]{\ifnum#1=9 %
\hatcurRVkeccenxxxxxA
\else
\ifnum#1=10 %
\hatcurRVkeccenxxxxxB
\else
??????\fi
\fi
}
\newcommand{\hatcurRVKeccen}[1]{\ifnum#1=9 %
\hatcurRVKeccenxxxxxA
\else
\ifnum#1=10 %
\hatcurRVKeccenxxxxxB
\else
??????\fi
\fi
}
\newcommand{\hatcurRVomegaeccen}[1]{\ifnum#1=9 %
\hatcurRVomegaeccenxxxxxA
\else
\ifnum#1=10 %
\hatcurRVomegaeccenxxxxxB
\else
??????\fi
\fi
}
\newcommand{\hatcurRVrheccen}[1]{\ifnum#1=9 %
\hatcurRVrheccenxxxxxA
\else
\ifnum#1=10 %
\hatcurRVrheccenxxxxxB
\else
??????\fi
\fi
}
\newcommand{\hatcurRVrkeccen}[1]{\ifnum#1=9 %
\hatcurRVrkeccenxxxxxA
\else
\ifnum#1=10 %
\hatcurRVrkeccenxxxxxB
\else
??????\fi
\fi
}
\newcommand{\hatcurRVtroneeccen}[1]{\ifnum#1=9 %
\hatcurRVtroneeccenxxxxxA
\else
\ifnum#1=10 %
\hatcurRVtroneeccenxxxxxB
\else
??????\fi
\fi
}
\newcommand{\hatcurRVtrtwoeccen}[1]{\ifnum#1=9 %
\hatcurRVtrtwoeccenxxxxxA
\else
\ifnum#1=10 %
\hatcurRVtrtwoeccenxxxxxB
\else
??????\fi
\fi
}
\newcommand{\hatcurSMEiiloggeccen}[1]{\ifnum#1=9 %
\hatcurSMEiiloggeccenxxxxxA
\else
\ifnum#1=10 %
\hatcurSMEiiloggeccenxxxxxB
\else
??????\fi
\fi
}
\newcommand{\hatcurSMEiiteffeccen}[1]{\ifnum#1=9 %
\hatcurSMEiiteffeccenxxxxxA
\else
\ifnum#1=10 %
\hatcurSMEiiteffeccenxxxxxB
\else
??????\fi
\fi
}
\newcommand{\hatcurSMEiivsineccen}[1]{\ifnum#1=9 %
\hatcurSMEiivsineccenxxxxxA
\else
\ifnum#1=10 %
\hatcurSMEiivsineccenxxxxxB
\else
??????\fi
\fi
}
\newcommand{\hatcurSMEiizfeheccen}[1]{\ifnum#1=9 %
\hatcurSMEiizfeheccenxxxxxA
\else
\ifnum#1=10 %
\hatcurSMEiizfeheccenxxxxxB
\else
??????\fi
\fi
}
\newcommand{\hatcurSMEiizfehshorteccen}[1]{\ifnum#1=9 %
\hatcurSMEiizfehshorteccenxxxxxA
\else
\ifnum#1=10 %
\hatcurSMEiizfehshorteccenxxxxxB
\else
??????\fi
\fi
}
\newcommand{\hatcurSMEiloggeccen}[1]{\ifnum#1=9 %
\hatcurSMEiloggeccenxxxxxA
\else
\ifnum#1=10 %
\hatcurSMEiloggeccenxxxxxB
\else
??????\fi
\fi
}
\newcommand{\hatcurSMEiteffeccen}[1]{\ifnum#1=9 %
\hatcurSMEiteffeccenxxxxxA
\else
\ifnum#1=10 %
\hatcurSMEiteffeccenxxxxxB
\else
??????\fi
\fi
}
\newcommand{\hatcurSMEivmaceccen}[1]{\ifnum#1=9 %
\hatcurSMEivmaceccenxxxxxA
\else
\ifnum#1=10 %
\hatcurSMEivmaceccenxxxxxB
\else
??????\fi
\fi
}
\newcommand{\hatcurSMEivmiceccen}[1]{\ifnum#1=9 %
\hatcurSMEivmiceccenxxxxxA
\else
\ifnum#1=10 %
\hatcurSMEivmiceccenxxxxxB
\else
??????\fi
\fi
}
\newcommand{\hatcurSMEivsineccen}[1]{\ifnum#1=9 %
\hatcurSMEivsineccenxxxxxA
\else
\ifnum#1=10 %
\hatcurSMEivsineccenxxxxxB
\else
??????\fi
\fi
}
\newcommand{\hatcurSMEizfeheccen}[1]{\ifnum#1=9 %
\hatcurSMEizfeheccenxxxxxA
\else
\ifnum#1=10 %
\hatcurSMEizfeheccenxxxxxB
\else
??????\fi
\fi
}
\newcommand{\hatcurSMEizfehshorteccen}[1]{\ifnum#1=9 %
\hatcurSMEizfehshorteccenxxxxxA
\else
\ifnum#1=10 %
\hatcurSMEizfehshorteccenxxxxxB
\else
??????\fi
\fi
}
\newcommand{\hatcurXAveccen}[1]{\ifnum#1=9 %
\hatcurXAveccenxxxxxA
\else
\ifnum#1=10 %
\hatcurXAveccenxxxxxB
\else
??????\fi
\fi
}
\newcommand{\hatcurXdisteccen}[1]{\ifnum#1=9 %
\hatcurXdisteccenxxxxxA
\else
\ifnum#1=10 %
\hatcurXdisteccenxxxxxB
\else
??????\fi
\fi
}
\newcommand{\hatcurXdistredeccen}[1]{\ifnum#1=9 %
\hatcurXdistredeccenxxxxxA
\else
\ifnum#1=10 %
\hatcurXdistredeccenxxxxxB
\else
??????\fi
\fi
}
\newcommand{\hatcurXEBVeccen}[1]{\ifnum#1=9 %
\hatcurXEBVeccenxxxxxA
\else
\ifnum#1=10 %
\hatcurXEBVeccenxxxxxB
\else
??????\fi
\fi
}
\newcommand{\hatcurXjhisoredeccen}[1]{\ifnum#1=9 %
\hatcurXjhisoredeccenxxxxxA
\else
\ifnum#1=10 %
\hatcurXjhisoredeccenxxxxxB
\else
??????\fi
\fi
}
\newcommand{\hatcurXjkisoredeccen}[1]{\ifnum#1=9 %
\hatcurXjkisoredeccenxxxxxA
\else
\ifnum#1=10 %
\hatcurXjkisoredeccenxxxxxB
\else
??????\fi
\fi
}
\newcommand{\hatcurXmhisoredeccen}[1]{\ifnum#1=9 %
\hatcurXmhisoredeccenxxxxxA
\else
\ifnum#1=10 %
\hatcurXmhisoredeccenxxxxxB
\else
??????\fi
\fi
}
\newcommand{\hatcurXmiisoredeccen}[1]{\ifnum#1=9 %
\hatcurXmiisoredeccenxxxxxA
\else
\ifnum#1=10 %
\hatcurXmiisoredeccenxxxxxB
\else
??????\fi
\fi
}
\newcommand{\hatcurXmjisoredeccen}[1]{\ifnum#1=9 %
\hatcurXmjisoredeccenxxxxxA
\else
\ifnum#1=10 %
\hatcurXmjisoredeccenxxxxxB
\else
??????\fi
\fi
}
\newcommand{\hatcurXmkisoredeccen}[1]{\ifnum#1=9 %
\hatcurXmkisoredeccenxxxxxA
\else
\ifnum#1=10 %
\hatcurXmkisoredeccenxxxxxB
\else
??????\fi
\fi
}
\newcommand{\hatcurXmvisoredeccen}[1]{\ifnum#1=9 %
\hatcurXmvisoredeccenxxxxxA
\else
\ifnum#1=10 %
\hatcurXmvisoredeccenxxxxxB
\else
??????\fi
\fi
}
\newcommand{\hatcurXsecdureccen}[1]{\ifnum#1=9 %
\hatcurXsecdureccenxxxxxA
\else
\ifnum#1=10 %
\hatcurXsecdureccenxxxxxB
\else
??????\fi
\fi
}
\newcommand{\hatcurXsecingdureccen}[1]{\ifnum#1=9 %
\hatcurXsecingdureccenxxxxxA
\else
\ifnum#1=10 %
\hatcurXsecingdureccenxxxxxB
\else
??????\fi
\fi
}
\newcommand{\hatcurXsecondaryeccen}[1]{\ifnum#1=9 %
\hatcurXsecondaryeccenxxxxxA
\else
\ifnum#1=10 %
\hatcurXsecondaryeccenxxxxxB
\else
??????\fi
\fi
}
\newcommand{\hatcurXsecphaseeccen}[1]{\ifnum#1=9 %
\hatcurXsecphaseeccenxxxxxA
\else
\ifnum#1=10 %
\hatcurXsecphaseeccenxxxxxB
\else
??????\fi
\fi
}
\newcommand{\hatcurXviisoredeccen}[1]{\ifnum#1=9 %
\hatcurXviisoredeccenxxxxxA
\else
\ifnum#1=10 %
\hatcurXviisoredeccenxxxxxB
\else
??????\fi
\fi
}
\newcommand{\hatcurXvkisoredeccen}[1]{\ifnum#1=9 %
\hatcurXvkisoredeccenxxxxxA
\else
\ifnum#1=10 %
\hatcurXvkisoredeccenxxxxxB
\else
??????\fi
\fi
}

\newcommand{\hatcurxxxxxA}{HATS-9}
\newcommand{\hatcurbxxxxxA}{HATS-9b}
\newcommand{\hatcurcxxxxxA}{HATS-9c}
\newcommand{\hatcurplanetnumxxxxxA}{9}

\newcommand{\hatcurRVgammaabsxxxxxA}{***TBD***}

\newcommand{\hatcurRVgammarelxxxxxA}{***TBD***}                           
\newcommand{\hatcurCCtassvixxxxxA}{***TBD***}                  
\newcommand{\hatcurSMEversionxxxxxA}{ii}                                      
\newcommand{\hatcurisoshortxxxxxA}{YY}
\newcommand{\hatcurisofullxxxxxA}{Yonsei-Yale (YY)}
\newcommand{\hatcurisocitexxxxxA}{yi:2001}
\newcommand{\hatcurlumindxxxxxA}{\arstar}
\newcommand{\hatcurjhkfilsetxxxxxA}{ESO}

\newcommand{\hatcurSMEteffxxxxxA}{\ifthenelse{\equal{\hatcurSMEversionxxxxxA}{i}}{\hatcurSMEiteff{\hatcurplanetnumxxxxxA}}{\hatcurSMEiiteff{\hatcurplanetnumxxxxxA}}}
\newcommand{\hatcurSMEzfehxxxxxA}{\ifthenelse{\equal{\hatcurSMEversionxxxxxA}{i}}{\hatcurSMEizfeh{\hatcurplanetnumxxxxxA}}{\hatcurSMEiizfeh{\hatcurplanetnumxxxxxA}}}
\newcommand{\hatcurSMEzfehshortxxxxxA}{\ifthenelse{\equal{\hatcurSMEversionxxxxxA}{i}}{\hatcurSMEizfehshort{\hatcurplanetnumxxxxxA}}{\hatcurSMEiizfehshort{\hatcurplanetnumxxxxxA}}}
\newcommand{\hatcurSMEloggxxxxxA}{\ifthenelse{\equal{\hatcurSMEversionxxxxxA}{i}}{\hatcurSMEilogg{\hatcurplanetnumxxxxxA}}{\hatcurSMEiilogg{\hatcurplanetnumxxxxxA}}}
\newcommand{\hatcurSMEvsinxxxxxA}{\ifthenelse{\equal{\hatcurSMEversionxxxxxA}{i}}{\hatcurSMEivsin{\hatcurplanetnumxxxxxA}}{\hatcurSMEiivsin{\hatcurplanetnumxxxxxA}}}
\newcommand{\hatcurSMEvmacxxxxxA}{\ifthenelse{\equal{\hatcurSMEversionxxxxxA}{i}}{\hatcurSMEivmac{\hatcurplanetnumxxxxxA}}{\hatcurSMEiivmac{\hatcurplanetnumxxxxxA}}}
\newcommand{\hatcurSMEvmicxxxxxA}{\ifthenelse{\equal{\hatcurSMEversionxxxxxA}{i}}{\hatcurSMEivmic{\hatcurplanetnumxxxxxA}}{\hatcurSMEiivmic{\hatcurplanetnumxxxxxA}}}

\newcommand{\hatcurxxxxxB}{HATS-10}
\newcommand{\hatcurbxxxxxB}{HATS-10b}
\newcommand{\hatcurcxxxxxB}{HATS-10c}
\newcommand{\hatcurplanetnumxxxxxB}{10}
\newcommand{\hatcurRVgammaabsxxxxxB}{***TBD***} 
\newcommand{\hatcurRVgammarelxxxxxB}{***TBD***} 
\newcommand{\hatcurCCtassvixxxxxB}{***TBD***}
\newcommand{\hatcurSMEversionxxxxxB}{ii} 
\newcommand{\hatcurisoshortxxxxxB}{YY}
\newcommand{\hatcurisofullxxxxxB}{Yonsei-Yale (YY)}
\newcommand{\hatcurisocitexxxxxB}{yi:2001}
\newcommand{\hatcurlumindxxxxxB}{\arstar}
\newcommand{\hatcurjhkfilsetxxxxxB}{ESO}

\newcommand{\hatcurSMEteffxxxxxB}{\ifthenelse{\equal{\hatcurSMEversionxxxxxB}{i}}{\hatcurSMEiteff{\hatcurplanetnumxxxxxB}}{\hatcurSMEiiteff{\hatcurplanetnumxxxxxB}}}
\newcommand{\hatcurSMEzfehxxxxxB}{\ifthenelse{\equal{\hatcurSMEversionxxxxxB}{i}}{\hatcurSMEizfeh{\hatcurplanetnumxxxxxB}}{\hatcurSMEiizfeh{\hatcurplanetnumxxxxxB}}}
\newcommand{\hatcurSMEzfehshortxxxxxB}{\ifthenelse{\equal{\hatcurSMEversionxxxxxB}{i}}{\hatcurSMEizfehshort{\hatcurplanetnumxxxxxB}}{\hatcurSMEiizfehshort{\hatcurplanetnumxxxxxB}}}
\newcommand{\hatcurSMEloggxxxxxB}{\ifthenelse{\equal{\hatcurSMEversionxxxxxB}{i}}{\hatcurSMEilogg{\hatcurplanetnumxxxxxB}}{\hatcurSMEiilogg{\hatcurplanetnumxxxxxB}}}
\newcommand{\hatcurSMEvsinxxxxxB}{\ifthenelse{\equal{\hatcurSMEversionxxxxxB}{i}}{\hatcurSMEivsin{\hatcurplanetnumxxxxxB}}{\hatcurSMEiivsin{\hatcurplanetnumxxxxxB}}}
\newcommand{\hatcurSMEvmacxxxxxB}{\ifthenelse{\equal{\hatcurSMEversionxxxxxB}{i}}{\hatcurSMEivmac{\hatcurplanetnumxxxxxB}}{\hatcurSMEiivmac{\hatcurplanetnumxxxxxB}}}
\newcommand{\hatcurSMEvmicxxxxxB}{\ifthenelse{\equal{\hatcurSMEversionxxxxxB}{i}}{\hatcurSMEivmic{\hatcurplanetnumxxxxxB}}{\hatcurSMEiivmic{\hatcurplanetnumxxxxxB}}}

\newcommand{\hatcur}[1]{\ifnum#1=9 %
\hatcurxxxxxA
\else
\ifnum#1=10 %
\hatcurxxxxxB
\else
??????\fi
\fi
}
\newcommand{\hatcurb}[1]{\ifnum#1=9 %
\hatcurbxxxxxA
\else
\ifnum#1=10 %
\hatcurbxxxxxB
\else
??????\fi
\fi
}
\newcommand{\hatcurc}[1]{\ifnum#1=9 %
\hatcurcxxxxxA
\else
\ifnum#1=10 %
\hatcurcxxxxxB
\else
??????\fi
\fi
}
\newcommand{\hatcurCCtassvi}[1]{\ifnum#1=9 %
\hatcurCCtassvixxxxxA
\else
\ifnum#1=10 %
\hatcurCCtassvixxxxxB
\else
??????\fi
\fi
}
\newcommand{\hatcurisocite}[1]{\ifnum#1=9 %
\hatcurisocitexxxxxA
\else
\ifnum#1=10 %
\hatcurisocitexxxxxB
\else
??????\fi
\fi
}
\newcommand{\hatcurisofull}[1]{\ifnum#1=9 %
\hatcurisofullxxxxxA
\else
\ifnum#1=10 %
\hatcurisofullxxxxxB
\else
??????\fi
\fi
}
\newcommand{\hatcurisoshort}[1]{\ifnum#1=9 %
\hatcurisoshortxxxxxA
\else
\ifnum#1=10 %
\hatcurisoshortxxxxxB
\else
??????\fi
\fi
}
\newcommand{\hatcurjhkfilset}[1]{\ifnum#1=9 %
\hatcurjhkfilsetxxxxxA
\else
\ifnum#1=10 %
\hatcurjhkfilsetxxxxxB
\else
??????\fi
\fi
}
\newcommand{\hatcurlumind}[1]{\ifnum#1=9 %
\hatcurlumindxxxxxA
\else
\ifnum#1=10 %
\hatcurlumindxxxxxB
\else
??????\fi
\fi
}
\newcommand{\hatcurplanetnum}[1]{\ifnum#1=9 %
\hatcurplanetnumxxxxxA
\else
\ifnum#1=10 %
\hatcurplanetnumxxxxxB
\else
??????\fi
\fi
}
\newcommand{\hatcurRVgammaabs}[1]{\ifnum#1=9 %
\hatcurRVgammaabsxxxxxA
\else
\ifnum#1=10 %
\hatcurRVgammaabsxxxxxB
\else
??????\fi
\fi
}
\newcommand{\hatcurRVgammarel}[1]{\ifnum#1=9 %
\hatcurRVgammarelxxxxxA
\else
\ifnum#1=10 %
\hatcurRVgammarelxxxxxB
\else
??????\fi
\fi
}
\newcommand{\hatcurSMElogg}[1]{\ifnum#1=9 %
\hatcurSMEloggxxxxxA
\else
\ifnum#1=10 %
\hatcurSMEloggxxxxxB
\else
??????\fi
\fi
}
\newcommand{\hatcurSMEteff}[1]{\ifnum#1=9 %
\hatcurSMEteffxxxxxA
\else
\ifnum#1=10 %
\hatcurSMEteffxxxxxB
\else
??????\fi
\fi
}
\newcommand{\hatcurSMEversion}[1]{\ifnum#1=9 %
\hatcurSMEversionxxxxxA
\else
\ifnum#1=10 %
\hatcurSMEversionxxxxxB
\else
??????\fi
\fi
}
\newcommand{\hatcurSMEvmac}[1]{\ifnum#1=9 %
\hatcurSMEvmacxxxxxA
\else
\ifnum#1=10 %
\hatcurSMEvmacxxxxxB
\else
??????\fi
\fi
}
\newcommand{\hatcurSMEvmic}[1]{\ifnum#1=9 %
\hatcurSMEvmicxxxxxA
\else
\ifnum#1=10 %
\hatcurSMEvmicxxxxxB
\else
??????\fi
\fi
}
\newcommand{\hatcurSMEvsin}[1]{\ifnum#1=9 %
\hatcurSMEvsinxxxxxA
\else
\ifnum#1=10 %
\hatcurSMEvsinxxxxxB
\else
??????\fi
\fi
}
\newcommand{\hatcurSMEzfeh}[1]{\ifnum#1=9 %
\hatcurSMEzfehxxxxxA
\else
\ifnum#1=10 %
\hatcurSMEzfehxxxxxB
\else
??????\fi
\fi
}
\newcommand{\hatcurSMEzfehshort}[1]{\ifnum#1=9 %
\hatcurSMEzfehshortxxxxxA
\else
\ifnum#1=10 %
\hatcurSMEzfehshortxxxxxB
\else
??????\fi
\fi
}

\newcounter{planetcounter}

\newboolean{emulateapj}
\setboolean{emulateapj}{true}

\newboolean{rvtablelong}
\setboolean{rvtablelong}{true}

\newboolean{astroph}
\setboolean{astroph}{true}

\shortauthors{Brahm et al.}
\shorttitle{\hatcur{9}\lowercase{b} and \hatcur{10}\lowercase{b}}
\ifthenelse{\boolean{emulateapj}}{
    \newcommand{\titledag}{$\dagger$}
}{
    \newcommand{\titledag}{\dagger}
}

\begin{document}

\title{
\hatcur{9}\lowercase{b} and \hatcur{10}\lowercase{b}:  Two compact hot
Jupiters in field 7 of the K2 mission\altaffilmark{\titledag}
}

\author{R. Brahm\altaffilmark{1,2}}
\author{A. Jord\'an\altaffilmark{1,2}}
\author{J. D. Hartman\altaffilmark{3}}
\author{G. \'A. Bakos\altaffilmark{3,*,**}}
\author{D. Bayliss\altaffilmark{4,5}}
\author{K. Penev\altaffilmark{3}}	
\author{G. Zhou\altaffilmark{5}}	
\author{S. Ciceri\altaffilmark{6}}	
\author{M. Rabus\altaffilmark{1,6}}	
\author{N. Espinoza\altaffilmark{1,2}}	
\author{L. Mancini\altaffilmark{6}}	
\author{M. de Val-Borro\altaffilmark{3}}
\author{W. Bhatti\altaffilmark{3}}	
\author{B. Sato\altaffilmark{7}}
\author{T. G. Tan\altaffilmark{8}}	
\author{Z. Csubry\altaffilmark{3}}	
\author{L. Buchhave\altaffilmark{9}}	
\author{T. Henning\altaffilmark{6}}
\author{B. Schmidt\altaffilmark{5}}	
\author{V. Suc\altaffilmark{1}}
\author{R. W. Noyes\altaffilmark{9}}	
\author{I. Papp\altaffilmark{10}}
\author{J. L\'az\'ar\altaffilmark{10}}	
\author{P. S\'ari\altaffilmark{10}}

\altaffiltext{1}{Instituto de Astrof\'isica, Facultad de F\'isica, Pontificia Universidad Cat\'olica de Chile, Av. Vicu\~na Mackenna 4860, 7820436 Macul, Santiago, Chile; rbrahm@astro.puc.cl}
\altaffiltext{2}{Millennium Institute of Astrophysics, Av. Vicu\~na Mackenna
4860, 7820436 Macul, Santiago, Chile}
\altaffiltext{3}{Department of Astrophysical Sciences, Princeton University, NJ 08544, USA}
\altaffiltext{*}{Alfred P. Sloan Research Fellow}
\altaffiltext{**}{Packard Fellow}
\altaffiltext{4}{Observatoire Astronomique de l'Universit\'e de Gen\`eve, 51
ch. des Maillettes, 1290 Versoix, Switzerland}
\altaffiltext{5}{The Australian National University, Canberra, Australia}
\altaffiltext{6}{Max Planck Institute for Astronomy, Koenigstuhl 17, 69117, Heidelberg, Germany}
\altaffiltext{7}{Department of Earth and Planetary Sciences, Tokyo Institute
of Technology, 2-12-1 Ookayama, Meguro-ku, Tokyo 152-
8551}
\altaffiltext{8}{Perth Exoplanet Survey Telescope, Perth, Australia}
\altaffiltext{9}{Harvard-Smithsonian Center for Astrophysics, Cambridge, MA 02138 USA}
\altaffiltext{10}{Hungarian Astronomical Association, Budapest, Hungary}

\altaffiltext{$\dagger$}{
 The HATSouth network is operated by a collaboration consisting of
Princeton University (PU), the Max Planck Institute for Astronomy
(MPIA), the Australian National University (ANU), and the Pontificia
Universidad Cat\'olica de Chile (PUC).  The station at Las Campanas
Observatory (LCO) of the Carnegie Institute is operated by PU in
conjunction with PUC, the station at the High Energy Spectroscopic
Survey (H.E.S.S.) site is operated in conjunction with MPIA, and the
station at Siding Spring Observatory (SSO) is operated jointly with
ANU.
Based in part on data collected at Subaru Telescope, which is operated
by the National Astronomical Observatory of Japan. Based in part on
observations made with the MPG~2.2\,m Telescope at the ESO Observatory
in La Silla.  This paper uses observations obtained with facilities of
the Las Cumbres Observatory Global Telescope. Based on observations
obtained with the Apache Point Observatory 3.5-meter telescope, which
is owned and operated by the Astrophysical Research Consortium.
}


\begin{abstract}

\setcounter{footnote}{10}
We report the discovery of two transiting extrasolar planets by the
HATSouth survey.  \hatcurb{9} orbits an old  ($10.8\pm1.5$ Gyr) V=13.3 \hatcurISOspec{9}\  dwarf
star, with a period $P\approx1.9153$\,d. The host star has a mass of
1.03\,\msun, radius of 1.503\rsun\, and effective
temperature \hatcurSMEteff{9}\,K.  The planetary companion has a mass
of 0.837\,\mjup, and radius of 1.065\,\rjup\
yielding a mean density of 0.85\,\gcmc.  \hatcurb{10}
orbits a V=13.1 \hatcurISOspec{10}\ dwarf star, with a period
$P\approx3.3128$\,d. The host star has a mass of
1.1\,\msun, radius of 1.11\,\rsun\, and
effective temperature \hatcurSMEteff{10}\,K. The planetary companion
has a mass of 0.53\,\mjup, and radius of
0.97\,\rjup\ yielding a mean density of
0.7\,\gcmc. Both planets are compact in comparison
with planets receiving similar irradiation from their host stars, and
lie in the nominal coordinates of Field 7 of K2 but only \hatcurb{9}
falls on working silicon. Future characterization of \hatcurb{9}
with the exquisite photometric precision of the {\em Kepler} telescope may provide
measurements of its reflected light signature.
\end{abstract}

\keywords{
    planetary systems ---
    stars: individual (
\setcounter{planetcounter}{1}
\hatcur{9},
\hatcurCCgsc{9}\loopcommanoperiod
\setcounter{planetcounter}{2}
\hatcur{10},
\hatcurCCgsc{10}\loopcommanoperiod
\setcounter{planetcounter}{3}
) 
    techniques: spectroscopic, photometric
}

\section{Introduction}
\label{sec:introduction}

\setcounter{footnote}{0}
Our current understanding of the structure and orbital evolution of
extrasolar giant planets has been, to a large degree, informed by the
characterization of transiting planetary systems. Besides the
determination of the planet radius, true mass, and bulk density,
follow-up studies of transiting extrasolar planets (TEPs) allow the
extraction of valuable information, like the spin-orbit
angle and the properties and composition of the planetary atmospheres,
that are not be easily recovered unless the orbital plane is favorably
oriented such that the planet eclipses its host star.

Detections of giant TEPs, mostly driven by transiting
ground based surveys like SuperWASP \citep{pollacco:2006} and HATNet
\citep{bakos:2004:hatnet}, have revealed a large number of systems in the
region of parameter space with $R_p$$>$0.8$R_J$, $M_p$$>$0.4$M_J$,
P$<$5 d and FGK-type host stars. The measured properties of these
systems, coupled with subsequent follow-up studies, have been
fundamental for testing formation and interior models of these giant
planets, which are known as hot Jupiters.

New ground-based transiting surveys like HATSouth \citep{bakos:2013:hatsouth}
have been designed with the goal of expanding the parameter space of
well characterized TEPs by detecting planets with smaller radii ($R_p
<0.4 R_J$) and/or longer periods ($P>10$ days).  In the process of
searching for these kinds of planets, new hot Jupiters are detected
which contribute to enlarging the sample of known systems. Even though
many hot Jupiters are already known, more are still needed to make
headway into understanding their physical properties, e.g. a firm understanding of the mechanism 
that causes some hot Jupiters to have inflated radii \citep[e.g. HAT-P-32b and HAT-P-33b,][]{hartman:2011:hat3233}.

New planet discoveries around bright stars, accessible by follow-up
facilities are especially valuable
given the wealth of detailed studies that they can be subject
to. Indeed, some of the most analyzed and characterized giant TEPs are
three planets (TrES-2b, HAT-P-7b and HAT-P-11b) that were detected by
ground-based surveys \citep{pal:2008:hat7,bakos:2010:hat11,
  odonovan:2006} and later observed by NASA's {\em Kepler} mission \citep{borucki:2010}. Even
though the primary goal of {\em Kepler} satellite was the detection
of planets near the habitable zone for estimating their frequency and
distribution in our galaxy, the high photometric precision of {\em
  Kepler} allowed very detailed studies of the small population of
giant planets on close orbits around moderately bright stars ($V<14$)
that fell in its field of view.

{\em Kepler} was able to detect secondary transits and phase
variations on TrES-2b and HAT-P-7b \citep{estevez:2013} which were
useful in the study of their atmospherical properties, such as the
determination of the geometric albedos and planetary phase curve
offsets. Doppler beaming and ellipsoidal variations measured with {\em
  Kepler} also constrained the mass of those planets. In the case of
HAT-P-11b, {\em Kepler} observations were useful in characterizing the
activity of the K-type host star; and the analysis of crossing stellar
spots allowed the determination of the spin-orbit misalignment of this
system \citep{deming:2011,sanchis:2011}. Simultaneous observations of
the transits of HAT-P-11b by {\em Kepler} and {\em Spitzer} allowed
also the detection of water vapor in the atmosphere of this Neptune-size
planet \citep{fraine:2014}.  Estimation of the planetary physical
parameters depend strongly on the estimated stellar properties. In
this regard, {\em Kepler} was also able to measure model independent
stellar properties by the use of asteroseismology on the three mentioned
systems \citep{christensen:2010}.

After the failure of two of its reaction wheels, the {\em Kepler} satellite
is still working, but with a new observation strategy
and a photometric precision within a factor of $\sim$2 of the nominal
Kepler mission performance \citep[e.g.,][]{ vanderburg:2014, aigrain:2015, foreman:2015,crossfield:2015}.
This new mission concept, called K2 \citep{howell:2014},
will observe 10 fields, each for a period of approximately 70 days, and some of
these fields lie in the Southern hemisphere.  One of the limitations
of K2 is that the number of stars that can be monitored in each field
is substantially lower than for the original {\em Kepler} mission. For
this reason, the pre-selection of targets, based on ground-based
observations of K2 fields is especially important for an efficient use of the
satellite.

In this work we present the discovery of \hatcur{9}\lowercase{b} and
\hatcur{10}\lowercase{b}, two hot Jupiters discovered by the HATSouth 
survey which are located in the nominal coordinates of
Field 7 of K2 mission. In \refsecl{obs} we summarize the observations
that allowed the discovery and confirmation of these planets.
In \refsecl{analysis} we show the global analysis of the spectroscopic and
photometric data that confirmed the planetary nature
of the transiting candidates and also rejected blend scenarios that
can mimic the photometric and radial velocity signals. Our findings
are discussed in \refsecl{discussion}.

\section{Observations}
\label{sec:obs}

\subsection{Photometric detection}
\label{sec:detection}

\hatcur{9} and \hatcur{10} were identified as transiting planetary
host candidates after obtaining $\sim$10000 images of the same field
with three stations on the three HATSouth observing sites.
The number of photometric
observations that were taken for each star on each of the HATSouth
stations is indicated in Table~\ref{tab:photobs}, where it can be seen
that in both cases $\sim$45$\%$ of the observations came from the
HATSouth station located at Las Campanas Observatory (LCO).

The HATSouth observations consist of four-minute Sloan $r$-band exposures
obtained with 24 Takahashi E180 astrographs (18cm aperture) coupled to
Apogee 4Kx4K U16M ALTA CCDs. Readout times are of the order of one minute
which results in a cadence of about 5 minutes.
Detailed descriptions of the image
processing steps and the candidate identification procedures of
HATSouth data can be found in \cite{bakos:2013:hatsouth} and
\cite{penev:2013:hats1}.  Briefly, after applying aperture photometry on the
images, the light curves generated are detrended using external
parameter decorrelation (EPD) and trend filtering algorithm
\citep[TFA][]{kovacs:2005:TFA}. Periodic transits on the detrended light curves are then
searched using Box-fitted Least Squares (BLS) algorithm
\citep{kovacs:2002:BLS}.

Figure~\ref{fig:hatsouth} shows the phase-folded detection light curves of
\hatcurb{9} and \hatcurb{10}, where a clear $\sim$10 mmag flat-bottom transit
can be observed in both cases.

\ifthenelse{\boolean{emulateapj}}{
    \begin{figure*}[!ht]
}{
    \begin{figure}[!ht]
}
\plottwo{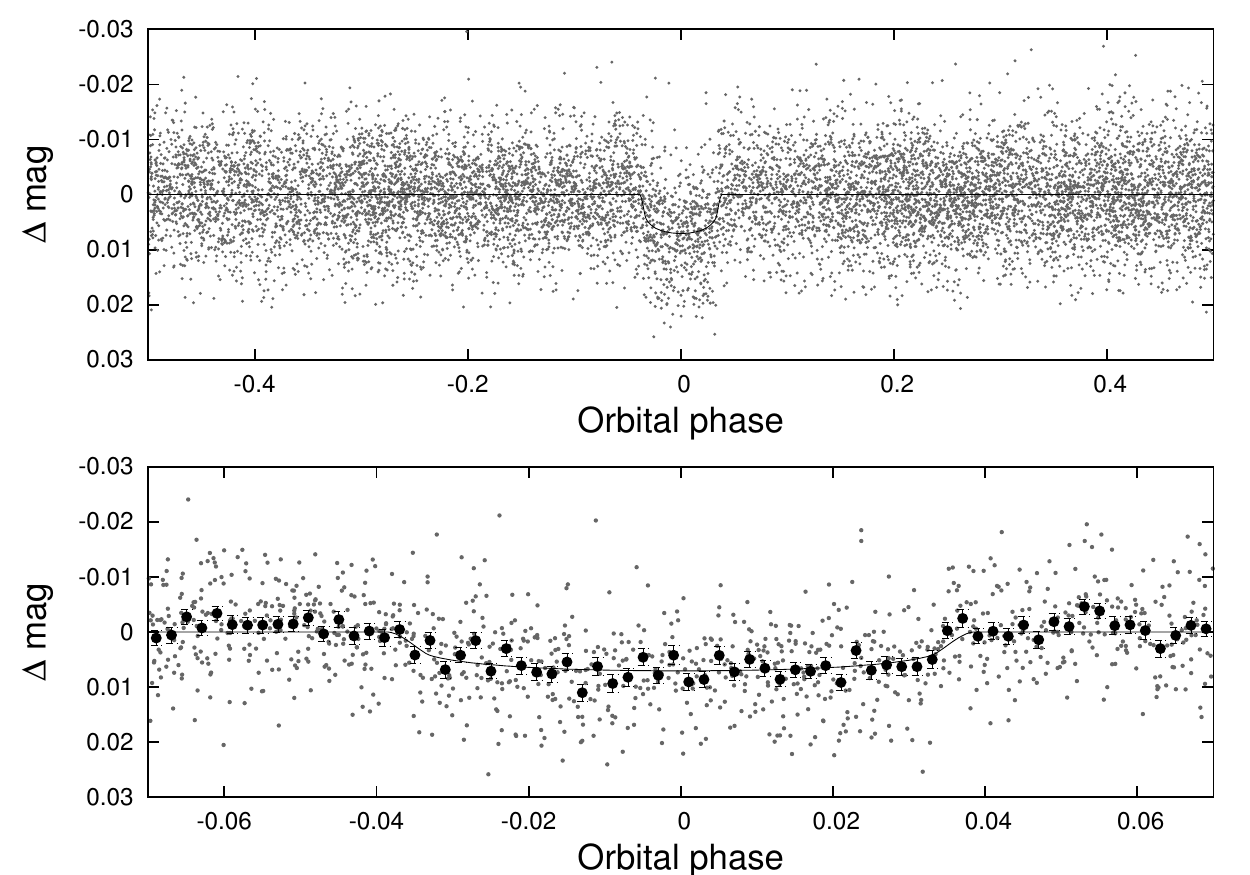}{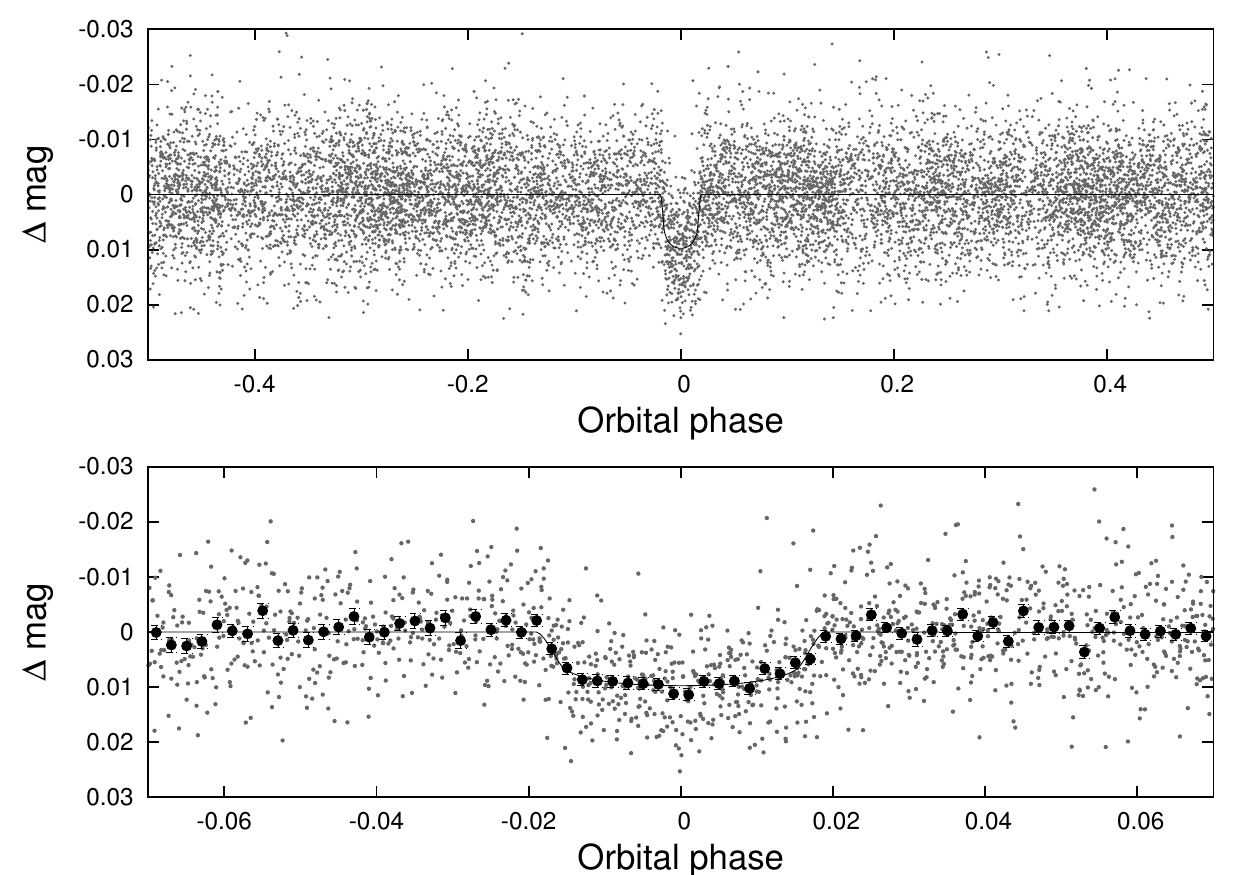}
\caption[]{
    Phase-folded unbinned HATSouth light curves for \hatcur{9} (left) and \hatcur{10} (right). In each case we show two panels. The
    top panel shows the full light curve, while the bottom panel shows
    the light curve zoomed-in on the transit. The solid lines show the
    model fits to the light curves. The dark filled circles in the
    bottom panels show the light curves binned in phase with a bin
    size of 0.002.
\label{fig:hatsouth}}
\ifthenelse{\boolean{emulateapj}}{
    \end{figure*}
}{
    \end{figure}
}

\ifthenelse{\boolean{emulateapj}}{
    \begin{deluxetable*}{llrrrr}
}{
    \begin{deluxetable}{llrrrr}
}
\tablewidth{0pc}
\tabletypesize{\scriptsize}
\tablecaption{
    Summary of photometric observations
    \label{tab:photobs}
}
\tablehead{
    \multicolumn{1}{c}{Instrument/Field\tablenotemark{a}} &
    \multicolumn{1}{c}{Date(s)} &
    \multicolumn{1}{c}{\# Images} &
    \multicolumn{1}{c}{Cadence\tablenotemark{b}} &
    \multicolumn{1}{c}{Filter} &
    \multicolumn{1}{c}{Precision\tablenotemark{c}} \\
    \multicolumn{1}{c}{} &
    \multicolumn{1}{c}{} &
    \multicolumn{1}{c}{} &
    \multicolumn{1}{c}{(sec)} &
    \multicolumn{1}{c}{} &
    \multicolumn{1}{c}{(mmag)}
}
\startdata

\sidehead{\textbf{\hatcur{9}}}
~~~~HS-1/G579 & 2010 Mar--2011 Aug & 4317 & 300 & \band{r} & 6.9 \\
~~~~HS-3/G579 & 2010 Mar--2011 Aug & 2138 & 303 & \band{r} & 7.6 \\
~~~~HS-5/G579 & 2010 Sep--2011 Aug & 2784 & 303 & \band{r} & 6.9 \\
~~~~FTS & 2013 Apr 11 & 134 & 80 & \band{i} & 1.4 \\
~~~~PEST & 2013 May 31 & 186 & 130 & \band{R_{C}} & 3.4 \\

\sidehead{\textbf{\hatcur{10}}}
~~~~HS-1/G579 & 2009 Sep--2011 Aug & 4389 & 301 & \band{r} & 7.3 \\
~~~~HS-3/G579 & 2010 Mar--2011 Aug & 2596 & 303 & \band{r} & 7.2 \\
~~~~HS-5/G579 & 2011 Mar--2011 Aug & 3297 & 303 & \band{r} & 7.8 \\
~~~~CTIO~0.9m & 2012 Aug 29 & 69 & 213 & \band{z} & 2.3 \\
~~~~FTS & 2013 Apr 05 & 142 & 63 & \band{i} & 4.3 \\
~~~~GROND & 2013 Jun 14 & 92 & 156 & \band{g} & 0.8 \\
~~~~GROND & 2013 Jun 14 & 88 & 156 & \band{r} & 1.3 \\
~~~~GROND & 2013 Jun 14 & 94 & 156 & \band{i} & 0.7 \\
~~~~GROND & 2013 Jun 14 & 89 & 156 & \band{z} & 0.8 \\
~~~~PEST & 2013 Jun 27 & 145 & 130 & \band{R_{C}} & 4.6 \\

\enddata
\tablenotetext{a}{
    For HATSouth data we list the HATSouth unit and field name from
    which the observations are taken. HS-1 and -2 are located at Las
    Campanas Observatory in Chile, HS-3 and -4 are located at the
    H.E.S.S. site in Namibia, and HS-5 and -6 are located at Siding
    Spring Observatory in Australia. Each field corresponds to one of
    838 fixed pointings used to cover the full 4$\pi$ celestial
    sphere. All data from a given HATSouth field are reduced together,
    while detrending through External Parameter Decorrelation (EPD) is
    done independently for each unique field+unit combination.
}
\tablenotetext{b}{
    The median time between consecutive images rounded to the nearest
    second. Due to weather, the day--night cycle, guiding and focus
    corrections, and other factors, the cadence is only approximately
    uniform over short timescales.
}
\tablenotetext{c}{
    The RMS of the residuals from the best-fit model.
}
\ifthenelse{\boolean{emulateapj}}{
    \end{deluxetable*}
}{
    \end{deluxetable}
}

\subsection{Spectroscopic Observations}
\label{sec:obsspec}

Transit-like light curves can be produced by different configurations of stellar binaries.
Spectroscopic observations are required to reject false positives and
to obtain the orbital parameters and masses of the true planets.  Due
to the great number of HATSouth candidates and the limited available
observing time on spectroscopic facilities, this follow-up is
performed in a two-step procedure as we now describe.  All
spectroscopic observations are summarized in \reftabl{specobs}.

First, initial spectra are acquired (with either low resolution, or low S/N)
to make a rough estimation of the stellar parameters, identifying spectra
composed of more than one star, and measuring RV variations produced
by stellar mass companions.  \hatcur{9} was observed with
WIFeS \citep{dopita:2007} on the ANU 2.3m telescope, obtaining $\teffstar$=5821
$\pm$ 300 K, $\loggstar$=3.9 $\pm$ 0.3 $\feh$=0.5 $\pm$ 0.5; and with
ARCES on the APO 3.5m obtaining $\teffstar$=5692 $\pm$ 50 K,
$\loggstar$=4.14 $\pm$ 0.1 $\feh$=0.5 $\pm$ 0.08. Both estimates of
stellar parameters were consistent with a G-type dwarf, but the
sub-solar surface gravity value points towards a slightly evolved
system. Details on the observing strategy, reduction methods and the
processing of the spectra for WIFeS can be found in
\cite{bayliss:2013:hats3}. The ARCES observation was carried out
using the $1\farcs 6 \times 3\farcs 2$ slit yielding an echelle spectrum
with 107 orders covering the wavelength range 3200--10000\AA\ at a
resolution of $\Delta\lambda/\lambda \sim 31500$. A single ThAr lamp
spectrum was obtained immediately following the science exposure with
the telescope still pointed toward \hatcur{9}. The science observation
was reduced to a wavelength-calibrated spectrum using the standard
IRAF echelle package\footnote{IRAF is distributed by the National Optical
Astronomy Observatories, which are operated by the Association of
Universities for Research in Astronomy, Inc., under cooperative agreement
with the National Science Foundation.}, and analyzed using the Spectral
Parameter Classification (SPC) program \citep{buchhave:2012} to determine
the radial velocity and stellar atmospheric parameters.

Reconnaissance spectroscopy was performed for
\hatcur{10} using the echelle spectrograph mounted on the du Pont 2.5m
telescope at Las Campanas Observatory.
One observation using the $1\farcs \times 4\farcs$ slit ($\Delta\lambda/\lambda \sim 40000$)
was enough to confirm that \hatcur{10} has a single lined spectrum with the
following stellar parameters: $\teffstar$=6100 $\pm$ 100 K,
$\loggstar$=4.6 $\pm$ 0.5 $\feh$=0.0 $\pm$ 0.5, $\vsini$=5.0 $\pm$ 2.0
km/s. This spectrum was reduced and analysed with an automated
pipeline developed to deal with data coming from a host of different echelle
spectrographs (Brahm et al. in prep.). The pipeline for DuPont is very
similar to the ones we have previously detailed for Coralie and FEROS
data in \cite{jordan:2014:hats4}.

Once both candidates were identified as single-lined late-type dwarfs,
spectra from high precision instruments were required to measure RV
variations with high precision ($< 30$ m/s) in order to measure the
mass of the substellar companions and obtain the orbital parameters.
\hatcur{9} and \hatcur{10} were observed several times with Coralie
\citep{queloz:2001} on the 1.2m Euler telescope, FEROS
\citep{kaufer:1998} on the 2.2m MPG telescope, and HDS on the 8m
Subaru telescope \citep{noguchi:2002}.  Coralie and FEROS data were
processed with the pipeline described in \cite{jordan:2014:hats4},
where RV values are obtained using the cross correlation technique
against a binary mask and bisector span (BS) measurements are computed
from the cross-correlation peak following \cite{queloz:2001}.
HDS RVs were measured using the procedure
detailed in Sato et al. (2002, 2012) which are in turn based on the
method of Butler et al. (1996) while BS values were obtained following
\cite{bakos:2007}.

Phased high-precision RV and BS measurements are shown for each system
in \reffigl{rvbis} and the data are listed in \reftabl{rvs}. Both candidates show RV variations in phase with
photometric ephemeris, however for \hatcur{10} the residuals are
higher than expected.  This deviation can be partly explained by
moonlight contamination in 5 spectra acquired with Coralie in August
2013 which are marked with crosses in \reffigl{rvbis}. There are no significant correlations between RV and BS
variations and thus we conclude the RV variations are not produced by
stellar activity. The 95\% confidence interval for the Pearson
correlation coefficient between RV and BS was computed for both
candidates using a bootstrap procedure. The confidence intervals are
[-0.57,0.07] and [-0.43, 0.37] for \hatcur{9} and \hatcur{10},
respectively.  The individual FEROS spectra were median combined for
both candidates to perform a precise estimation of the stellar
parameters.

\ifthenelse{\boolean{emulateapj}}{
    \begin{deluxetable*}{llrrrrr}
}{
    \begin{deluxetable}{llrrrrrrrr}
}
\tablewidth{0pc}
\tabletypesize{\scriptsize}
\tablecaption{
    Summary of spectroscopy observations    \label{tab:specobs}
}
\tablehead{
    \multicolumn{1}{c}{Instrument}          &
    \multicolumn{1}{c}{UT Date(s)}             &
    \multicolumn{1}{c}{\# Spec.}   &
    \multicolumn{1}{c}{Res.}          &
    \multicolumn{1}{c}{S/N Range\tablenotemark{a}}           &
    \multicolumn{1}{c}{$\gamma_{\rm RV}$\tablenotemark{b}} &
    \multicolumn{1}{c}{RV Precision\tablenotemark{c}} \\
    &
    &
    &
    \multicolumn{1}{c}{$\Delta \lambda$/$\lambda$/1000} &
    &
    \multicolumn{1}{c}{(\kms)}              &
    \multicolumn{1}{c}{(\ms)}
}
\startdata

\sidehead{\textbf{\hatcur{9}}}

APO~3.5\,m/ARCES & 2012 Aug 25 & 1 & 31.5 & 27 & -11.5 & 500 \\
ANU~2.3\,m/WiFeS & 2012 Sep 8 & 1 & 3 & 140 & $\cdots$ & $\cdots$ \\
Euler~1.2\,m/Coralie & 2012 Nov 6--10 & 4 & 60 & 14--20 & -10.634 & 37 \\
MPG~2.2\,m/FEROS & 2012 Aug--2013 May & 9 & 48 & 32--76 & -10.653 & 32 \\
Subaru 8\,m/HDS & 2012 Sep 19 & 3 & 60 & 100--114 & $\cdots$ & $\cdots$ \\
Subaru 8\,m/HDS+I$_{2}$ & 2012 Sep 20--22 & 9 & 60 & 60--100 & $\cdots$ & 11 \\
\sidehead{\textbf{\hatcur{10}}}

du~Pont~2.5\,m/Echelle & 2013 Aug 21 & 1 & 40 & 48 & -29.2 & 500 \\
Euler~1.2\,m/Coralie & 2012 Aug--2013 Aug & 12 & 60 & 17--23 & -28.131 & 68 \\
MPG~2.2\,m/FEROS & 2013 Mar--Jul & 5 & 48 & 29--85 & -28.044 & 50 \\
Subaru 8\,m/HDS & 2012 Sep 22 & 3 & 60 & 74--94 & $\cdots$ & $\cdots$ \\
Subaru 8\,m/HDS+I$_{2}$ & 2012 Sep 19--21 & 9 & 60 & 41--99 & $\cdots$ & 14 \\
\enddata 
\tablenotetext{a}{
    S/N per resolution element near 5180\,\AA.
}
\tablenotetext{b}{
  For Coralie and FEROS this is the systemic RV from fitting an orbit
  to the observations in \refsecl{globmod}. For ARCES and the du~Pont
  Echelle it is the measured RV of the single observation. We do not
  provide this quantity for instruments for which only relative RVs
  are measured, or for WiFeS which was only used to measure stellar
  atmospheric parameters.
}
\tablenotetext{c}{
    For High-precision RV observations included in the orbit
    determination this is the RV residuals from the best-fit orbit,
    for other instruments used for reconnaissance spectroscopy this is
    an estimate of the precision.
}
\ifthenelse{\boolean{emulateapj}}{
    \end{deluxetable*}
}{
    \end{deluxetable}
}

\setcounter{planetcounter}{1}
%
\ifthenelse{\boolean{emulateapj}}{
    \begin{figure*} [ht]
}{
    \begin{figure}[ht]
}
\plottwo{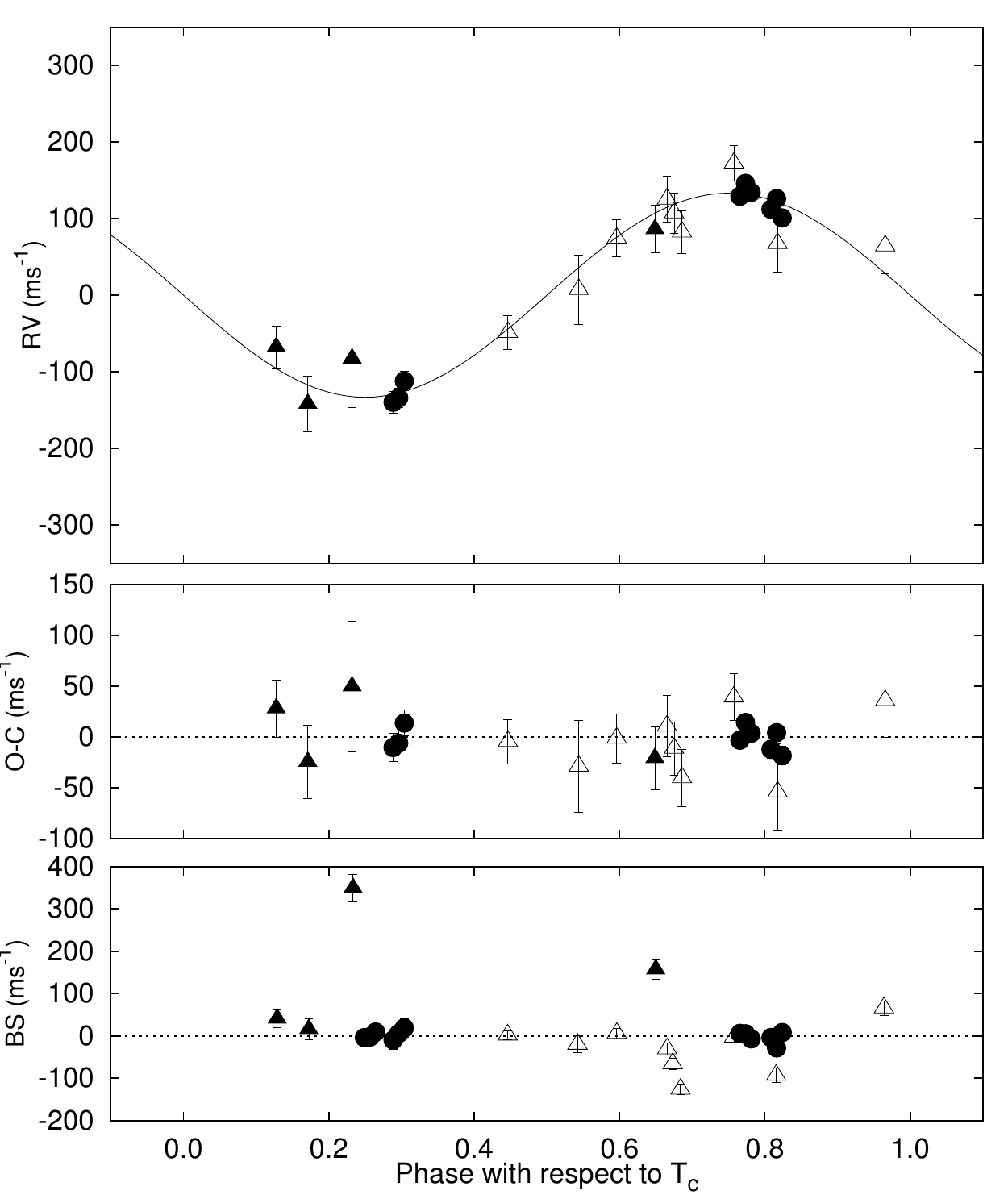}{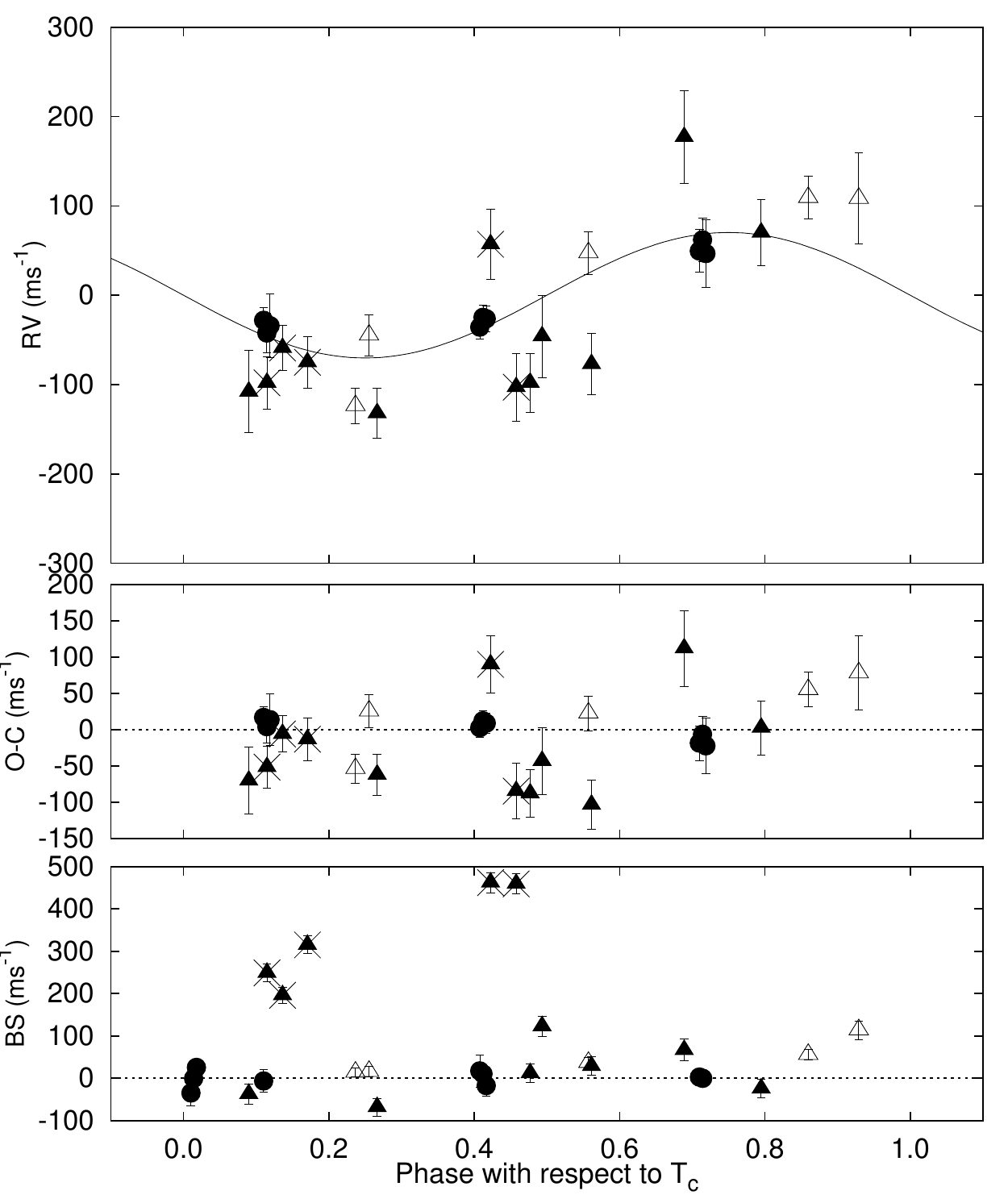}
\caption{
Phased high-precision RV measurements for \hbox{\hatcur{9}{}} (left),
and \hbox{\hatcur{10}{}} (right) from HDS (filled circles), FEROS
(open triangles), and Coralie (filled triangles). In each case we show
three panels. The top panel shows the phased measurements together
with our best-fit model (see \reftabl{planetparam}) for each
system. Zero-phase corresponds to the time of mid-transit. The
center-of-mass velocity has been subtracted. The second panel shows
the velocity $O\!-\!C$ residuals from the best fit. The error bars
include the jitter terms listed in \reftabl{planetparam} added in
quadrature to the formal errors for each instrument. The third panel
shows the bisector spans (BS), with the mean value subtracted. Note
the different vertical scales of the panels. RV measurements highly contaminated
with moonlight are marked with crosses.
}
\label{fig:rvbis}
\ifthenelse{\boolean{emulateapj}}{
    \end{figure*}
}{
    \end{figure}
}

\ifthenelse{\boolean{emulateapj}}{
    \begin{deluxetable*}{lrrrrrl}
}{
    \begin{deluxetable}{lrrrrrl}
}
\tablewidth{0pc}
\tablecaption{
    Relative radial velocities and bisector spans for \hatcur{9} and
    \hatcur{10}.
    \label{tab:rvs}
}
\tablehead{
    \colhead{BJD} &
    \colhead{RV\tablenotemark{a}} &
    \colhead{\ensuremath{\sigma_{\rm RV}}\tablenotemark{b}} &
    \colhead{BS} &
    \colhead{\ensuremath{\sigma_{\rm BS}}} &
    \colhead{Phase} &
    \colhead{Instrument}\\
    \colhead{\hbox{(2,456,000$+$)}} &
    \colhead{(\ms)} &
    \colhead{(\ms)} &
    \colhead{(\ms)} &
    \colhead{(\ms)} &
    \colhead{} &
    \colhead{}
}
\startdata
\multicolumn{7}{c}{\bf HATS-9} \\
\hline\\
$ 169.62456 $ & $    82.24 $ & $    28.00 $ & $ -126.0 $ & $   13.0 $ & $   0.686 $ & FEROS \\
$ 171.51911 $ & $   107.24 $ & $    26.00 $ & $  -66.0 $ & $   13.0 $ & $   0.675 $ & FEROS \\
$ 173.70707 $ & $    67.24 $ & $    37.00 $ & $  -93.0 $ & $   17.0 $ & $   0.817 $ & FEROS \\
$ 189.85616 $ & \nodata      & \nodata      & $   -3.6 $ & $   13.0 $ & $   0.249 $ & Subaru \\
$ 189.87089 $ & \nodata      & \nodata      & $   -2.3 $ & $   11.0 $ & $   0.257 $ & Subaru \\
$ 189.88561 $ & \nodata      & \nodata      & $    9.6 $ & $    9.2 $ & $   0.264 $ & Subaru \\
$ 190.84576 $ & $   129.27 $ & $     7.39 $ & $    6.2 $ & $   13.4 $ & $   0.766 $ & Subaru \\
$ 190.86048 $ & $   146.04 $ & $     8.17 $ & $    5.3 $ & $   10.0 $ & $   0.773 $ & Subaru \\
$ 190.87520 $ & $   134.32 $ & $     7.39 $ & $   -6.7 $ & $   13.6 $ & $   0.781 $ & Subaru \\
$ 191.84685 $ & $  -139.82 $ & $    13.91 $ & $   -9.6 $ & $   21.4 $ & $   0.288 $ & Subaru \\
$ 191.86157 $ & $  -133.96 $ & $    12.35 $ & $    6.0 $ & $   13.5 $ & $   0.296 $ & Subaru \\
$ 191.87633 $ & $  -112.04 $ & $    12.85 $ & $   19.1 $ & $   21.9 $ & $   0.304 $ & Subaru \\
$ 192.84272 $ & $   112.22 $ & $     8.57 $ & $   -3.7 $ & $   18.5 $ & $   0.808 $ & Subaru \\
$ 192.85745 $ & $   126.06 $ & $    10.89 $ & $  -28.3 $ & $   16.3 $ & $   0.816 $ & Subaru \\
$ 192.87217 $ & $   100.81 $ & $     9.05 $ & $    8.0 $ & $    8.8 $ & $   0.824 $ & Subaru \\
$ 205.55663 $ & $   -48.76 $ & $    22.00 $ & $    2.0 $ & $   11.0 $ & $   0.446 $ & FEROS \\
$ 213.50471 $ & $    74.24 $ & $    24.00 $ & $    6.0 $ & $   12.0 $ & $   0.596 $ & FEROS \\
$ 215.55235 $ & $   125.24 $ & $    30.00 $ & $  -31.0 $ & $   14.0 $ & $   0.665 $ & FEROS \\
$ 219.55921 $ & $   172.24 $ & $    23.00 $ & $   -3.0 $ & $   12.0 $ & $   0.757 $ & FEROS \\
$ 237.50625 $ & $   -67.84 $ & $    28.00 $ & $   41.0 $ & $   22.0 $ & $   0.128 $ & Coralie \\
$ 238.50417 $ & $    86.16 $ & $    31.00 $ & $  157.0 $ & $   24.0 $ & $   0.649 $ & Coralie \\
$ 239.50502 $ & $  -141.84 $ & $    36.00 $ & $   16.0 $ & $   24.0 $ & $   0.171 $ & Coralie \\
$ 241.53678 $ & $   -82.84 $ & $    64.00 $ & $  349.0 $ & $   32.0 $ & $   0.232 $ & Coralie \\
$ 424.89567 $ & $    64.24 $ & $    36.00 $ & $   66.0 $ & $   17.0 $ & $   0.965 $ & FEROS \\
$ 427.91875 $ & $     7.24 $ & $    45.00 $ & $  -20.0 $ & $   20.0 $ & $   0.544 $ & FEROS \\
\cutinhead{\bf HATS-10}
$ 160.60805 $ & $  -132.00 $ & $    28.00 $ & $  -68.0 $ & $   21.0 $ & $   0.267 $ & Coralie \\
$ 161.58511 $ & $   -77.00 $ & $    34.00 $ & $   29.0 $ & $   22.0 $ & $   0.561 $ & Coralie \\
$ 164.61796 $ & $   -98.00 $ & $    33.00 $ & $   12.0 $ & $   22.0 $ & $   0.477 $ & Coralie \\
$ 189.90597 $ & $   -27.97 $ & $    14.29 $ & $   -6.1 $ & $   27.3 $ & $   0.110 $ & Subaru \\
$ 189.92069 $ & $   -42.32 $ & $    22.31 $ & \nodata      & \nodata      & $   0.115 $ & Subaru \\
$ 189.93541 $ & $   -34.14 $ & $    35.70 $ & \nodata      & \nodata      & $   0.119 $ & Subaru \\
$ 190.89115 $ & $   -35.74 $ & $    13.59 $ & $   17.7 $ & $   38.3 $ & $   0.408 $ & Subaru \\
$ 190.90587 $ & $   -24.73 $ & $    14.10 $ & $   10.6 $ & $   18.6 $ & $   0.412 $ & Subaru \\
$ 190.92060 $ & $   -26.35 $ & $    14.32 $ & $  -17.0 $ & $   25.1 $ & $   0.416 $ & Subaru \\
$ 191.89225 $ & $    49.44 $ & $    23.96 $ & $    3.2 $ & $   15.3 $ & $   0.710 $ & Subaru \\
$ 191.90697 $ & $    62.06 $ & $    24.16 $ & $    0.2 $ & $   17.3 $ & $   0.714 $ & Subaru \\
$ 191.92170 $ & $    46.62 $ & $    38.06 $ & \nodata      & \nodata      & $   0.719 $ & Subaru \\
$ 192.88724 $ & \nodata      & \nodata      & $  -34.5 $ & $   31.0 $ & $   0.010 $ & Subaru \\
$ 192.89965 $ & \nodata      & \nodata      & $   -0.2 $ & $   23.1 $ & $   0.014 $ & Subaru \\
$ 192.91206 $ & \nodata      & \nodata      & $   25.9 $ & $   15.7 $ & $   0.018 $ & Subaru \\
$ 237.55535 $ & $   -46.00 $ & $    46.00 $ & $  123.0 $ & $   24.0 $ & $   0.493 $ & Coralie \\
$ 238.55388 $ & $    70.00 $ & $    37.00 $ & $  -24.0 $ & $   22.0 $ & $   0.795 $ & Coralie \\
$ 239.53128 $ & $  -108.00 $ & $    46.00 $ & $  -37.0 $ & $   24.0 $ & $   0.090 $ & Coralie \\
$ 241.51608 $ & $   177.00 $ & $    52.00 $ & $   67.0 $ & $   26.0 $ & $   0.689 $ & Coralie \\
$ 375.90634 $ & $   -44.76 $ & $    23.00 $ & $   16.0 $ & $   12.0 $ & $   0.255 $ & FEROS \\
$ 376.90643 $ & $    47.24 $ & $    24.00 $ & $   37.0 $ & $   12.0 $ & $   0.557 $ & FEROS \\
$ 377.90816 $ & $   109.24 $ & $    24.00 $ & $   56.0 $ & $   12.0 $ & $   0.860 $ & FEROS \\
$ 427.83173 $ & $   108.24 $ & $    51.00 $ & $  114.0 $ & $   22.0 $ & $   0.929 $ & FEROS \\
$ 491.79404 $ & $  -123.76 $ & $    20.00 $ & $   14.0 $ & $   10.0 $ & $   0.236 $ & FEROS \\
$ 524.51947^c $ & $   -98.00 $ & $    29.00 $ & $  249.0 $ & $   21.0 $ & $   0.115 $ & Coralie \\
$ 524.59002^c  $ & $   -59.00 $ & $    25.00 $ & $  196.0 $ & $   19.0 $ & $   0.136 $ & Coralie \\
$ 524.70382^c  $ & $   -75.00 $ & $    29.00 $ & $  315.0 $ & $   21.0 $ & $   0.170 $ & Coralie \\
$ 525.53878^c  $ & $    57.00 $ & $    39.00 $ & $  462.0 $ & $   24.0 $ & $   0.422 $ & Coralie \\
$ 525.65573^c  $ & $  -103.00 $ & $    38.00 $ & $  459.0 $ & $   24.0 $ & $   0.458 $ & Coralie \\
\enddata
\tablenotetext{a}{
    The zero-point of these velocities is arbitrary. An overall offset
    $\gamma_{\rm rel}$ fitted independently to the velocities from
    each instrument has been subtracted.
}
\tablenotetext{b}{
    Internal errors excluding the component of astrophysical jitter
    considered in \refsecl{globmod}.
}
\tablenotetext{b}{
    Coralie observations acquired in August 2013 were contaminated with moonlight.
}
\ifthenelse{\boolean{rvtablelong}}{
    \tablecomments{
        Note that for the iodine-free template exposures we do not
        measure the RV but do measure the BS and S index.  Such
        template exposures can be distinguished by the missing RV
        value. The Subaru/HDS observations of \hatcur{10} without BS
        measurements have too low S/N in the I$_{2}$-free blue
        spectral region to pass our quality threshold for calculating accurate
        BS values.
    }
}{
    \tablecomments{
        Note that for the iodine-free template exposures we do not
        measure the RV but do measure the BS and S index.  Such
        template exposures can be distinguished by the missing RV
        value.  This table is presented in its entirety in the
        electronic edition of the Astrophysical Journal.  A portion is
        shown here for guidance regarding its form and content.
    }
} 
\ifthenelse{\boolean{emulateapj}}{
    \end{deluxetable*}
}{
    \end{deluxetable}
}

\subsection{Photometric follow-up observations}
\label{sec:phot}

\setcounter{planetcounter}{1}

\begin{figure*}[!ht]
\plotone{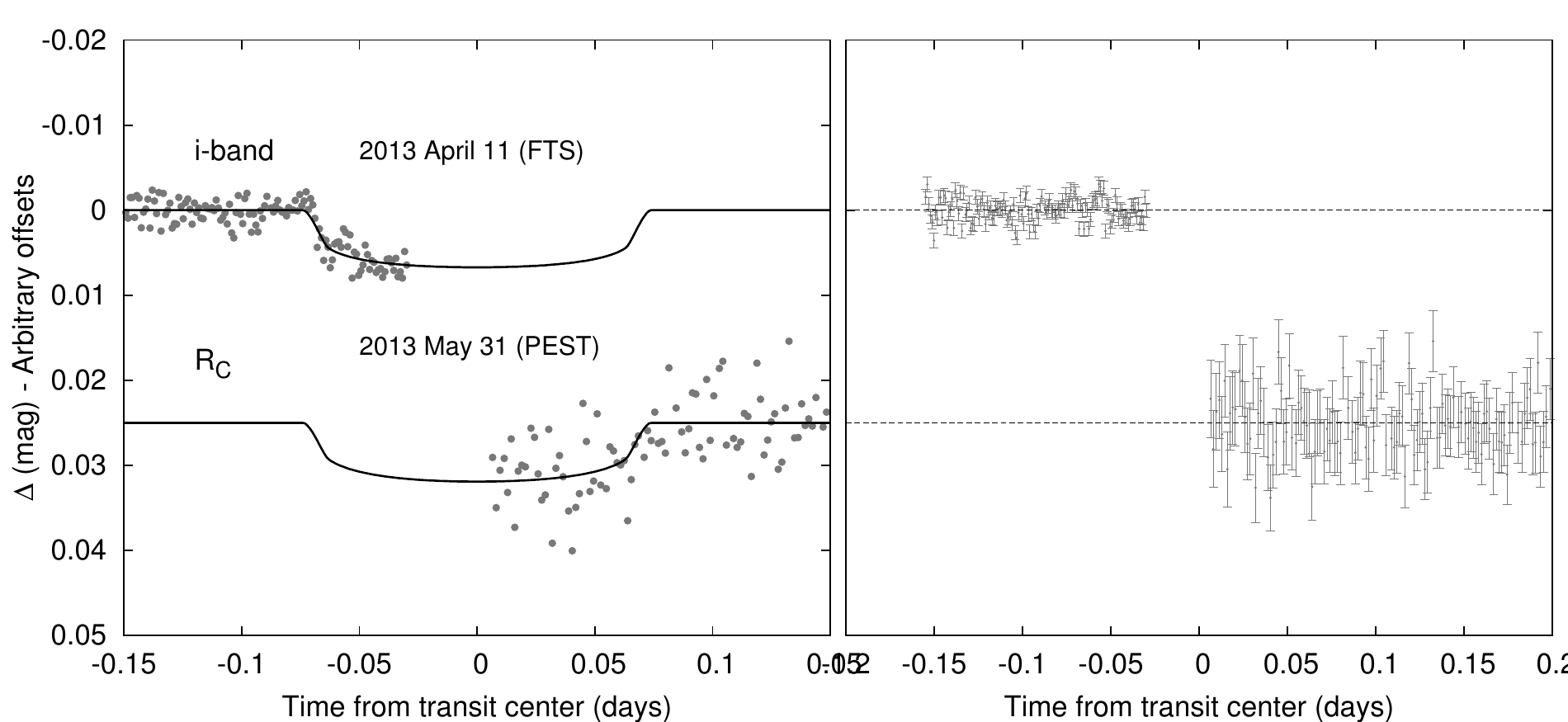}
\caption{
    Left: Unbinned transit \lcs{} for \hatcur{9}.  The light curves have
    been corrected for quadratic trends in time fitted simultaneously
    with the transit model.
    The dates of the events, filters and instruments used are indicated.  The
    second curve is displaced vertically for clarity.  Our best fit
    from the global modeling described in \refsecl{globmod} is shown
    by the solid lines.  Right: residuals from the fits are displayed
    in the same order as the left curves.  The error bars represent
    the photon and background shot noise, plus the readout noise.
}
\label{fig:lc9}
\end{figure*}
\setcounter{planetcounter}{2}

\begin{figure*}[!ht]
\plotone{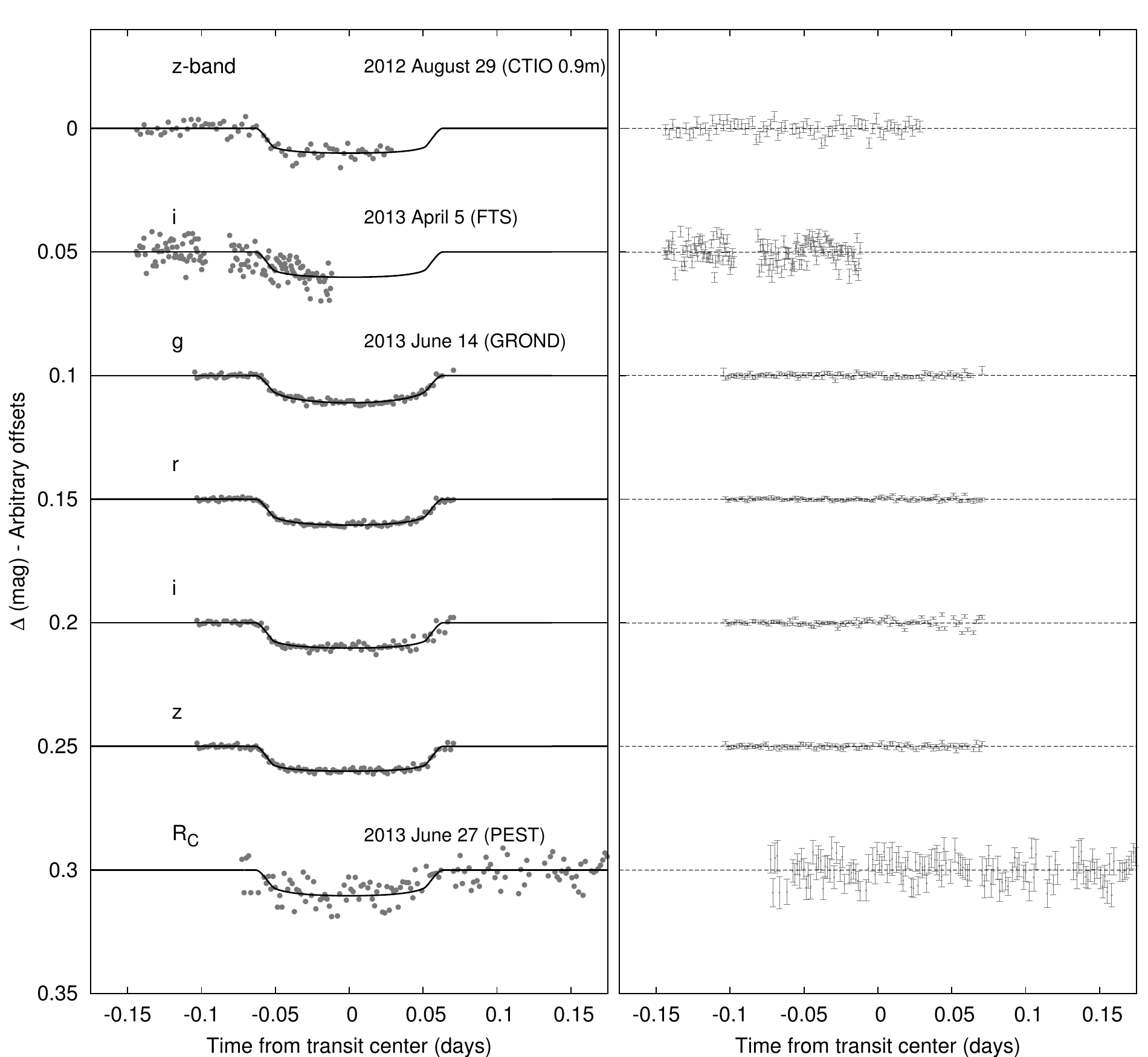}
\caption{
    Similar to \reffigl{lc9}; here we show the follow-up
    \lcs{} for \hatcur{10}.
}
\label{fig:lc10}
\end{figure*}


In order to confirm the occurrence of the transits and to better
constrain the orbital and physical parameters of the companions,
higher precision light curves for both candidates were acquired using
several telescopes around the globe. Table~\ref{tab:photobs}
summarizes the key aspects of this photometric follow-up, including
the dates of the observations, the cadence and the filter.

Two partial transits of \hatcur{9} were detected using the 0.3m Perth
Exoplanet Survey Telescope (PEST) and the Spectral camera on the 2m
Faulkes Telescope South (FTS), part of Las Cumbres Observatory Global Telescope
(LCOGT). Results of these observations are presented in \reftabl{phfu}
and shown in Figure~\ref{fig:lc9}.  Two partial transits of
\hatcur{10} were observed with FTS and the CTIO 0.9m
telescope. Another two full transits were measured with PEST and the GROND instrument on the MPG~2.2\,m.
These \hatcur{10} light curves are shown in
Figure~\ref{fig:lc10}.  All the facilities used for high precision
photometric follow-up have been previously used by HATSouth; the
instrument specifications, observation strategies and reduction
procedures adopted can be found in \cite{bayliss:2013:hats3},
\cite{zhou:2014:mebs}, \cite{hartman:2014:hats6} and
\cite{mohlerfischer:2013:hats2} for FTS, PEST, CTIO 0.9m, and GROND,
respectively.

\clearpage

\ifthenelse{\boolean{emulateapj}}{
    \begin{deluxetable*}{llrrrrl}
}{
    \begin{deluxetable}{llrrrrl}
}
\tablewidth{0pc}
\tablecaption{
    Light curve data for \hatcur{9} and \hatcur{10}\label{tab:phfu}.
}
\tablehead{
    \colhead{Object\tablenotemark{a}} &
    \colhead{BJD\tablenotemark{b}} & 
    \colhead{Mag\tablenotemark{c}} & 
    \colhead{\ensuremath{\sigma_{\rm Mag}}} &
    \colhead{Mag(orig)\tablenotemark{d}} & 
    \colhead{Filter} &
    \colhead{Instrument} \\
    \colhead{} &
    \colhead{\hbox{~~~~(2,400,000$+$)~~~~}} & 
    \colhead{} & 
    \colhead{} &
    \colhead{} & 
    \colhead{} &
    \colhead{}
}
\startdata
HATS-9 & $ 55744.07098 $ & $  -0.00037 $ & $   0.00552 $ & $   0.00000 $ & $ r$ &         HS\\
HATS-9 & $ 55749.81701 $ & $  -0.00018 $ & $   0.00572 $ & $   0.00000 $ & $ r$ &         HS\\
HATS-9 & $ 55780.46237 $ & $   0.00906 $ & $   0.00604 $ & $   0.00000 $ & $ r$ &         HS\\
HATS-9 & $ 55767.05534 $ & $  -0.01086 $ & $   0.00553 $ & $   0.00000 $ & $ r$ &         HS\\
HATS-9 & $ 55696.18926 $ & $   0.01787 $ & $   0.00581 $ & $   0.00000 $ & $ r$ &         HS\\
HATS-9 & $ 55657.88321 $ & $  -0.00168 $ & $   0.00549 $ & $   0.00000 $ & $ r$ &         HS\\
HATS-9 & $ 55726.83440 $ & $   0.00055 $ & $   0.00619 $ & $   0.00000 $ & $ r$ &         HS\\
HATS-9 & $ 55680.86732 $ & $   0.01840 $ & $   0.00534 $ & $   0.00000 $ & $ r$ &         HS\\
HATS-9 & $ 55788.12454 $ & $   0.00811 $ & $   0.00716 $ & $   0.00000 $ & $ r$ &         HS\\
HATS-9 & $ 55776.63287 $ & $  -0.00411 $ & $   0.00550 $ & $   0.00000 $ & $ r$ &         HS\\
\enddata
\tablenotetext{a}{
    Either HATS-9, or HATS-10.
}
\tablenotetext{b}{
    Barycentric Julian Date is computed directly from the UTC time
    without correction for leap seconds.
}
\tablenotetext{c}{
    The out-of-transit level has been subtracted. For observations
    made with the HATSouth instruments (identified by ``HS'' in the
    ``Instrument'' column) these magnitudes have been corrected for
    trends using the EPD and TFA procedures applied {\em prior} to
    fitting the transit model. This procedure may lead to an
    artificial dilution in the transit depths. For \hatcur{9} our fit
    is consistent with no dilution, for \hatcur{10} the HATSouth
    transit depth is $\sim 93$\% that of the true depth. For
    observations made with follow-up instruments (anything other than
    ``HS'' in the ``Instrument'' column), the magnitudes have been
    corrected for a quadratic trend in time fit simultaneously with
    the transit.
}
\tablenotetext{d}{
    Raw magnitude values without correction for the quadratic trend in
    time. These are only reported for the follow-up observations.
}
\tablecomments{
    This table is available in a machine-readable form in the online
    journal.  A portion is shown here for guidance regarding its form
    and content.
}
\ifthenelse{\boolean{emulateapj}}{
    \end{deluxetable*}
}{
    \end{deluxetable}
}

\section{Analysis}
\label{sec:analysis}

\subsection{Properties of the parent star}
\label{sec:stelparam}

We determine precise stellar parameters for \hatcur{9} and
\hatcur{10} using a new code called ZASPE (Zonal Atmospherical Stellar
Parameter Estimator) on median combined FEROS spectra. The detailed
structure and performance of ZASPE will be presented elsewhere (Brahm et al.~in preparation),
but in summary ZASPE is a Python-based code that computes the $\chi^2$
between the observed spectra and the PHOENIX grid of synthetic spectra
\citep{husser:2013} only on the most sensitive spectral zones to each
stellar parameter. The optimal set of stellar parameters
($\teffstar$,$\loggstar$, $\feh$ and $\vsini$) is found iteratively
and the sensitive zones are determined in each iteration.  One of the
most novel features of ZASPE is that the errors on the stellar
parameters are computed from the data itself and include the
systematic mismatches between the observations and the best fitted
model.  We have validated the results of ZASPE against a set of stars
with interferometrically determined stellar parameters \citep{boyajian:2012:afg} and which have
publicly available FEROS spectra. Results of this comparison are shown
in Figure~\ref{zaspe}.  The resulting parameters for \hatcur{9} are:
$\teffstar$=5363 $\pm$ 90 K, $\loggstar$=3.97 $\pm$ 0.2, $\feh$=0.33
$\pm$ 0.09, $\vsini$=4.67 $\pm$ 0.5 km/s; while for \hatcur{10} we
get: $\teffstar$=5974 $\pm$ 110 K, $\loggstar$=4.44 $\pm$ 0.13,
$\feh$=0.19 $\pm$ 0.07, $\vsini$=5.66 $\pm$ 0.5 km/s.

\ifthenelse{\boolean{emulateapj}}{
    \begin{figure*}[!ht]
}{
    \begin{figure}[!ht]
}
\plottwo{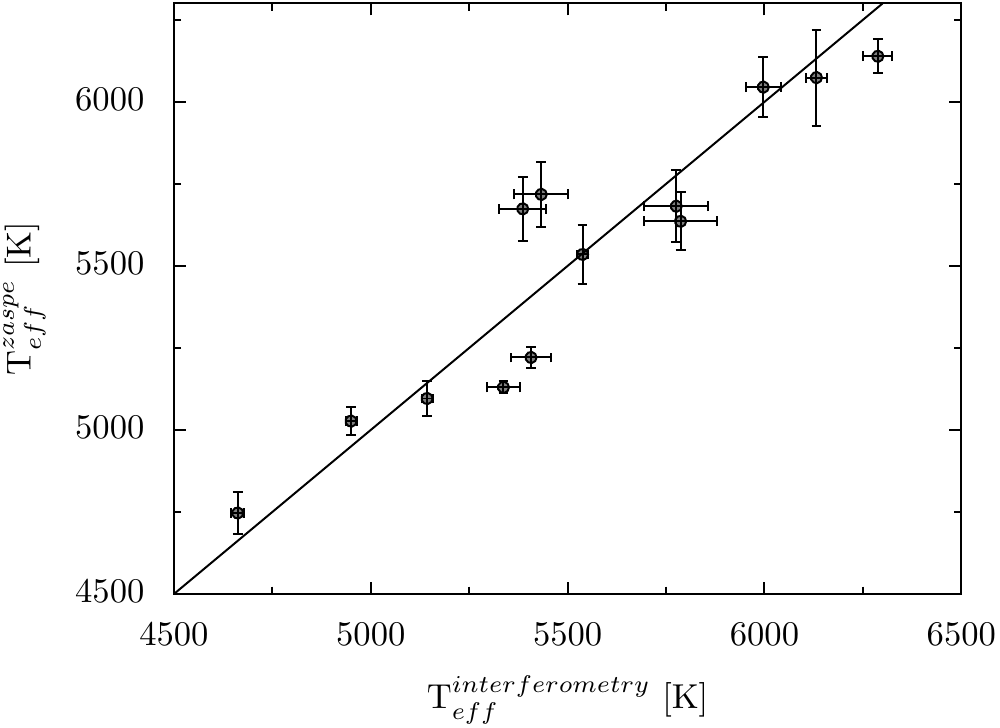}{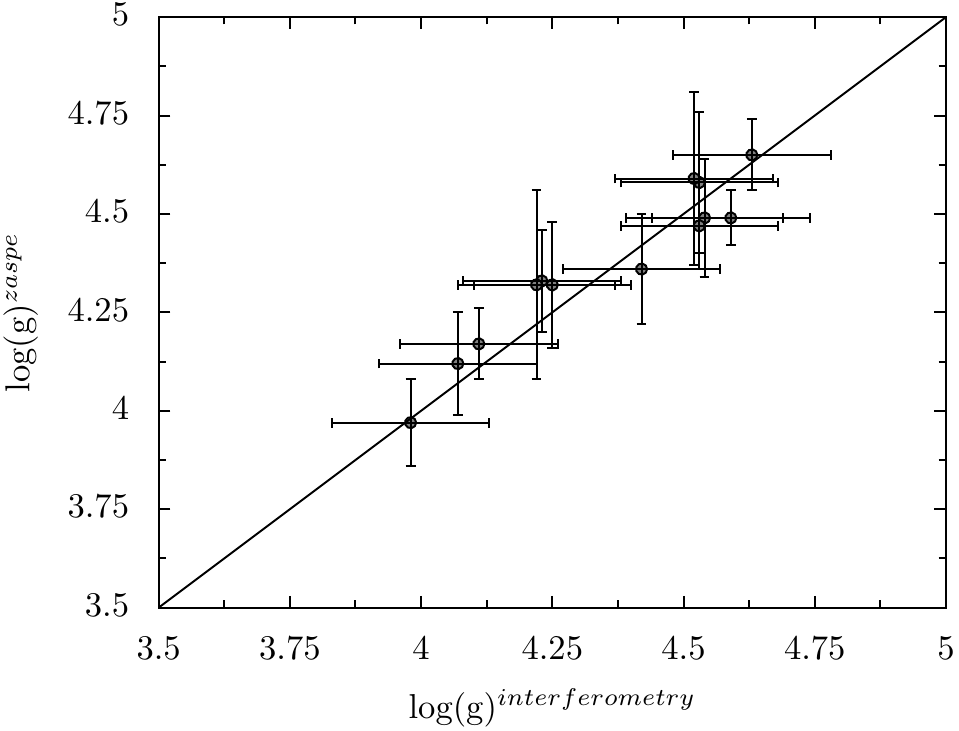}
\caption{
    (Left): Comparison between T$_{eff}$ values obtained with ZASPE and the ones derived from
    interferometric information. (Right): Comparison between log(g) values obtained with ZASPE and the ones derived from interferometric information. 
}
\label{zaspe}
\ifthenelse{\boolean{emulateapj}}{
    \end{figure*}
}{
    \end{figure}
}

These sets of stellar parameters were refined using the information
contained in the transit light-curves. The stellar mean density
($\rho_{\star}$) can be computed directly from one of the light-curve
model parameters ($a/R_{\star}$) and the period and eccentricity of the orbit by using
Kepler's third law with only a slight dependence on the stellar
parameters through the limb-darkening coefficients
\citep{sozzetti:2007}.  The spectroscopically determined $\teffstar$
and $\feh$ were coupled with $\rho_{\star}$ and Yonsei-Yale stellar
evolution models \citep[Y2;][]{yi:2001} to determine the stellar
physical parameters ($\rstar$, $\mstar$ and the age of the star),
which were used to compute a new and more precise estimation of
$\loggstar$ for \hatcur{9} ($\loggstar$=4.12 $\pm$ 0.04) and
\hatcur{10} ($\loggstar$=4.38 $\pm$ 0.03).  A new set of $\teffstar$,
$\feh$, $\vsini$ was determined using ZASPE with $\loggstar$ fixed to
the precise values obtained by modeling the light curves, followed by
a new estimation of $\rho_{\star}$ and a new modeling of stellar
isochrones. The new set of stellar parameters fixing $\loggstar$,
which are the ones we adopted for further analysis, were consistent
with the initial values quoted in the previous paragraph and are
listed in Table~\ref{tab:stellar}, where distances are determined by
comparing the measured broad-band photometry listed in that table to
the predicted magnitudes in each filter from the isochrones. We assume
a $R_{V} = 3.1$ extinction law from \citet{cardelli:1989} to determine
the extinction.  The 1$\sigma$ and 2$\sigma$ confidence ellipsoids in
$\teffstar$ and $\rho_{\star}$ are plotted in \reffigl{iso} for both
planet hosts, along with the Y2 isochrones for the ZASPE determined
$\feh$.  We find that \hatcur{9} is a \hatcurISOmlong{9} $M_{\odot}$, \hatcurISOrlong{9} $R_{\odot}$,
quite evolved (\hatcurISOage{9} Gyr) star, while \hatcur{10} is a \hatcurISOmlong{10} $M_{\odot}$,
\hatcurISOrlong{10} $R_{\odot}$ main-sequence star.

We attempted to measure the Lithium absorption line at 6707.8 \AA  \ for
testing the age estimation of \hatcur{9} but the quality of our spectra
was only enough to rule out a strong absorption feature.


\ifthenelse{\boolean{emulateapj}}{
    \begin{figure*}[!ht]
}{
    \begin{figure}[!ht]
}
\plottwo{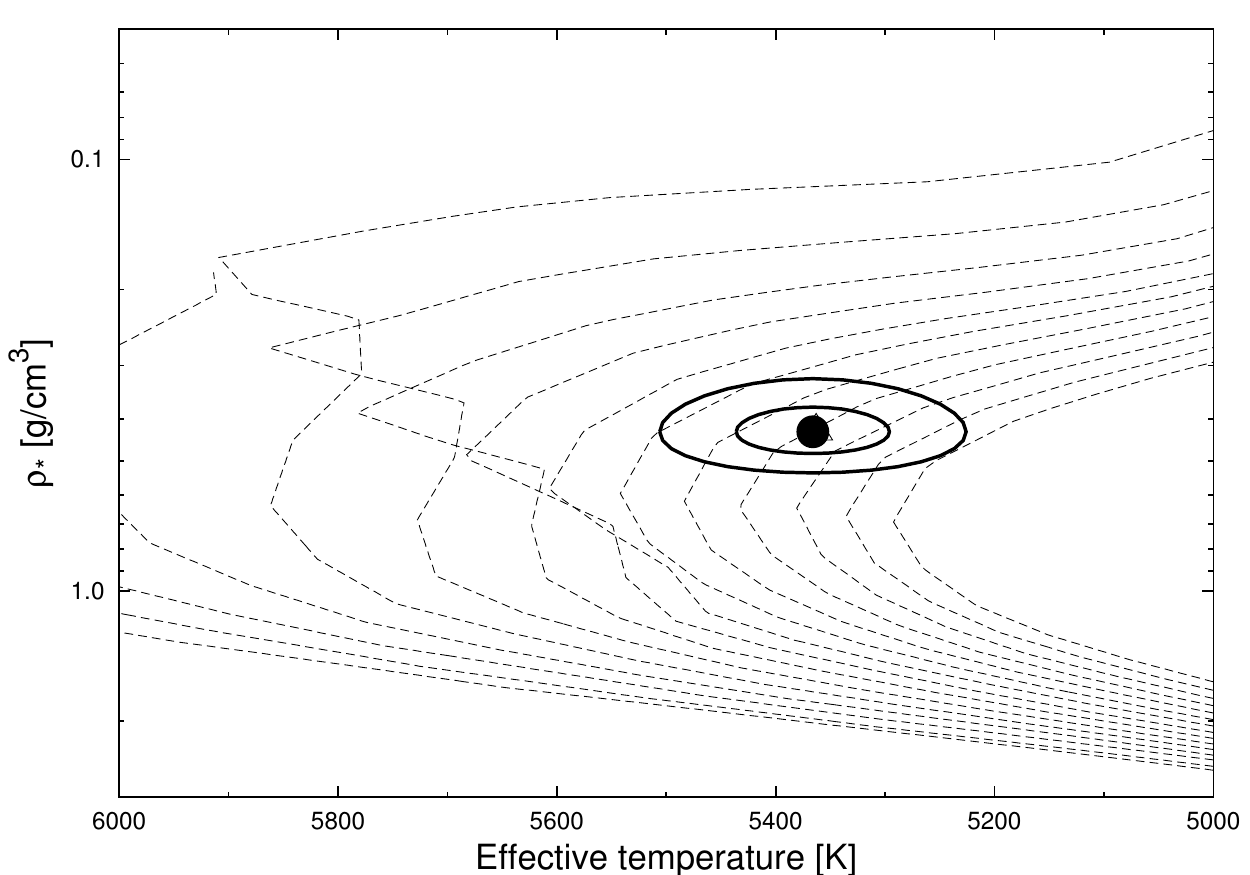}{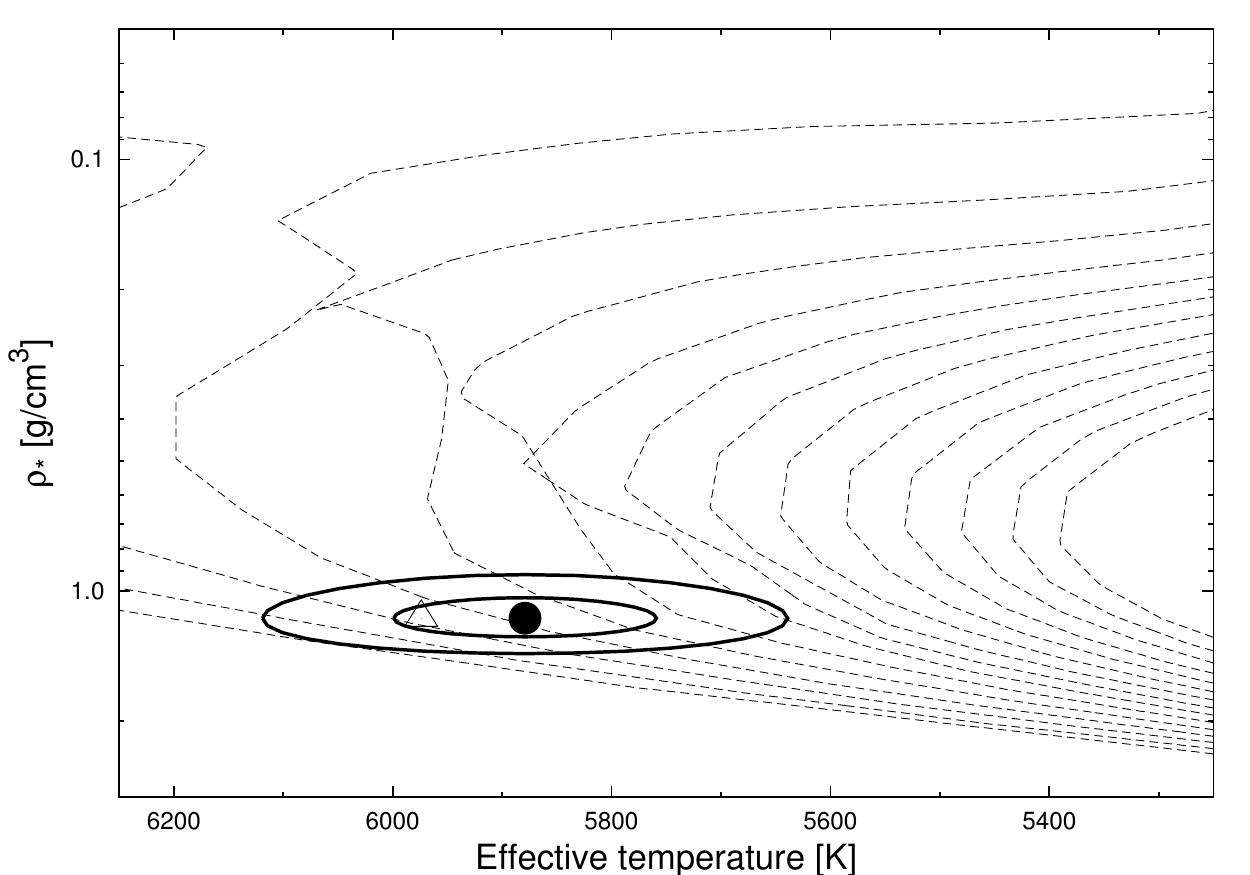}
\caption{
    Model isochrones from \cite{\hatcurisocite{9}} for the measured
    metallicities of \hatcur{9} (left) and \hatcur{10} (right). In each case we show models for ages of 0.2\,Gyr and 1.0 to 14.0\,Gyr in 1.0\,Gyr increments (ages increasing from left to right). The
    adopted values of $\teffstar$ and \rhostar\ are shown together with
    their 1$\sigma$ and 2$\sigma$ confidence ellipsoids.  The initial
    values of \teffstar\ and \rhostar\ from the first ZASPE and \lc\
    analyzes are represented with a triangle.
}
\label{fig:iso}
\ifthenelse{\boolean{emulateapj}}{
    \end{figure*}
}{
    \end{figure}
}

\ifthenelse{\boolean{emulateapj}}{
    \begin{deluxetable*}{lccl}
}{
    \begin{deluxetable}{lccl}
}
\tablewidth{0pc}
\tabletypesize{\scriptsize}
\tablecaption{
    Stellar parameters for \hatcur{9} and \hatcur{10}
    \label{tab:stellar}
}
\tablehead{
    \multicolumn{1}{c}{} &
    \multicolumn{1}{c}{\bf HATS-9} &
    \multicolumn{1}{c}{\bf HATS-10} &
    \multicolumn{1}{c}{} \\
    \multicolumn{1}{c}{~~~~~~~~Parameter~~~~~~~~} &
    \multicolumn{1}{c}{Value}                     &
    \multicolumn{1}{c}{Value}                     &
    \multicolumn{1}{c}{Source}
}
\startdata
\noalign{\vskip -3pt}
\sidehead{Astrometric properties and cross-identifications}

~~~~2MASS-ID\dotfill               & \hatcurCCtwomass{9}  & \hatcurCCtwomass{10} & \\
~~~~GSC-ID\dotfill                 & \hatcurCCgsc{9}      & \hatcurCCgsc{10}     & \\
~~~~R.A. (J2000)\dotfill            & \hatcurCCra{9}       & \hatcurCCra{10}    & 2MASS\\
~~~~Dec. (J2000)\dotfill            & \hatcurCCdec{9}      & \hatcurCCdec{10}   & 2MASS\\
~~~~$\mu_{\rm R.A.}$ (\masy)              & \hatcurCCpmra{9}     & \hatcurCCpmra{10}  & UCAC4\\
~~~~$\mu_{\rm Dec.}$ (\masy)              & \hatcurCCpmdec{9}    & \hatcurCCpmdec{10} & UCAC4\\
\sidehead{Spectroscopic properties}

~~~~$\teffstar$ (K)\dotfill         &  \hatcurSMEteff{9}   & \hatcurSMEteff{10} & ZASPE\tablenotemark{a}\\
~~~~$\feh$\dotfill                  &  \hatcurSMEzfeh{9}   & \hatcurSMEzfeh{10} & ZASPE                 \\
~~~~$\vsini$ (\kms)\dotfill         &  \hatcurSMEvsin{9}   & \hatcurSMEvsin{10} & ZASPE                 \\
~~~~$\vmac$ (\kms)\dotfill          &  4.6   & 3.8 & Assumed\tablenotemark{b}              \\
~~~~$\vmic$ (\kms)\dotfill          &  1.0   & 1.0 & Assumed\tablenotemark{c}              \\
~~~~$\gamma_{\rm RV}$ (\kms)\dotfill&  $-10.644 \pm 0.013$ & $-28.088 \pm 0.024$ & Coralie,FEROS  \\
\sidehead{Photometric properties}

~~~~$B$ (mag)\dotfill               &  \hatcurCCtassmB{9}  & \hatcurCCtassmB{10} & APASS\tablenotemark{d} \\
~~~~$V$ (mag)\dotfill               &  \hatcurCCtassmv{9}  & \hatcurCCtassmv{10} & APASS\tablenotemark{d} \\
~~~~$g$ (mag)\dotfill               &  \hatcurCCtassmg{9}  & \hatcurCCtassmg{10} & APASS\tablenotemark{d} \\
~~~~$r$ (mag)\dotfill               &  \hatcurCCtassmr{9}  & \hatcurCCtassmr{10} & APASS\tablenotemark{d} \\
~~~~$i$ (mag)\dotfill               &  \hatcurCCtassmi{9}  & \hatcurCCtassmi{10} & APASS\tablenotemark{d} \\
~~~~$J$ (mag)\dotfill               &  \hatcurCCtwomassJmag{9} & \hatcurCCtwomassJmag{10} & 2MASS           \\
~~~~$H$ (mag)\dotfill               &  \hatcurCCtwomassHmag{9} & \hatcurCCtwomassHmag{10} & 2MASS           \\
~~~~$K_s$ (mag)\dotfill             &  \hatcurCCtwomassKmag{9} & \hatcurCCtwomassKmag{10} & 2MASS           \\
\sidehead{Derived properties}
~~~~$\mstar$ ($\msun$)\dotfill      &  \hatcurISOmlong{9}   & \hatcurISOmlong{10} & YY+$\rhostar$+ZASPE \tablenotemark{e}\\
~~~~$\rstar$ ($\rsun$)\dotfill      &  \hatcurISOrlong{9}   & \hatcurISOrlong{10} & YY+$\rhostar$+ZASPE         \\
~~~~$\loggstar$ (cgs)\dotfill       &  \hatcurISOlogg{9}    & \hatcurISOlogg{10} & YY+$\rhostar$+ZASPE         \\
~~~~$\rhostar$ (\gcmc)\dotfill       &  \hatcurISOrho{9}    & \hatcurISOrho{10} & YY+$\rhostar$+ZASPE \tablenotemark{f}         \\
~~~~$\lstar$ ($\lsun$)\dotfill      &  \hatcurISOlum{9}     & \hatcurISOlum{10} & YY+$\rhostar$+ZASPE         \\
~~~~$M_V$ (mag)\dotfill             &  \hatcurISOmv{9}      & \hatcurISOmv{10} & YY+$\rhostar$+ZASPE         \\
~~~~$M_K$ (mag,\hatcurjhkfilset{9})\dotfill &  \hatcurISOMK{9} & \hatcurISOMK{10} & YY+$\rhostar$+ZASPE         \\
~~~~Age (Gyr)\dotfill               &  \hatcurISOage{9}     & \hatcurISOage{10} & YY+$\rhostar$+ZASPE         \\
~~~~$A_{V}$ (mag)\dotfill               &  \hatcurXAv{9}     & \hatcurXAv{10} & YY+$\rhostar$+ZASPE         \\
~~~~Distance (pc)\dotfill           &  \hatcurXdistred{9}\phn  & \hatcurXdistred{10} & YY+$\rhostar$+ZASPE\\ [-1.5ex]
\enddata
\tablenotetext{a}{
    ZASPE = **** routine for the analysis of high-resolution spectra
    (***cite***), applied to the FEROS spectra of \hatcur{9} and
    \hatcur{10}. These parameters rely primarily on ZASPE, but have a
    small dependence also on the iterative analysis incorporating the
    isochrone search and global modeling of the data, as described in
}
\tablenotetext{b}{
    Computed following \cite{valenti:2005}.
}
\tablenotetext{c}{
    \cite{husser:2013}.
}
\tablenotetext{d}{
    From APASS DR6 for \hatcur{9}, \hatcur{10} as listed in the UCAC 4 catalog \citep{zacharias:2012:ucac4}..
}
\tablenotetext{e}{
    \hatcurisoshort{9}+\rhostar+ZASPE = Based on the \hatcurisoshort{9}
    isochrones \citep{\hatcurisocite{9}}, \rhostar\ as a luminosity
    indicator, and the ZASPE results.
}
\tablenotetext{f}{
    In the case of $\rhostar$ the parameter is primarily determined
    from the global fit to the light curves and RV data. The value
    shown here also has a slight dependence on the stellar models and
    ZASPE parameters due to restricting the posterior distribution to
    combinations of $\rhostar$+$\teffstar$+$\feh$ that match to a
    \hatcurisoshort{9} stellar model.
}
\ifthenelse{\boolean{emulateapj}}{
    \end{deluxetable*}
}{
    \end{deluxetable}
}

\subsection{Excluding blend scenarios}
\label{sec:blend}

In order to exclude blend scenarios we carried out a blend analysis of the observations following \citet{hartman:2012:hat39hat41}.  For \hatcur{9} we find that scenarios involving blends between a stellar eclipsing binary and a foreground or background star can be ruled out with greater than 5$\sigma$ confidence based on the photometric data alone. The primary constraint in this case is the lack of out-of-transit variations seen in the HATSouth light curve. Due to the short orbital period, the best-fit blend model which reproduces the shape of the transit has a $\sim 1$\,mmag amplitude ellipsoidal variation, and a $\sim 0.5$\,mmag deep secondary eclipse, neither of which are detected in the HATSouth observations.  Moreover the Subaru/HDS observations of \hatcur{9} show no significant bisector span variation (the RMS scatter of the BS measurements is 12\,\ms) providing further evidence that the system is not a blended eclipsing binary.  For \hatcur{10} the photometric observations can be fit by a G+M star eclipsing binary blended with another G star that is slightly brighter than the primary in the eclipsing system. Based on the difference in $\chi^2$, this model is indistinguishable from a single G star with a transiting planet. We simulated spectra for blend models that could plausibly fit the photometric observations, finding that it all cases the blended systems would have easily been detected as having composite spectra. They also would produce RV and BS variations of several \kms, whereas the observed RV variation is \hatcurRVK{10}\,\ms, and the Subaru/HDS BS scatter is only 18\,\ms. We conclude that neither \hatcur{9} nor \hatcur{10} is a blended eclipsing binary system. As is often the case, however, we are not able to rule out the possibility that either transiting planet system has a fainter stellar-mass companion. For both systems a stellar companion of any mass, up to the mass of the planet-hosting star, is possible. If a massive stellar companion is present in a given system, the true planet radius would be up to $\sim 60$\% larger than inferred here. The planet mass would also be larger. High resolution adaptive optics imaging, and/or long-term RV observations are needed to determine whether either system has a stellar companion \citep[e.g.][]{howell:2011, horch:2014, everett:2015}.

\subsection{Global modeling of the data}
\label{sec:globmod}

We modeled the HATSouth photometry, the follow-up photometry, and the
high-precision RV measurements following
\citet{pal:2008:hat7,bakos:2010:hat11,hartman:2012:hat39hat41}. We fit
\citet{mandel:2002} transit models to the light curves, allowing for a
dilution of the HATSouth transit depth as a result of blending from
neighboring stars and over-correction by the trend-filtering
method. For the follow-up light curves we include a quadratic trend in
time in our model for each event to correct for systematic errors in
the photometry. We fit Keplerian orbits to the RV curves allowing the
zero-point for each instrument to vary independently in the fit, and
allowing for RV jitter which we we also vary as a free parameter for
each instrument. 

We used a Differential Evolution Markov Chain Monte
Carlo procedure \citep{terBraak:2006,eastman:2013} to explore the fitness landscape and to determine the
posterior distribution of the parameters.

The resulting parameters for each system are listed in
\reftabl{planetparam}.
\hatcurb{9} has a radius of $\hatcurPPrlong{9}$ $\rjup$
and a mass of $\hatcurPPmlong{9}$ $\mjup$, while \hatcurb{10}
has a radius of $\hatcurPPrlong{10}$ $\rjup$
and a mass of $\hatcurPPmlong{10}$ $\mjup$.
Both planets have bulk densities slightly lower than the one of Jupiter
($\hatcurPPrho{9}$ g cm$^{-3}$ and  $\hatcurPPrho{10}$ g cm$^{-3}$, respectively)

\ifthenelse{\boolean{emulateapj}}{
    \begin{deluxetable*}{lcc}
}{
    \begin{deluxetable}{lcc}
}
\tabletypesize{\scriptsize}
\tablecaption{Orbital and planetary parameters for \hatcurb{9} and \hatcurb{10}\label{tab:planetparam}}
\tablehead{
    \multicolumn{1}{c}{} &
    \multicolumn{1}{c}{\bf HATS-9b} &
    \multicolumn{1}{c}{\bf HATS-10b}\\
    \multicolumn{1}{c}{~~~~~~~~~~~~~~~Parameter~~~~~~~~~~~~~~~} &
    \multicolumn{1}{c}{Value} &
    \multicolumn{1}{c}{Value}
}
\startdata
\noalign{\vskip -3pt}
\sidehead{\Lc{} parameters}
~~~$P$ (days)             \dotfill    & $\hatcurLCP{9}$ & $\hatcurLCP{10}$ \\
~~~$T_c$ (${\rm BJD}$)    
      \tablenotemark{a}   \dotfill    & $\hatcurLCT{9}$ & $\hatcurLCT{10}$ \\
~~~$T_{14}$ (days)
      \tablenotemark{a}   \dotfill    & $\hatcurLCdur{9}$ & $\hatcurLCdur{10}$ \\
~~~$T_{12} = T_{34}$ (days)
      \tablenotemark{a}   \dotfill    & $\hatcurLCingdur{9}$ & $\hatcurLCingdur{10}$ \\
~~~$\arstar$              \dotfill    & $\hatcurPPar{9}$ & $\hatcurPPar{10}$ \\
~~~$\zrstar$ \tablenotemark{b}             \dotfill    & $\hatcurLCzeta{9}$\phn & $\hatcurLCzeta{10}$\phn \\
~~~$\rpl/\rstar$          \dotfill    & $\hatcurLCrprstar{9}$ & $\hatcurLCrprstar{10}$ \\
~~~$b^2$                  \dotfill    & $\hatcurLCbsq{9}$ & $\hatcurLCbsq{10}$ \\
~~~$b \equiv a \cos i/\rstar$
                          \dotfill    & $\hatcurLCimp{9}$ & $\hatcurLCimp{10}$ \\
~~~$i$ (deg)              \dotfill    & $\hatcurPPi{9}$\phn & $\hatcurPPi{10}$\phn \\

\sidehead{Limb-darkening coefficients \tablenotemark{c}}
~~~$c_1,g$ (linear term)    \dotfill    & $\cdots$ & $\hatcurLBig{10}$             \\
~~~$c_2,g$ (quadratic term) \dotfill    & $\cdots$ & $\hatcurLBiig{10}$            \\
~~~$c_1,r$                  \dotfill    & $\hatcurLBir{9}$ & $\hatcurLBir{10}$            \\
~~~$c_2,r$                  \dotfill    & $\hatcurLBiir{9}$ & $\hatcurLBiir{10}$            \\
~~~$c_1,i$                  \dotfill    & $\hatcurLBii{9}$ & $\hatcurLBii{10}$ \\
~~~$c_2,i$                  \dotfill    & $\hatcurLBiii{9}$ & $\hatcurLBiii{10}$ \\
~~~$c_1,z$                  \dotfill    & $\cdots$ & $\hatcurLBiz{10}$             \\
~~~$c_2,z$                  \dotfill    & $\cdots$ & $\hatcurLBiiz{10}$            \\
~~~$c_1,R$                  \dotfill    & $\hatcurLBiR{9}$ & $\hatcurLBiR{10}$             \\
~~~$c_2,R$                  \dotfill    & $\hatcurLBiiR{9}$ & $\hatcurLBiiR{10}$           \\

\sidehead{RV parameters}
~~~$K$ (\ms)              \dotfill    & $\hatcurRVK{9}$\phn\phn & $\hatcurRVK{10}$\phn\phn \\
~~~$e$ \tablenotemark{d}               \dotfill    & $\hatcurRVeccentwosiglimeccen{9}$ & $\hatcurRVeccentwosiglimeccen{10}$ \\

~~~RV jitter HDS (\ms) \tablenotemark{e}       \dotfill    & \hatcurRVjitterA{9} & \hatcurRVjitterC{10} \\
~~~RV jitter FEROS (\ms)        \dotfill    & \hatcurRVjitterB{9} & \hatcurRVjitterB{10} \\
~~~RV jitter Coralie (\ms)        \dotfill    & \hatcurRVjitterC{9} & \hatcurRVjitterA{10} \\

\sidehead{Planetary parameters}
~~~$\mpl$ ($\mjup$)       \dotfill    & $\hatcurPPmlong{9}$ & $\hatcurPPmlong{10}$ \\
~~~$\rpl$ ($\rjup$)       \dotfill    & $\hatcurPPrlong{9}$ & $\hatcurPPrlong{10}$ \\
~~~$C(\mpl,\rpl)$
    \tablenotemark{f}     \dotfill    & $\hatcurPPmrcorr{9}$ & $\hatcurPPmrcorr{10}$ \\
~~~$\rhopl$ (\gcmc)       \dotfill    & $\hatcurPPrho{9}$ & $\hatcurPPrho{10}$ \\
~~~$\log g_p$ (cgs)       \dotfill    & $\hatcurPPlogg{9}$ & $\hatcurPPlogg{10}$ \\
~~~$a$ (AU)               \dotfill    & $\hatcurPParel{9}$ & $\hatcurPParel{10}$ \\
~~~$T_{\rm eq}$ (K)        \dotfill   & $\hatcurPPteff{9}$ & $\hatcurPPteff{10}$ \\
~~~$\Theta$ \tablenotemark{g} \dotfill & $\hatcurPPtheta{9}$ & $\hatcurPPtheta{10}$ \\
~~~$\log_{10}\langle F \rangle$ (cgs) \tablenotemark{h}
                          \dotfill    & $\hatcurPPfluxavglog{9}$ & $\hatcurPPfluxavglog{10}$ \\ [-1.5ex]
\enddata
\tablenotetext{a}{
    Times are in Barycentric Julian Date calculated directly from UTC {\em without} correction for leap seconds.
    \ensuremath{T_c}: Reference epoch of
    mid transit that minimizes the correlation with the orbital
    period.
    \ensuremath{T_{14}}: total transit duration, time
    between first to last contact;
    \ensuremath{T_{12}=T_{34}}: ingress/egress time, time between first
    and second, or third and fourth contact.
}
\tablenotetext{b}{
   Reciprocal of the half duration of the transit used as a jump parameter in our MCMC analysis in place of $\arstar$. It is related to $\arstar$ by the expression $\zrstar = \arstar(2\pi(1+e\sin\omega))/(P\sqrt{1-b^2}\sqrt{1-e^2})$ \citep{bakos:2010:hat11}.
}
\tablenotetext{c}{
    Values for a quadratic law, adopted from the tabulations by
    \cite{claret:2004} according to the spectroscopic (ZASPE) parameters
    listed in \reftabl{stellar}.
}
\tablenotetext{d}{
    As discussed in \refsecl{globmod} the adopted parameters for all
    four systems are determined assuming circular orbits. We also list
    the 95\% confidence upper limit on the eccentricity determined
    when $\sqrt{e}\cos\omega$ and $\sqrt{e}\sin\omega$ are allowed to
    vary in the fit.
}
\tablenotetext{e}{
    Term added in quadrature to the formal RV uncertainties for each
    instrument. This is treated as a free parameter in the fitting
    routine.
}
\tablenotetext{f}{
    Correlation coefficient between the planetary mass \mpl\ and radius
    \rpl\ estimated from the posterior parameter distribution.
}
\tablenotetext{g}{
    The Safronov number is given by $\Theta = \frac{1}{2}(V_{\rm
    esc}/V_{\rm orb})^2 = (a/\rpl)(\mpl / \mstar )$
    \citep[see][]{hansen:2007}.
}
\tablenotetext{h}{
    Incoming flux per unit surface area, averaged over the orbit.
}
\ifthenelse{\boolean{emulateapj}}{
    \end{deluxetable*}
}{
    \end{deluxetable}
}

\section{Discussion}
\label{sec:discussion}
  
We have presented the discovery of two new transiting planets which
are shown on mass-radius and equilibrium temperature versus radius
diagrams in Figure~\ref{mrd}.  From the mass-radius diagram,
\hatcurb{9} and \hatcurb{10} can be classified as typical non-inflated
hot Jupiters. \hatcurb{9} is slightly less massive than Jupiter (0.84
$\mjup$) and has almost the same radius. Its orbital period of $P=1.9$
days is rather short compared to the period distribution of known hot
Jupiters. \hatcurb{10} has a mass in the range between Saturn and
Jupiter (0.53 $\mjup$), a radius consistent with that of Jupiter
and a period of $P=3.3$ days, which is close to the mean period of
known hot Jupiters.

The equilibrium temperature versus radius diagram shows that both
planets tend to depart from the known correlation between the planet
radius and its degree of irradiation. This correlation, first proposed
in \cite{guillot:2005}, indicates that the inflated radius of some hot
Jupiters can be at least partially explained by the enhanced
insolation from their parent star.  \hatcurb{9} has a moderately high
equilibrium temperature ($T_{\rm eq}$=$\hatcurPPteff{9}$K) due to the
small star-planet separation coupled to the large stellar radius,
while \hatcurb{10} has a more typical equilibrium temperature for a
hot Jupiter ($T_{\rm eq}$=$\hatcurPPteff{10}$K).  According to the
empirical relations proposed in \cite{enoch:2012}, which give the
radius of a giant planet from its equilibrium temperature and
semi-major axis, \hatcurb{9} and \hatcurb{10} should have radii of
1.36$\rjup$ and 1.22$\rjup$, respectively. The observed radii are
$3\sigma$ and $5\sigma$ below these values, which indicates that these
planets are very compact given their irradiation levels and that thus
additional variables must be responsible of setting the radii of short
period giant planets.

One possible explanation is that \hatcurb{9} and \hatcurb{10} may have
significant amounts of heavy elements in their cores. According to the
interior models of \cite{fortney:2007}, both planets will require a
core mass of $\sim$60 M$_\oplus$ to explain their radii based on their
masses, stellar host masses and orbital periods for an age of 4.5
Gyr. This explanation can be further motivated by the relatively high
metallicity of their parent stars (\hatcurSMEzfeh{9} dex and
\hatcurSMEzfeh{10} dex, respectively).  Several works
\citep{guillot:2006, burrows:2007, enoch:2011, enoch:2012} have proposed a
correlation between the inferred core mass of giant planets and the
metallicity of the parent star. The principal idea behind the proposed
correlation is that a more metal rich proto-planetary disk will be
more efficient in creating massive cores following the core-accretion
scenario of planetary formation. Even though this process is expected
to occur in the formation and migration steps, the final relation
between the stellar metallicity and the radius of giant planets is not
at all clear and other phenomena can act in the opposite direction. As
shown by \cite{burrows:2007}, the presence of heavy elements in the
atmosphere of young giant planets will increase its opacity, slowing the
contraction and making the planetary radius more inflated than
expected. Moreover, the validity of the proposed correlation has been
put into question by the analysis \cite{zhou:2014:hats5} who find no
significant correlation between $R_p$ and [Fe/H] for the
complete sample of detected giant TEPs.

The age of the system may be another important variable, since the
radius of giant planets should undergo Kelvin-Helmholtz contraction as
they age, controlled by their upper radiative atmosphere
\citep{hubbard:1977}.
Figure~\ref{fig:irradiated} presents the mass-radius diagram of transiting
hot Jupiters having similar insolation levels to \hatcurb{9} 
(1750K $<$ $T_{\rm eq}$ $<$ 1900K).
This figure shows that in general the bloating of the atmosphere
of strongly irradiated planets is prevented for more massive hot Jupiters.
This correlation presents some outliers, with \hatcurb{9} the most extreme one.
A peculiarity of \hatcurb{9} is the advanced
age of the system ($\sim$11 Gyr) contrasted with the ages of the rest of
the planets in Figure~\ref{fig:irradiated} ($<$5 Gyr).
Among the complete sample of well characterized hot Jupiters,
\hatcurb{9} and CoRoT-17 b (10.7 $\pm$ 1.0 Gyr) are the oldest
systems known to have an age uncertainty better than 20\%.
Figure \ref{age} shows the radius as function of age for hot Jupiters
with 0.5$\mjup<\mpl<$2$\mjup$ and orbital period P $<$ 10 days
having age uncertainties smaller than 40\%. Systems older than
3 Gyr exhibit the expected contraction of the envelope through time
but most of them are systematically more inflated than expected
from theoretical models of structure and evolution.
By fitting a straight line through the planets with ages higher than 3 Gyr
we obtain an empirical contraction function for hot Jupiters:
$\rpl=1.45-0.03t$, where $t$ is the age of the system
in Gyr. The difference between the theoretical function and the empirical relation
decreases with the age of the system and for the case of \hatcurb{9}
both functions are consistent with the observed values.
The proposed empirical relation between the age of the system and
the radius of the planet shown in Figure \ref{age} supports the study of \cite{burrows:2007} where
for young giant planets the higher opacity produced by heavy elements delays
the contraction, while at later ages the higher mean molecular weight dominates
and leads to smaller radii. However, in order to perform a precise study
of the evolution of the the radii of giant extrasolar planets,
particular models with the properties of each system should be constructed.

A possible confusing factor in Figures 7, 8 and 9 is the
assumption of zero albedo and complete heat redistribution. The
measurement of secondary transits on these systems in different
wavelengths will be informative for explaining the departure of
\hatcurb{9} from the correlation.  A more precise determination of the
radius of \hatcurb{9} is also required. The somewhat larger uncertainty in the
radius is a result of the incomplete photometric follow-up for this
system. The errors in the planet radius are governed at this point by
the light-curve data, but future precise measurements of the transit
of \hatcurb{9} will be able to lower this uncertainty until it becomes
dominated by the uncertainties on the stellar parameters.

Future precise RV measurements of \hatcurb{10} are required to
determine a more precise mass of the planet and to explain the high
jitter measured with FEROS and Coralie with respect to Subaru/HDS.
One possible explanation may be the presence of another planetary
companion.  Subaru/HDS observations, which don't seem to show enhanced
jitter, were performed in three continuous days, while Coralie and
FEROS observations were separated by months and in this case the
influence of a second more distant companion should be stronger.
The jitter values quoted in Table~\ref{tab:planetparam} refer to
RV uncertainties for each instrument  that have to be added in quadrature
to the formal RV errors in order for them to be consistent with the RV signal computed
with the orbital parameters of the system.

\subsection{K2 possibilities}

Even thought \hatcurb{9} and \hatcurb{10} are located in the nominal
coordinates of field 7 of K2, only \hatcurb{9} falls on working
silicon. A proposal to observe this star in short cadence was recently
submitted. The high photometric precision of K2 will allow
us to estimate a much more precise radius for \hatcurb{9}, which will
help us in determining if this planet is a true outlier in the
correlation between planet radius, equilibrium temperature and planet
mass. The high insolation of this planet makes it a very good target
for measuring secondary transits and phase curve variations with K2,
which will allow us to estimate the albedo and provide a more reliable
estimate of its equilibrium temperature. Figure~\ref{fig:K2} shows a
measure of the reflected light signature , $(R_P/a)^2$, for hot
Jupiters observed by {\em Kepler} as a function of planetary radius.
From this Figure we can see that the potential of detecting reflected
light signatures of \hatcurb{9} is high and its amplitude should be similar
to the one of the giant planets observed by {\em Kepler} so far. 
Other subtle photometric effects, like ellipsoidal variations, Doppler beaming and the
measurement of asteroseismological frequencies, if present, will also
be very valuable for the detailed characterization of this particular
planet.

\ifthenelse{\boolean{emulateapj}}{
    \begin{figure*}[!ht]
}{
    \begin{figure}[!ht]
}
\plottwo{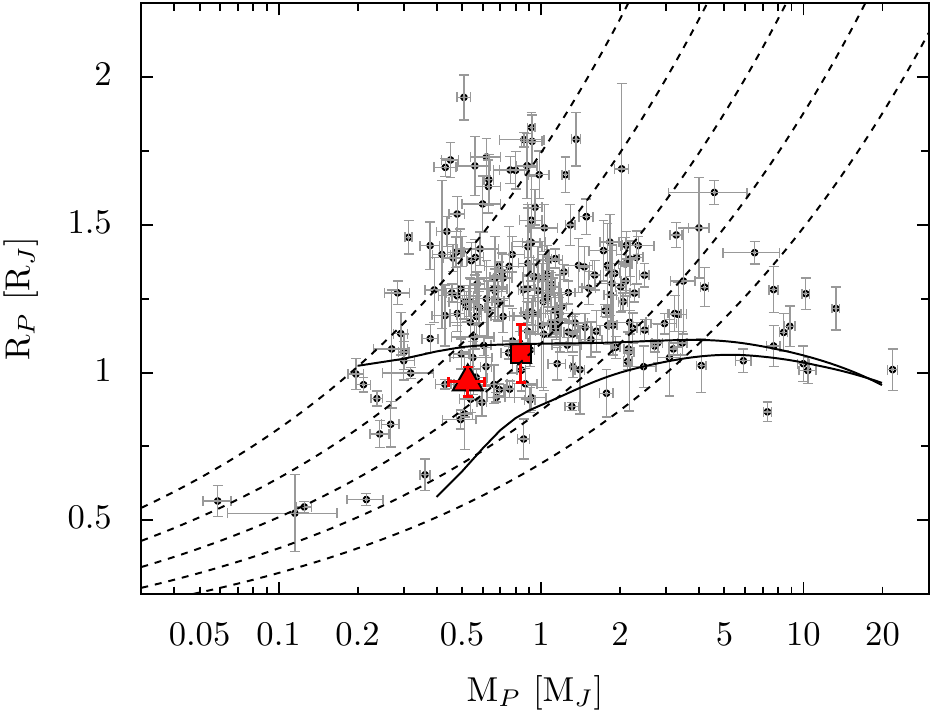}{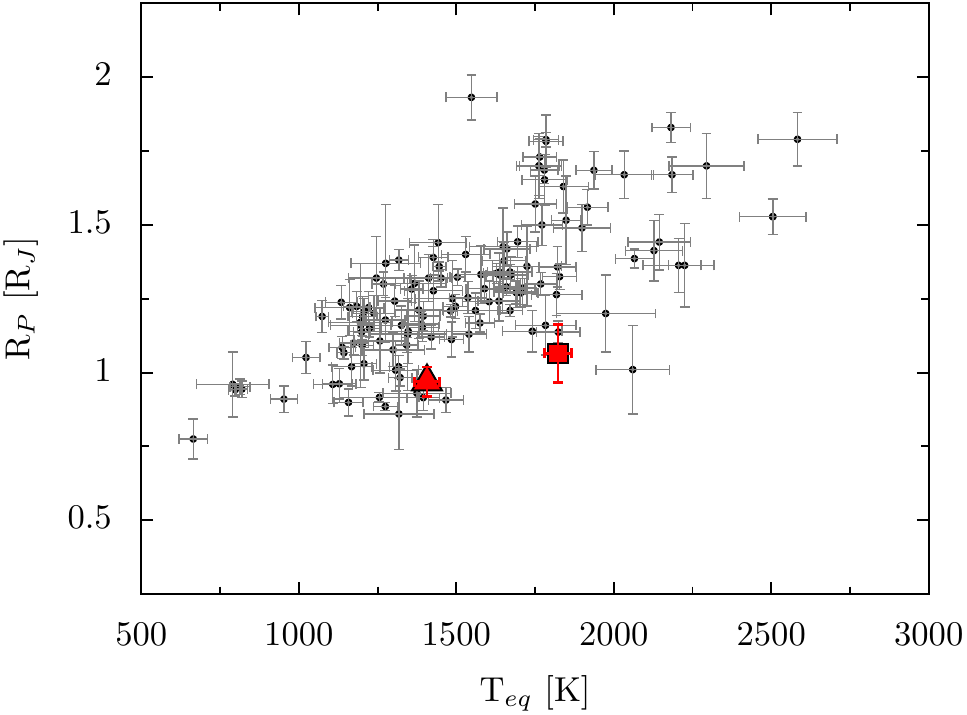}
\caption{
    (Left): Mass-radius diagram of giant TEPs. \hatcurb{9} is marked with a filled square and
    \hatcurb{10} with a filled triangle. Isodensity curves are plotted with dashed lines
    for $\rho_P=\{0.25, 0.5, 1.0, 2.0,4.0\}$ gr cm$^3$ and the 4.5 Gyr isochrones \citep{fortney:2007}
    for core masses of 0 and 100 M$_{\oplus}$ with solid lines. (Right): Equilibrium temperature
    versus radius diagram for giant TEPs. Again, \hatcurb{9} is marked with a filled square and
    \hatcurb{10} with a filled triangle. 
}
\label{mrd}
\ifthenelse{\boolean{emulateapj}}{
    \end{figure*}
}{
    \end{figure}
}

\ifthenelse{\boolean{emulateapj}}{
    \begin{figure*}[!ht]
}{
    \begin{figure}[!ht]
}
\plotone{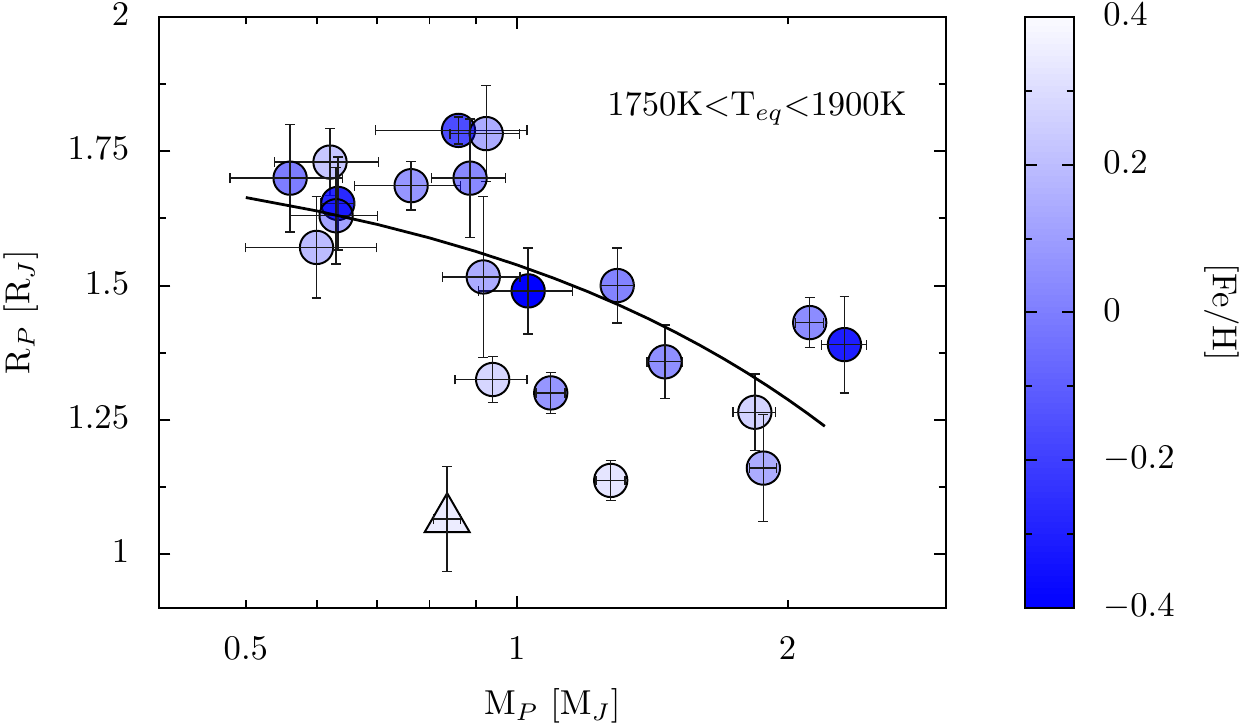}
\caption{
    Mass-radius diagram of giant TEPs having similar insolation levels to \hatcurb{9} (1750K$<$ T$_{eq}$$<$1900K). \hatcurb{9} is marked with a triangle. Filled symbols are colored according to the metallicity of the host star. \hatcurb{9} does not follow the correlation formed by the other hot Jupiters with similar irradiation levels.
}
\label{fig:irradiated}
\ifthenelse{\boolean{emulateapj}}{
    \end{figure*}
}{
    \end{figure}
}

\ifthenelse{\boolean{emulateapj}}{
    \begin{figure*}[!ht]
}{
    \begin{figure}[!ht]
}
\plotone{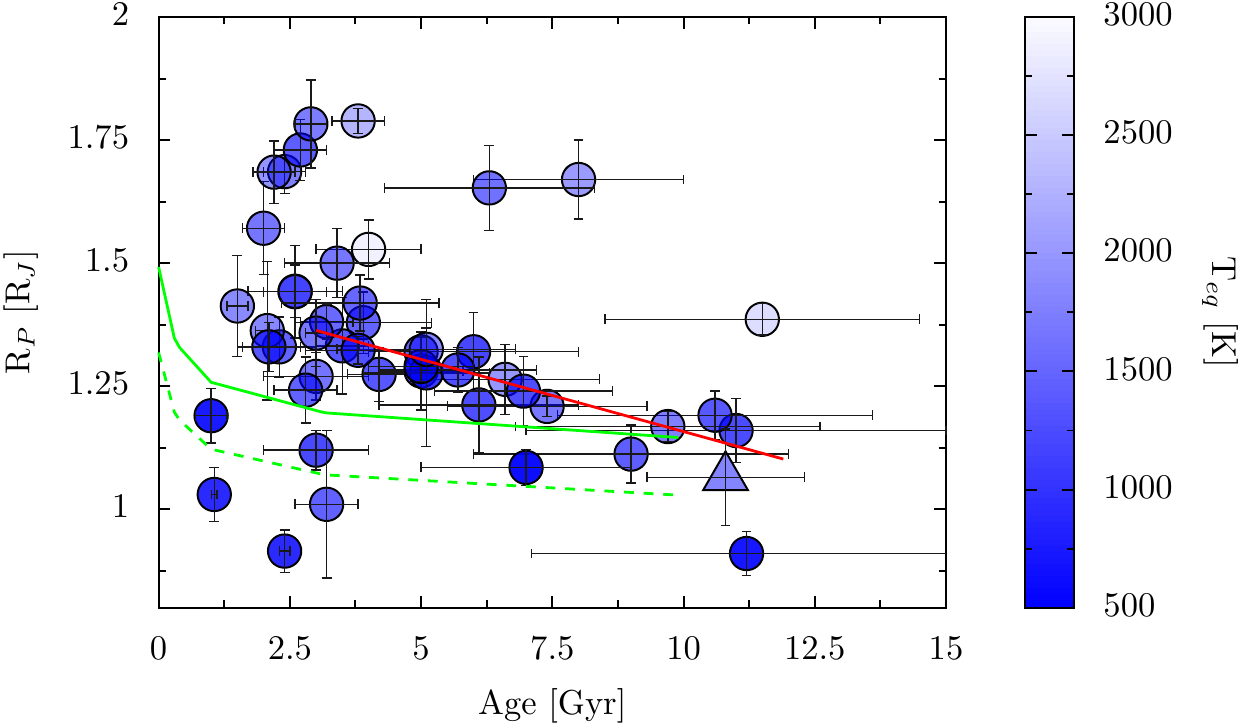}
\caption{
    Radius as function of the age of the system for hot Jupiters having $0.5\mjup<\mpl<2\mjup$, P$<$10 days and
    an age estimation with a precision better than 40\%. Green lines are the theoretical models of \cite{fortney:2007} 
    for $\mpl=1\mjup$, $a = 0.02$AU and a core mass of 0 (dashed) and 50 (solid) times the mass of the earth. The 
    red line is an empirical relation computed with these data points.  \hatcurb{9} is marked with a triangle. Hot Jupiters
    older than 3 Gyr follow the contraction of their radius over time but the observed contraction rate is steeper than 
    the one predicted from the theoretical models. The theoretical radii for Hot Jupiters with ages greater than ~10 Gyr
    (like \hatcurb{9}) is consistent with the observations. 
}
\label{age}
\ifthenelse{\boolean{emulateapj}}{
    \end{figure*}
}{
    \end{figure}
}

\ifthenelse{\boolean{emulateapj}}{
    \begin{figure*}[!ht]
}{
    \begin{figure}[!ht]
}
\plotone{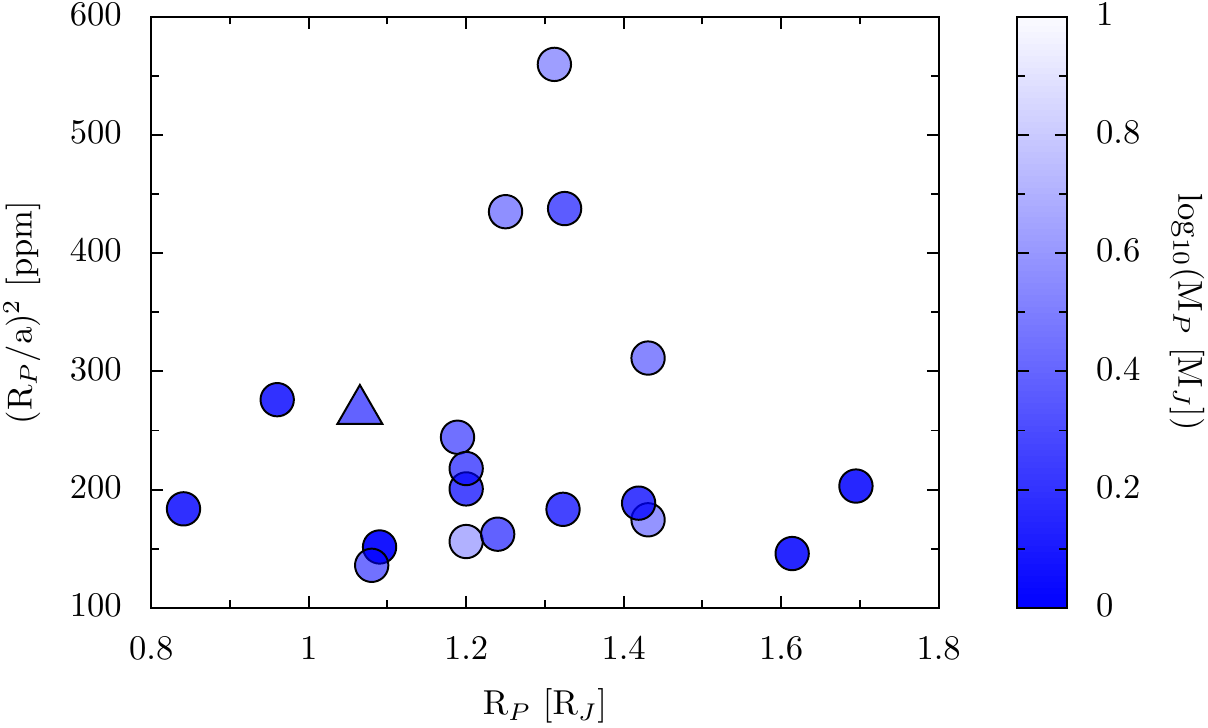}
\caption{
    Reflected light signature as function of the planetary radius for the hot Jupiters observed with Kepler. The symbols are colored according to the planetary mass. \hatcurb{9} is marked with a triangle. Given that the photometric precision of K2 is similar to the one of the original {\em Kepler} mission, phase curve variations and secondary
the secondary transit of  \hatcurb{9} should be measured by K2.
}
\label{fig:K2}
\ifthenelse{\boolean{emulateapj}}{
    \end{figure*}
}{
    \end{figure}
}


\acknowledgements 

Development of the HATSouth project was funded by NSF MRI grant
NSF/AST-0723074, operations have been supported by NASA grants
NNX09AB29G / NNX12AH91H and internal Princeton funds. Follow-up observations receive partial
support from grant NSF/AST-1108686.
A.J.\ acknowledges support from FONDECYT project 1130857, BASAL CATA
PFB-06, and project IC120009 ``Millennium Institute of Astrophysics
(MAS)'' of the Millenium Science Initiative, Chilean Ministry of
Economy. R.B.\ and N.E.\ are supported by CONICYT-PCHA/Doctorado
Nacional. R.B.\ and N.E.\ acknowledge additional support from project
IC120009 ``Millenium Institute of Astrophysics (MAS)'' of the
Millennium Science Initiative, Chilean Ministry of Economy.  V.S.\
acknowledges support form BASAL CATA PFB-06. 
K.P. acknowledges support from NASA grant NNX13AQ62G.
This work is based on observations made with ESO Telescopes at the La
Silla Observatory.
This paper also uses observations obtained with facilities of the Las
Cumbres Observatory Global Telescope.
Work at the Australian National University is supported by ARC Laureate
Fellowship Grant FL0992131.
Operations at the MPG~2.2\,m Telescope are jointly performed by the
Max Planck Gesellschaft and the European Southern Observatory.  The
imaging system GROND has been built by the high-energy group of MPE in
collaboration with the LSW Tautenburg and ESO\@.  We thank R\'egis
Lachaume for his technical assistance during the observations at the
MPG~2.2\,m Telescope. We thank Helmut Steinle and Jochen Greiner for
supporting the GROND observations presented in this manuscript.
We are grateful to P.Sackett for her help in the early phase of the
HATSouth project.
We thank Adam Burrows for his useful comments regarding the evolutionary models of hot Jupiters.

\clearpage
\bibliographystyle{apj}
\bibliography{hatsbib}

\clearpage
\LongTables


\end{document}